\documentclass[aps,prb,reprint,groupedaddress]{revtex4-2}

\usepackage{graphicx}
\usepackage{amsmath,amssymb}
\usepackage{mathrsfs}

\begin{document}



\title{Nonequilibrium nonlinear response theory of amplitude-dependent dissipative conductivity in disordered superconductors}


\author{Takayuki Kubo}
\email[]{kubotaka@post.kek.jp}
\affiliation{High Energy Accelerator Research Organization (KEK), Tsukuba, Ibaraki 305-0801, Japan}
\affiliation{The Graduate University for Advanced Studies (Sokendai), Hayama, Kanagawa 240-0193, Japan}



\begin{abstract}
This work investigates amplitude-dependent nonlinear corrections to the dissipative conductivity in superconductors, using the Keldysh--Usadel theory of nonequilibrium superconductivity, which captures the nonequilibrium dynamics of both quasiparticles and the pair potential.
Our rigorous formulation naturally incorporates both the direct nonlinear action of the photon field and indirect contributions mediated by nonequilibrium variations in the pair potential, namely the Eliashberg effect and the Higgs mode.
The third-harmonic current, often regarded as a hallmark of the Higgs mode in disordered superconductors, arises from both the direct photon action and the Higgs mode.  
Our numerical results are in excellent agreement with previous studies.  
In contrast, the first-harmonic current, and consequently the dissipative conductivity, receives contributions from all three mechanisms: the direct photon action, the Higgs mode, and the Eliashberg effect.
It is shown that that the nonlinear correction to dissipative conductivity in dirty-limit superconductors can serve as a fingerprint of the Higgs mode, appearing as a resonance peak at a frequency near the superconducting gap \( \Delta \).  
In addition, our results provide microscopic insight into amplitude-dependent dissipation at frequencies well below \( \Delta \), which is particularly relevant for applied superconducting devices.  
In particular, the long-standing issue concerning the frequency dependence of the amplitude-dependent quality factor is explained as originating from the direct nonlinear action of the photon field, rather than from contributions by the Higgs mode and the Eliashberg effect.
Our practical and explicit expression for the nonlinear conductivity formula makes our results accessible to a broad range of researchers.
\end{abstract}

\maketitle


\section{Introduction}\label{intro}

The nonlinear electromagnetic response in superconductors has attracted considerable interest from a fundamental physics standpoint, particularly due to its connection with the Higgs mode.  
This mode, \( \delta \Delta(t) \), represents collective oscillations in the amplitude of the superconducting order parameter \( \Delta \)~\cite{Anderson, Volkov_Kogan, Shimano_review, Tsuji_review, 2013_Matsunaga, 2014_Matsunaga, Tsuji_Aoki, 2018_Jujo, Silaev, Murotani, Tsuji_Nomura, Seibold, Eremin, Dzero}, and couples to electromagnetic fields via a term of the form \( {\bf A} \cdot {\bf A} \, \delta \Delta \).  
This interaction is analogous to the coupling of the Higgs boson \( h \) to two \( Z \) bosons through the \( Z Z h \) vertex in particle physics.  
As a result, observation of the Higgs mode typically requires either a sufficiently strong ac field that drives the system into the nonlinear regime~\cite{2013_Matsunaga, 2014_Matsunaga}, or a dc bias current that induces a vertex of the form \( {\bf A}_{\rm dc} \cdot {\bf A} \, \delta \Delta \), enabling linear excitation of the mode~\cite{Moor, Nakamura, Jujo, 2024_Kubo, 2025_Kubo, 2025_Wang}.

The Higgs mode, when driven by a sufficiently strong ac field, induces a nonlinear current response of the form \( \sim A\, \delta \Delta(A^2) \), which constitutes a third-order nonlinear contribution, \( \mathcal{O}(A^3) \), to the current.  
This term gives rise to a third-harmonic component in the current, \( J_{3\rm H} \), which exhibits a resonance-like peak near \( \hbar\omega_{\rm ac} \simeq \Delta \) in disordered superconductors and is widely regarded as a hallmark of the Higgs mode (note that in the clean limit, the contribution of the Higgs mode is significantly suppressed, so the observed peak cannot be attributed solely to the Higgs mode~\cite{Seibold,Silaev}). 
For this reason, the nonlinear electromagnetic response at high frequencies, with \( \hbar \omega_{\rm ac} \) on the order of \( \Delta \), has been a major focus of interest in the superconductivity community over the past decade.

On the other hand, from an applied perspective, the nonlinear electromagnetic response in superconductors at low frequencies, $\hbar\omega_{\rm ac}\ll \Delta$, has attracted attention for a very different reason: practical necessity.  
Superconducting resonators made from conventional \( s \)-wave superconductors exhibit extremely low electromagnetic dissipation at frequencies $\hbar\omega_{\rm ac}\ll \Delta$, and state-of-the-art devices have achieved ultra-high quality factors exceeding \( 10^{11} \) (see e.g., Refs.~\cite{Kubo_R1, 2020_Romanenko}). 
This technology has also been adopted in three-dimensional circuit quantum electrodynamics architectures for quantum information processing (e.g., Refs.~\cite{Milul, Takenaka_Kubo}).
Despite this remarkable performance, the microscopic mechanisms underlying electromagnetic dissipation are not yet fully understood.
In particular, the nonlinear evolution of the quality factor as a function of the amplitude of the applied ac field remains poorly understood.  
This contrasts with the linear-response regime, where theoretical frameworks are well established. These include not only the classical theories~\cite{MB, Nam, Zimmermann}, but also recent advances tailored to realistic device materials~\cite{Gurevich_Kubo}, supported by surface characterization studies~\cite{Iavarone}.
To gain insight into the amplitude dependence and to identify possible routes for further improving the quality factor,  
it is essential to develop a theoretical framework that captures the relevant nonlinear phenomena beyond the standard linear-response formalism.

A nonlinear theory of ac-amplitude-dependent dissipative conductivity in the low-frequency limit \( \hbar\omega_{\rm ac} \to 0 \) was developed in Refs.~\cite{2014_Gurevich, Kubo_Gurevich}, incorporating a {\it nonperturbative} effect of a slowly varying ac field \( A \) on the superconducting order parameter \( \Delta \).  
This theory predicts that increasing the ac amplitude leads to a logarithmic suppression of the real part of the complex conductivity \( \sigma_1 \), which originates from the broadening of the quasiparticle density of states (DOS) peak at the gap edge caused by the well-established current-induced weak pair breaking~\cite{Maki, Maki_book}.
Consequently, the electromagnetic dissipation decreases, and the quality factor \( Q \) of a superconducting resonator reaches a maximum at a certain field strength.  
This nonmonotonic behavior and the logarithmic dependence have been confirmed experimentally~\cite{2014_Ciovati}.
However, the theory has a notable limitation: it fails to account for experimental observations indicating that the amplitude-dependent enhancement of the quality factor \( Q \) becomes more pronounced at higher resonator frequencies~\cite{Martinello, Martinello650, Dhakal}. 
Resolving this discrepancy has remained a long-standing challenge in the field; see Ref.~\cite{Dhakal} for a detailed discussion.

In this paper, we conduct a rigorous analysis of the ac-amplitude dependence of dissipative conductivity in disordered superconductors, employing the Keldysh-Usadel theory of nonequilibrium superconductivity~\cite{Larkin, Kopnin, Rammer, Sauls}. 
We perturbatively incorporate nonlinear corrections to the current response up to order \( \mathcal{O}(A^3) \), where \( A \) denotes the amplitude of the ac field.  
Within this framework, both the Eliashberg effect and the Higgs mode naturally arise as \( \mathcal{O}(A^2) \) nonequilibrium variations of the superconducting pair potential.  
The resulting nonlinear current response consists of two components: the third-harmonic current and the correction to the first-harmonic current.  
The third-harmonic current comprises two contributions: the direct nonlinear action of the photon field and the contribution mediated by the Higgs mode, which is often regarded as a hallmark of the Higgs resonance in disordered superconductors.
In contrast, the correction to the first-harmonic current, which has received less attention, contains all three contributions: the direct nonlinear photon action, the Higgs-mode-mediated term, and the Eliashberg contribution.  
Importantly, the nonlinear correction to the dissipative conductivity originates from this first-harmonic current correction.  
Our analysis reveals that the direct nonlinear photon action plays a central role in explaining the experimentally observed trend that the ac-amplitude-induced enhancement of the quality factor \( Q \) becomes more pronounced at higher resonator frequencies.

It should be noted that a limitation of this \textit{perturbative} approach is that it cannot capture \textit{nonperturbative} effects of the ac field, such as those on the quasiparticle spectrum~\cite{2014_Gurevich, Kubo_Gurevich, Semenov}.
As a result, the theory breaks down at high ac field intensities, where ac-induced pair-breaking leads to significant modifications of the DOS, which in turn dominate the correction to the dissipative conductivity.  
A nonperturbative extension of the present theory remains an important future challenge for establishing a nonlinear \( Q \) theory valid up to the depairing ac current.

On the other hand, as long as the ac current (or ac field) remains well below the depairing current (or the thermodynamic critical field), the theory developed in this work remains fully valid and provides a robust framework for understanding nonlinear dissipation and the evolution of the \( Q \)-factor in superconducting resonators.  
This condition is typically satisfied in most practical superconducting devices~\cite{Zmuidzinas, cQED}.  
An exception is superconducting accelerator cavities, which often operate at more than 50\% of the depairing current or the thermodynamic critical field (see, e.g., Refs.~\cite{Gurevich_Review, Kubo_RRR}). 
Nonetheless, their low-field behavior can still be described within the scope of the present theory.

The remainder of this paper is organized as follows.  
In Section~II, we first introduce the Keldysh-Usadel equations~\cite{Larkin, Kopnin, Rammer, Sauls}, which have been widely applied in the literature to study nonequilibrium superconductivity~\cite{Eremin, Dzero, Moor, 2024_Kubo, 2025_Kubo, Dzero_2}. 
We then solve these equations under a monochromatic ac field and explicitly derive the nonlinear current response formula in terms of the equilibrium Green's functions.
Although the resulting formulas are lengthy, they are straightforward to implement and accessible to a broad range of researchers.  
In Section~III, we apply this framework to analyze the nonlinear correction to the dissipative conductivity and show that the theory accounts for the experimentally observed trend in which the ac-amplitude-induced enhancement of the quality factor \( Q \) becomes more pronounced at higher resonator frequencies.  
Finally, Section~IV summarizes our findings and discusses their broader implications.

\section{Theoretical framework} \label{formulation}

\subsection{Keldysh-Usadel equations}

We investigate the nonlinear response of a thin-film superconductor in the dirty limit, subjected to an ac field \( \mathbf{A}(t) \), or equivalently, to the associated superfluid momentum \( \hbar \mathbf{q} = \hbar \nabla \chi - 2e \mathbf{A} \), taking into account contributions up to third order in the ac amplitude.  
To analyze the nonequilibrium dynamics in this regime, we employ the Keldysh-Usadel formalism, a widely used theoretical framework for describing nonequilibrium diffusive superconductors (see, e.g., Refs.~\cite{Kopnin, Rammer, Sauls, Rainer_Sauls}). 
Within this formalism, the quasiclassical Green's function is represented in Keldysh space as
\begin{eqnarray}
\begin{pmatrix}
\hat{g}^{R} & \hat{g}^{K} \\
0           & \hat{g}^{A}
\end{pmatrix} ,
\end{eqnarray}
where each of \( \hat{g}^{R}(\epsilon, t) \), \( \hat{g}^{A}(\epsilon, t) \), and \( \hat{g}^{K}(\epsilon, t) \) is a \( 2 \times 2 \) matrix in Nambu space.

The equations governing the retarded (R) and advanced (A) components of the Green's functions, along with their normalization conditions, are given by
\begin{eqnarray}
&&i\hbar D \hat{\partial} \circ \big\{ \hat{g}^{r} \circ (\hat{\partial} \circ \hat{g}^{r}) \big\}
=[\epsilon \hat{\tau}_3 + \hat{\Delta} , \hat{g}^{r}]_{\circ} , \label{RA1} \\
&&\hat{g}^r \circ \hat{g}^r =  1 , \label{normRA1}
\end{eqnarray}
where $D$ is the electron diffusion constant and $r = R, A$. 
Here, $\hat{\tau}_3$ denotes the third Pauli matrix, and $\hat{\Delta}$ represents the superconducting gap.
The Keldysh component $\hat{g}^K$ satisfies a similar set of equations:
\begin{eqnarray}
&&i\hbar D \hat{\partial} \circ \big\{ \hat{g}^{R} \circ (\hat{\partial} \circ \hat{g}^{K}) + \hat{g}^K \circ  (\hat{\partial} \circ \hat{g}^{A})  \big\} \nonumber \\
&&=  [\epsilon \hat{\tau}_3 + \hat{\Delta} , \hat{g}^{K}]_{\circ} ,  \label{K1} \\
&&\hat{g}^R \circ \hat{g}^K + \hat{g}^K \circ  \hat{g}^A=0 . \label{normK1}
\end{eqnarray}
Here, the $\circ$ product denotes the Moyal product, defined as
$X \circ Y = \exp\left[(i\hbar/2)\left(\partial_\epsilon^X \partial_t^Y - \partial_\epsilon^Y \partial_t^X\right)\right] XY$,
and $[X, Y]_{\circ}$ represents the Moyal commutator, given by $X\circ Y - Y\circ X$. 
Under the influence of a gauge-invariant superfluid momentum $\hbar {\bf q}$, the operator $\hat{\partial} \circ$ becomes:
$\hat{\partial} \circ \hat{g}^{R, A, K} =  (i/2) [ \hat{\tau}_3 {\bf q} , \hat{g}]_{\circ}$. 
The gap equation governing the superconducting order parameter $\Delta$ is obtained from the Keldysh component and is given by
\begin{eqnarray}
\Delta(t) = -\frac{{\mathscr G}}{8}\int d\epsilon {\rm Tr} [(-i\tau_2) \hat{g}^K(\epsilon, t)]  , \label{gap1}
\end{eqnarray}
where ${\mathscr G}$ is the BCS coupling constant. 
The current density ${\bf J}(t)$ in the superconductor is given by
\begin{eqnarray}
&& {\bf J}(t) = \frac{\sigma_n}{8e}\int \!\!d\epsilon {\bf S}(\epsilon, t) , \label{general_J} \\
&& {\bf S}(\epsilon, t)=  {\rm Tr} \bigl[ \hat{\tau}_3 
\bigl\{ \hat{g}^{R} \circ (\hat{\partial} \circ \hat{g}^{K}) + \hat{g}^K \circ  (\hat{\partial} \circ \hat{g}^{A}) \bigr\} \bigr]  ,  \label{general_S} 
\end{eqnarray}
where $\sigma_n=2e^2 N_0 D$ is the normal conductivity and $N_0$ is the normal density of states at the Fermi level.

Eqs.~(\ref{RA1})-(\ref{general_S}) constitute the general framework of the Keldysh-Usadel formalism for describing nonequilibrium superconductivity. 
Within this framework, we investigate the nonlinear response to an ac field, considering contributions up to third order in the ac field.

\subsection{Keldysh-Usadel Equations under a Monochromatic ac Field}

Our strategy for solving the Keldysh--Usadel equations is to incorporate nonlinear corrections perturbatively up to third order in the photon field~\cite{Eremin}.  
We carry out the analysis in the \( (\epsilon, t) \) space and its Fourier counterpart, the \( (\epsilon, \omega) \) space, following the approach used in Refs.~\cite{Rainer_Sauls, Sauls, 2024_Kubo, 2025_Kubo}.  
Note that here \( \omega \) denotes the Fourier frequency, whereas \( \omega_{\rm ac} \) represents the physical frequency of the ac field in what follows.

We consider a monochromatic ac field ${\bf q}(t) = {\bf q}_0 \cos (\omega_{\rm ac} t)$. 
The Green's functions are decomposed as $\hat{g}^{R,A,K}(\epsilon, t) = \hat{g}^{R,A,K}_e (\epsilon) + \delta \hat{g}^{R,A,K}(\epsilon, t)$, where $\hat{g}^{R,A,K}_e(\epsilon)$ denotes the equilibrium component in the absence of the ac field, and $\delta \hat{g}^{R,A,K}(\epsilon, t)$ represents the nonequilibrium corrections induced by the ac field. 
Similarly, the superconducting order parameter is expressed as $\hat{\Delta}(t) = \hat{\Delta}_e + \delta \hat{\Delta}(t)$, where $\hat{\Delta}_e$ is the equilibrium value in the absence of the ac field, and $\delta \hat{\Delta}(t)$ denotes the time-dependent fluctuation of the order parameter.

For the calculations that follow, it is convenient to express $\cos(\omega_{\rm ac} t)$ using a dummy variable $\eta$: ${\bf q}(t)= ({\bf q}_0/2) \sum_{\eta=\pm \omega_{\rm ac}} e^{i \eta t}$. 
Utilizing this form, we obtain
\begin{eqnarray}
&&\hat{\partial} \circ \hat{X} = \frac{i{\bf q}_0}{2} [\hat{\tau}_3 \cos\omega_{\rm ac} t , \hat{X}]_{\circ} \nonumber \\
&&= \frac{i{\bf q}_0}{4} \sum_{\eta=\pm \omega_{\rm ac}} e^{i \eta t} 
\Bigl\{ \hat{\tau}_3 \hat{X} \Bigl(\epsilon+\frac{\hbar \eta}{2}, t\Bigr)
 - \hat{X}\Bigl(\epsilon-\frac{\hbar \eta}{2}, t\Bigr) \hat{\tau}_3 \Bigr\} , \nonumber \\ \label{dX}
\end{eqnarray}
where $\hat{X}(\epsilon, t)$ represents either $\hat{g}_e^{R, A, K}(\epsilon)$ or $\delta \hat{g}^{R, A, K}(\epsilon, t)$.

For notational convenience, we define
\begin{eqnarray}
s := \frac{\hbar D}{2} q_0^2 = \biggl( \frac{q_0}{q_{\xi}} \biggr)^2 \Delta_0,
\end{eqnarray}
which characterizes the strength of the ac superflow. 
Here, $q_{\xi}:=\xi^{-1}=\sqrt{2\Delta_0/\hbar D}$. 
Note that \( s \) is an \( \mathcal{O}(q_0^2) \) quantity and plays a central role in the subsequent perturbative expansion.

We now formulate the Keldysh-Usadel equations under a monochromatic ac field. 
Applying Eq.~(\ref{dX}) and discarding terms of $\mathcal{O}(q_0^4)$, Eq.~(\ref{RA1}) leads to the equations governing the $R$ and $A$ components in Fourier space:
\begin{eqnarray}
&&-i\frac{s}{8} \sum_{\eta=\pm\omega_{\rm ac}} \sum_{\eta'=\pm \omega_{\rm ac}} 2\pi \delta(\omega+ \eta + \eta') \times \nonumber \\
&&\Bigl[ \hat{\tau_3} \hat{g}_e^r \Bigl( \epsilon - \frac{\hbar \eta}{2} + \frac{\hbar \eta'}{2} \Bigr) \hat{\tau}_3 \hat{g}_e^r \Bigl( \epsilon + \frac{\hbar \eta}{2} + \frac{\hbar \eta'}{2} \Bigr) 
\nonumber \\
&&- \hat{g}_e^r \Bigl( \epsilon - \frac{\hbar \eta}{2} - \frac{\hbar \eta'}{2} \Bigr) \hat{\tau}_3 \hat{g}_e^r \Bigl( \epsilon + \frac{\hbar \eta}{2} - \frac{\hbar \eta'}{2} \Bigr) \hat{\tau}_3 \Bigr] 
\nonumber \\
&=& \Bigl( \epsilon +\frac{\hbar \omega}{2} \Bigr) \hat{\tau}_3 \delta \hat{g}^r (\epsilon,\omega) 
- \Bigl( \epsilon -\frac{\hbar \omega}{2} \Bigr) \delta \hat{g}^r (\epsilon,\omega) \hat{\tau}_3 
\nonumber \\
&& + \delta \hat{\Delta}(\omega) \hat{g}_e^r \Bigl(\epsilon-\frac{\hbar \omega}{2} \Bigr) 
- \hat{g}_e^r \Bigl(\epsilon+\frac{\hbar \omega}{2} \Bigr) \delta \hat{\Delta}(\omega) \nonumber \\
&&+ [ \hat{\Delta}_e , \delta \hat{g}^r(\epsilon,\omega) ]  . \label{RA2}
\end{eqnarray}
Eq.~(\ref{normRA1}) yields the normalization conditions for the $R$ and $A$ components in Fourier space: 
\begin{eqnarray}
\hat{g}^{r}_{e} \Bigl(\epsilon + \frac{\hbar\omega}{2} \Bigr) \delta \hat{g}^{r}(\epsilon, \omega) 
+ \delta \hat{g}^{r}(\epsilon, \omega) \hat{g}^{r}_{e}\Bigl(\epsilon- \frac{\hbar\omega}{2}\Bigr) =0 .  \label{normRA2}
\end{eqnarray}
Eq.~(\ref{K1}) yields the equations governing the $K$ components in Fourier space, utilizing Eq.~(\ref{dX}) and neglecting terms of $\mathcal{O}(q_0^4)$:
\begin{eqnarray}
&&-i\frac{s}{8} \sum_{\eta=\pm\omega_{\rm ac}} \sum_{\eta'=\pm\omega_{\rm ac}} 2\pi \delta(\omega+ \eta + \eta') \times \nonumber \\
&&\Bigl[ 
\hat{\tau_3} \hat{g}_e^R \Bigl( \epsilon - \frac{\hbar \eta}{2} + \frac{\hbar \eta'}{2} \Bigr) \hat{\tau}_3 \hat{g}_e^K \Bigl( \epsilon + \frac{\hbar \eta}{2} + \frac{\hbar \eta'}{2} \Bigr) 
\nonumber \\
&&- \hat{g}_e^R \Bigl( \epsilon - \frac{\hbar \eta}{2} - \frac{\hbar \eta'}{2} \Bigr) \hat{\tau}_3 \hat{g}_e^K \Bigl( \epsilon + \frac{\hbar \eta}{2} - \frac{\hbar \eta'}{2} \Bigr) \hat{\tau}_3 
\nonumber \\
&&+\hat{\tau_3} \hat{g}_e^K \Bigl( \epsilon - \frac{\hbar \eta}{2} + \frac{\hbar \eta'}{2} \Bigr) \hat{\tau}_3 \hat{g}_e^A \Bigl( \epsilon + \frac{\hbar \eta}{2} + \frac{\hbar \eta'}{2} \Bigr) 
\nonumber \\
&&- \hat{g}_e^K \Bigl( \epsilon - \frac{\hbar \eta}{2} - \frac{\hbar \eta'}{2} \Bigr) \hat{\tau}_3 \hat{g}_e^A \Bigl( \epsilon + \frac{\hbar \eta}{2} - \frac{\hbar \eta'}{2} \Bigr) \hat{\tau}_3 
\Bigr] 
\nonumber \\
&=& \Bigl( \epsilon +\frac{\hbar \omega}{2} \Bigr) \hat{\tau}_3 \delta \hat{g}^K (\epsilon,\omega) 
- \Bigl( \epsilon -\frac{\hbar \omega}{2} \Bigr) \delta \hat{g}^K (\epsilon,\omega) \hat{\tau}_3 
\nonumber \\
&& + \delta \hat{\Delta}(\omega) \hat{g}_e^K \Bigl(\epsilon-\frac{\hbar \omega}{2} \Bigr) 
- \hat{g}_e^K \Bigl(\epsilon+\frac{\hbar \omega}{2} \Bigr) \delta \hat{\Delta}(\omega) \nonumber \\
&&+ [ \hat{\Delta}_e , \delta \hat{g}^K(\epsilon,\omega) ]  . \label{K2}
\end{eqnarray}
Eq.~(\ref{normK1}) yields the normalization condition for the $K$ component in Fourier space:  
\begin{eqnarray}
&&\hat{g}_{e}^R\Bigl(\epsilon +\frac{\hbar\omega}{2}\Bigr) \delta \hat{g}^K (\epsilon, \omega) 
+ \delta \hat{g}^K  (\epsilon, \omega) \hat{g}_{e}^A\Bigl(\epsilon -\frac{\hbar\omega}{2}\Bigr)  \nonumber \\
&&+ \hat{g}_{e}^K\Bigl(\epsilon +\frac{\hbar\omega}{2}\Bigr)  \delta \hat{g}^A (\epsilon, \omega)  
+\delta \hat{g}^R (\epsilon, \omega) \hat{g}_{e}^K\Bigl(\epsilon -\frac{\hbar\omega}{2}\Bigr)   =0 .\nonumber \\
 \label{normK2}
\end{eqnarray}
The gap equation for the pair-potential variation reduces to
\begin{eqnarray}
\delta \Delta(\omega) = -\frac{{\mathscr G}}{8}\int d\epsilon {\rm Tr} [(-i\tau_2) \delta \hat{g}^K(\epsilon, \omega)]  . \label{gap2}
\end{eqnarray}
To solve these equations, it is useful to express $\delta \hat{g}^K$ in terms of the {\it anomalous} term $\delta \hat{g}^a$, following Eliashberg's approach~\cite{Gorkov, Eliashberg}:  
\begin{eqnarray}
\delta \hat{g}^K (\epsilon, \omega) = \delta \hat{g}^R \mathcal{T}\Bigl(\epsilon-\frac{\hbar\omega}{2}\Bigr) - \delta \hat{g}^A \mathcal{T}\Bigl(\epsilon+\frac{\hbar\omega}{2}\Bigr) + \delta \hat{g}^a , 
\nonumber \\ \label{gAno1}
\end{eqnarray}
where $\mathcal{T}(\epsilon) := \tanh \left( \epsilon / 2k_B T \right) = 1 - 2f_{\rm FD}(\epsilon)$ is the equilibrium distribution function, with $f_{\rm FD}(\epsilon)$ representing the Fermi-Dirac distribution.

With this formulation, all the necessary equations are now in place. 
The set of equations to be solved comprises Eqs.~(\ref{RA2})-(\ref{gAno1}), which will be addressed in the following subsection.

\subsection{Analytical solutions} \label{analytical_solution}

In Eqs.~(\ref{RA2}) and (\ref{K2}), the dummy frequency variables \( (\eta, \eta') \) take the values \( (\pm \omega_{\rm ac}, \pm \omega_{\rm ac}) \), which lead to frequency components at \( \omega = 0 \) and \( \omega = \pm 2\omega_{\rm ac} \) through the factor \( \delta(\omega + \eta + \eta') \).  
Accordingly, \( \delta \hat{g}^{R,A,K} \) and \( \delta \hat{\Delta} \) can be decomposed into the sum of a zero harmonic (time-independent) component and second harmonic components as follows:
\begin{eqnarray}
\delta \hat{g}^{R,A,K}(\epsilon, \omega) 
&=& \delta \hat{g}^{R,A,K}_{\rm 0H} 2\pi \delta(\omega) 
+ \delta \hat{g}^{R,A,K}_{\rm 2H} 2\pi \delta(\omega -2 \omega_{\rm ac}) \nonumber \\
&&+ \delta \hat{g}^{R,A,K}_{\rm -2H} 2\pi \delta(\omega+2 \omega_{\rm ac})  , \label{dg0H2H}
\end{eqnarray}
and
\begin{eqnarray}
\delta \hat{\Delta}(\omega) 
&=& \delta \hat{\Delta}_{\rm 0H} 2\pi \delta(\omega) 
+ \delta \hat{\Delta}_{\rm 2H}  2\pi \delta(\omega -2 \omega_{\rm ac}) \nonumber \\
&& + \delta \hat{\Delta}_{\rm -2H}  2\pi \delta(\omega+2 \omega_{\rm ac})  . \label{deltaDelta_spectrum}
\end{eqnarray}
Here, the zero harmonic $\delta \Delta_{\rm 0H}$ corresponds to the time-averaged pair-potential variation under ac irradiation, representing the (positive and negative) {\it Eliashberg effect}, while the second harmonics $\delta \Delta_{\rm \pm 2H}$ correspond to the {\it Higgs mode} driven by the nonlinear ac irradiation.

It should be noted that this decomposition can be understood by observing that \( {\bf q}^2 = {\bf q}_0^2 \cos^2(\omega_{\rm ac} t) \propto 1 + \cos(2\omega_{\rm ac} t) \) contains both zero-harmonic (\( \omega = 0 \)) and second-harmonic (\( \omega = \pm 2\omega_{\rm ac} \)) components.

In the following, we solve Eqs.~(\ref{RA2})-(\ref{gAno1}) using Eqs.~(\ref{dg0H2H}) and (\ref{deltaDelta_spectrum}) and explicitly express the solutions in terms of the matrix components of the Green's functions.  
This formulation is intended to make the results more accessible to a broader research community, including those who may not be familiar with the present theoretical formalism.  
To this end, we express the matrix components as 
$\hat{\Delta} = i\hat{\tau}_2 \Delta$, 
$\hat{g}^R= \hat{\tau}_3 G +  i\hat{\tau}_2 F$, 
and $\hat{g}^A=-\hat{\tau}_3\hat{g}^{R\dagger} \hat{\tau}_3$.

\subsubsection{Zero harmonic component}

We begin by summarizing the zero harmonic solutions (see also Appendix~\ref{appendix_1}). 
The equations governing the $R$ and $A$ components are derived from Eq.~(\ref{RA2}) along with the normalization condition in Eq.~(\ref{normRA2}) [see also Eqs.~(\ref{RA_0H}) and (\ref{normRA_0H})]. 
These yield the following relations:
\begin{eqnarray}
&& \delta G_{\rm 0H} = \frac{F_2 + F_{-2}}{G_2 + G_{-2}} \delta F_{\rm 0H}, \label{dG0H}\\
&& \delta F_{\rm 0H} = \zeta_{\rm 0H} \delta \Delta_{\rm 0H} + s \kappa_{\rm 0H} , \\
&& \zeta_{0H} = \frac{G^2}{\epsilon G - \Delta_e F} , \\
&& \kappa_{0H} = -\frac{i}{8} \frac{ G \{ (F_2 + F_{-2})G + (G_2 + G_{-2})F \} }{\epsilon G - \Delta_e F} .
\end{eqnarray}
Here, \( G(\epsilon) = (\epsilon + i\Gamma) / \sqrt{(\epsilon + i\Gamma)^2 - \Delta^2} \) and \( F(\epsilon) = \Delta / \sqrt{(\epsilon + i\Gamma)^2 - \Delta^2} \) denote the equilibrium Green's functions.  
The damping parameter \( \Gamma \) is treated as a small constant, and throughout the main text we take \( \Gamma / \Delta_0 = 10^{-3} \).
We also define \( G_{\pm n} := G(\epsilon \pm n \hbar \omega_{\rm ac} / 2) \) and \( F_{\pm n} := F(\epsilon \pm n \hbar \omega_{\rm ac} / 2) \) for notational convenience, with \( n \in \mathbb{N} \).
Similarly, the equations governing the $K$ component are obtained from Eq.~(\ref{K2}) and the normalization condition in Eq.~(\ref{normK2}) [see also Eqs.~(\ref{K_0H})-(\ref{gAno_0H})]. 
These yield the following expressions:
\begin{eqnarray}
&& \delta G^a_{\rm 0H} = \frac{F_2 - F_{-2}^*}{G_2 - G_{-2}^*} \delta F^a_{\rm 0H}, \\
&& \delta F^a_{\rm 0H} = \zeta^a_{\rm 0H} \delta \Delta_{\rm 0H} + s \kappa^a_{\rm 0H} , \\
&& \zeta^a_{0H} = 0 , \\
&& \kappa^a_{0H} = -\frac{1}{4} \frac{{\rm Im}G}{\epsilon {\rm Im}G - \Delta_e {\rm Im}F} 
 \nonumber \\
&& \times \Bigl\{  ({\rm Im}G \, {\rm Re}F_2 + {\rm Im}F \, {\rm Re}G_2) (\mathcal{T} - \mathcal{T}_2)  \nonumber \\
&&+ ({\rm Im}G \, {\rm Re}F_{-2} + {\rm Im}F \, {\rm Re}G_{-2}) (\mathcal{T} - \mathcal{T}_{-2}) \Bigr\} . \label{kappaAno0H}
\end{eqnarray}
Here, $\mathcal{T}_{\pm n}:= \mathcal{T}(\epsilon \pm n \hbar \omega_{\rm ac} / 2)$ with \( n \in \mathbb{N} \). 
The zero harmonic pair-potential variation $\delta\Delta_{\rm 0H}$ is obtained by substituting the solutions into the gap equation [Eq.~(\ref{gap2})]: 
\begin{eqnarray}
\delta \Delta_{\rm 0H} = s \Psi_{\rm 0H}  , \label{deltaDelta0H} 
\end{eqnarray}
where
\begin{eqnarray}
\Psi_{\rm 0H} &=& \frac{\Xi_{\rm 0H}}{1-\Pi_{\rm 0H}} , \label{Psi0H} \\
\Xi_{\rm 0H}&=& -({\mathscr G}/4)\int \{ (\kappa_{\rm 0H} + \kappa_{\rm 0H}^*) \mathcal{T} + \kappa_{\rm 0H}^a \} d\epsilon , \\
\Pi_{\rm 0H}&=& -({\mathscr G}/4)\int \{  (\zeta_{\rm 0H} + \zeta_{\rm 0H}^*) \mathcal{T} \}  d\epsilon .
\end{eqnarray}
This represents the time-averaged pair-potential variation and the (positive and negative) Eliashberg effect.

\begin{figure}[tb]
   \begin{center}
   \includegraphics[width=0.49\linewidth]{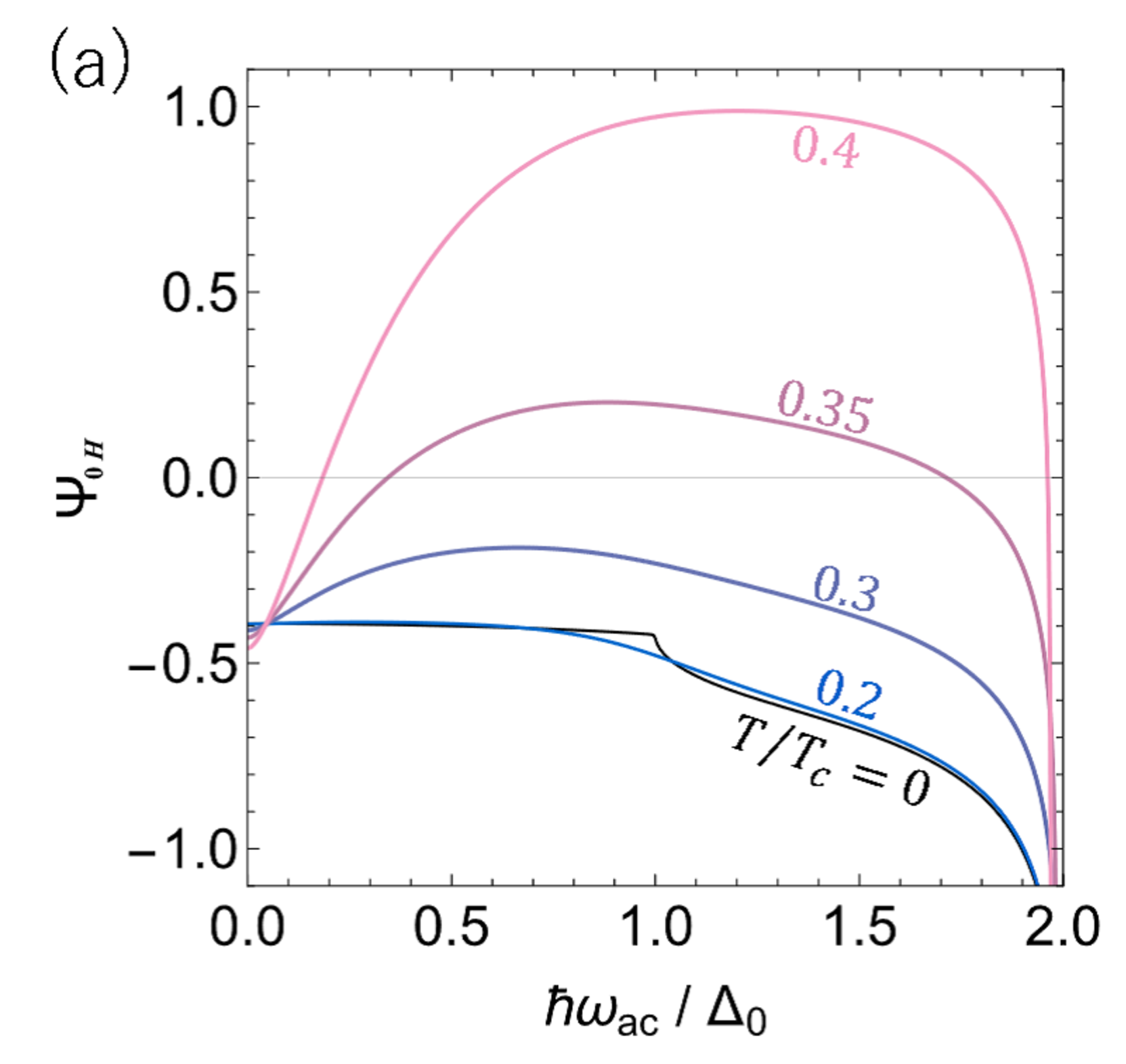}
   \includegraphics[width=0.49\linewidth]{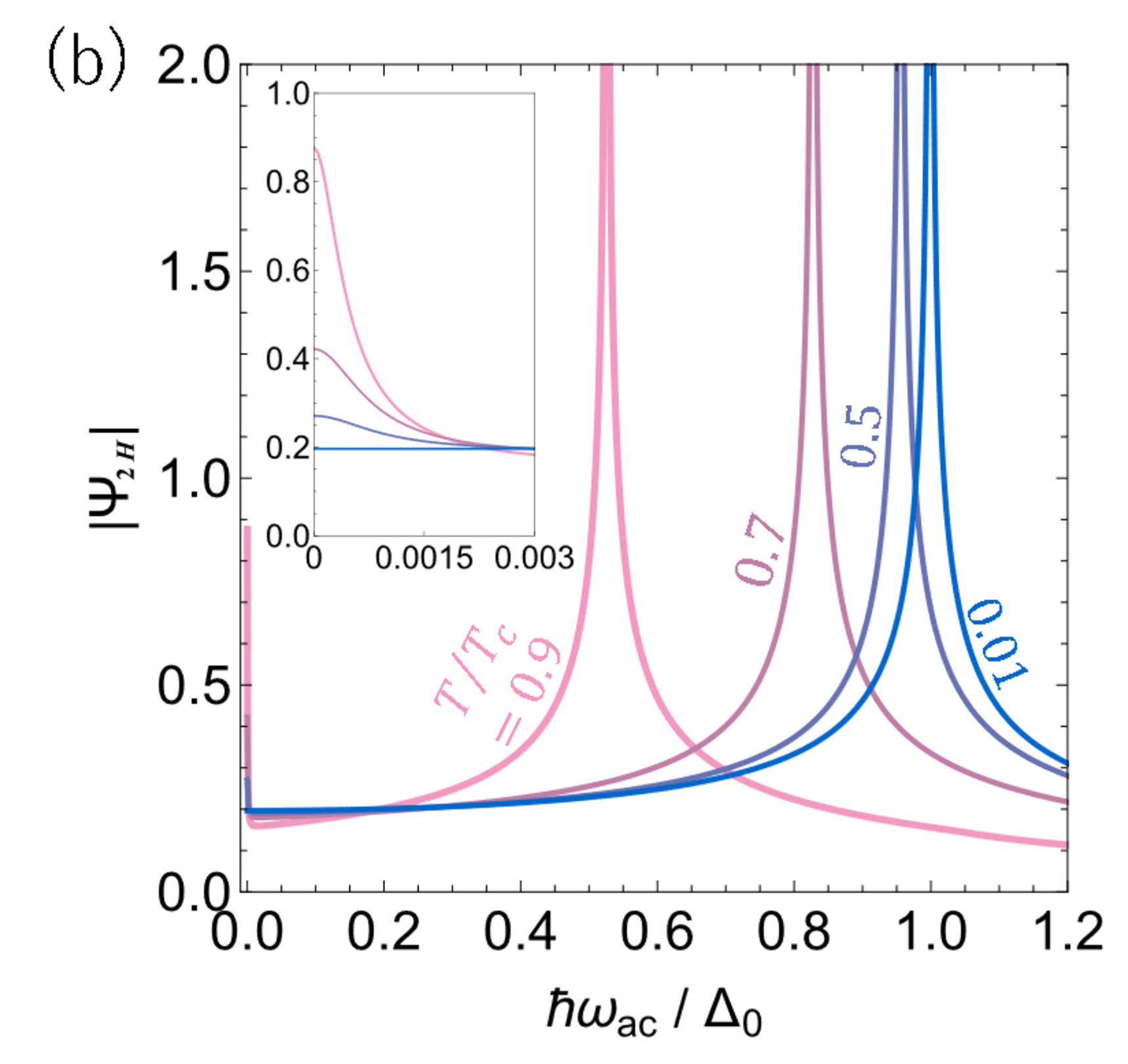}
   \end{center}\vspace{0 cm}
   \caption{
(a) Zero-harmonic (time-averaged) pair-potential variation \( \Psi_{\rm 0H} \) as a function of \( \omega_{\rm ac} \), illustrating the positive (\( \Psi_{\rm 0H} > 0 \)) and negative (\( \Psi_{\rm 0H} < 0 \)) Eliashberg effects.  
(b) Second-harmonic pair-potential variation \( |\Psi_{\rm 2H} | \) as a function of \( \omega_{\rm ac} \), showing the Higgs-mode resonance near \( \omega_{\rm ac} \simeq \Delta(T) \). 
See also Appendix~\ref{appendix_check_deltaDelta} for comparison with previous studies. 
   }\label{fig1}
\end{figure}

Figure~\ref{fig1}(a) shows \( \Psi_{\rm 0H} \) as a function of \( \omega_{\rm ac} \) for various temperatures.  
At higher temperatures, \( \Psi_{\rm 0H} \) becomes positive within a certain frequency range, indicating the emergence of the positive Eliashberg effect.  
In this regime, a large number of thermally excited quasiparticles are present. These quasiparticles can be further excited by the ac irradiation, resulting in an enhancement of the superconducting order parameter.  
In contrast, at lower temperatures, \( \Psi_{\rm 0H} \) remains negative for all frequencies.  
This is because only a small number of quasiparticles exist, and their excitation by the irradiation is insufficient to enhance the order parameter.
For \( \hbar \omega_{\rm ac} \simeq 2\Delta \), \( \Psi_{\rm 0H} \) exhibits a sharp negative dip regardless of temperature.  
This behavior originates from photon-induced pair breaking, which always suppresses the time-averaged pair potential when the photon energy is close to \( 2\Delta \). 
For \( T/T_c = 0 \), the small sharp feature observed at \( \hbar \omega_{\rm ac} = 0 \) originates from the finite damping factor \( \Gamma/\Delta_0 = 10^{-3} \).  
This feature becomes smoothed out at \( T/T_c = 0.2 \), where the thermal energy exceeds the scale set by \( \Gamma \).

As a consistency check, Appendix~\ref{appendix_check_deltaDelta} shows the region in the \( (T, \omega_{\rm ac}) \) plane where the Eliashberg effect is positive, i.e., where \( \delta \Delta_{\rm 0H} > 0 \).  
The result demonstrates good agreement with previous studies~\cite{Eremin, 2018_Klapwijk}.

\subsubsection{Second harmonic component}

Next, we summarize the second harmonic solutions (see also Appendix~\ref{appendix_1}). 
By solving the equations for the $R$ and $A$ components, Eqs.~(\ref{RA2}) and (\ref{normRA2}) [see also Eqs.~(\ref{RA_2H}) and (\ref{normRA_2H})], we obtain:
\begin{eqnarray}
&& \delta G_{\rm 2H} = \frac{F_2 + F_{-2}}{G_2 + G_{-2}} \delta F_{\rm 2H}, \label{dG2H}\\
&& \delta F_{\rm 2H} = \zeta_{\rm 2H} \delta \Delta_{\rm 2H} + s \kappa_{\rm 2H} , \\
&& \zeta_{2H} = -\frac{G_2 + G_{-2}}{F_2 + F_{-2}} \frac{F_2 - F_{-2}}{2\hbar \omega_{\rm ac}} , \label{zeta2H}\\
&& \kappa_{2H} = \frac{i}{8}  \frac{G_2 + G_{-2}}{F_2 + F_{-2}} \frac{ (G_2 - G_{-2})G + (F_2 - F_{-2})F }{2\hbar \omega_{\rm ac}} \label{kappa2H} . 
\end{eqnarray}
Similarly, for the $K$ component, solving Eqs.~(\ref{K2}) and (\ref{normK2}) [see also Eqs.~(\ref{K_2H})-(\ref{gAno_2H})] yields
\begin{eqnarray}
&& \delta G_{\rm 2H}^a = \frac{F_2 - F_{-2}^*}{G_2 - G_{-2}^*} \delta F_{\rm 2H}^a, \\
&& \delta F_{\rm 2H}^a = \zeta_{\rm 2H}^a \delta \Delta_{\rm 2H} + \kappa_{\rm 2H}^a s , \\
&& \zeta^a_{2H} = \frac{G_2 -G_{-2}^*}{F_2 -F_{-2}^*} \frac{F_2 + F_{-2}^*}{2\hbar \omega_{\rm ac}} (\mathcal{T}_{-2} - \mathcal{T}_2 ) , \label{zetaAno2H} \\
&& \kappa^a_{2H} = \frac{i}{16\hbar \omega_{\rm ac}} \frac{G_2 -G_{-2}^*}{F_2 -F_{-2}^*}  \nonumber \\
&&\times \Bigl[ \bigl\{ (G_2 + G_{-2}^*)G + (F_2 + F_{-2}^*)F \bigr\} (\mathcal{T} - \mathcal{T}_{-2} ) \nonumber \\
&& + \bigl\{ (G_2 + G_{-2}^*)G^* + (F_2 + F_{-2}^*)F^* \bigr\} (\mathcal{T} - \mathcal{T}_2) \Bigr] 
. \label{kappaAno2H}
\end{eqnarray}
The solutions for $\omega = -2\omega_{\rm ac}$ are obtained by substituting $2\omega_{\rm ac} \to -2\omega_{\rm ac}$ in the expressions above: $\zeta_{\rm -2H}=\zeta_{\rm 2H}$, $\kappa_{\rm -2H}=\kappa_{\rm 2H}$, $\zeta_{\rm -2H}^a=\zeta_{\rm 2H}^{a*}$, and $\kappa_{\rm -2H}^a=\kappa_{\rm 2H}^{a*}$. 
The second harmonic component of the pair-potential variation, $\delta\Delta_{\rm 2H}$, is obtained by substituting the solutions into the gap equation [Eq.~(\ref{gap2})]. 
This results in: 
\begin{eqnarray}
\delta \Delta_{\rm 2H}  = s \Psi_{\rm 2H} , \label{deltaDelta2H} 
\end{eqnarray}
where
\begin{eqnarray}
\Psi_{\rm 2H} &=& \frac{\Xi_{\rm 2H}}{1-\Pi_{\rm 2H}} , \label{Psi2H} \\
\Xi_{\rm 2H}&=& -({\mathscr G}/4)\int (\kappa_{\rm 2H} \mathcal{T}_- + \kappa_{\rm 2H}^* \mathcal{T}_+ + \kappa_{\rm 2H}^a) d\epsilon , \\ 
\Pi_{\rm 2H} &=& -({\mathscr G}/4)\int ( \zeta_{\rm 2H} \mathcal{T}_- + \zeta_{\rm 2H}^* \mathcal{T}_+ + \zeta_{\rm 2H}^a) d\epsilon , 
\end{eqnarray}
and $\delta \Delta_{\rm -2H}=\delta \Delta_{\rm 2H}^*$. 
This result describes the second harmonic pair-potential variation driven by the nonlinear interaction with the ac field, corresponding to the Higgs mode in the superconducting state.

Figure~\ref{fig1}(b) shows \( \delta\Delta_{\rm 2H} \) as a function of \( \omega_{\rm ac} \).  
A pronounced resonance associated with the Higgs mode appears at \( \hbar \omega_{\rm ac} = \Delta(T) \).  
As the temperature \( T \) increases, the superconducting gap \( \Delta(T) \) decreases, resulting in a downward shift of the resonance frequency.  
As a consistency check, Appendix~\ref{appendix_check_deltaDelta} presents a comparison with the previous study~\cite{Eremin}, showing good agreement.

Another notable feature is the emergence of a peak in the low-frequency regime as \( \omega_{\rm ac} \to 0 \), which becomes increasingly prominent near \( T_c \) [see the inset of Fig.~\ref{fig1}(b)]. 
Although this low-frequency behavior may initially appear counterintuitive, similar anomalous peaks have been independently reported by various authors~\cite{Silaev, Eremin}. This anomaly originates from the rapid variation of \( \Pi_{\rm 2H} \) in the limit \( \omega_{\rm ac} \to 0 \) at elevated temperatures~\cite{Eremin}.
The peak width is determined by the damping factor \( \Gamma/\Delta_0 = 10^{-3} \).  
To illustrate this dependence, Appendix~\ref{appendix_check_deltaDelta} shows the result for a larger damping value \( \Gamma/\Delta_0 = 0.02 \), where the corresponding peak exhibits a broader width on the order of \( \Gamma \).  
Similar broadening effects are observed in previous studies~\cite{Silaev, Eremin}, which used \( \Gamma/\Delta_0 = 0.1 \) and 0.02, respectively.
Moreover, the behavior of \( \delta\Delta \) in the \( \omega_{\rm ac} \to 0 \) limit agrees with the pair-potential response under a dc supercurrent, \( \delta\Delta_{\rm dc} \), as discussed in Appendix~\ref{appendix_dc_deltaDelta}.

\subsection{Linear and nonlinear current response}

Having obtained the solutions of the Keldysh-Usadel equations in terms of the equilibrium (zero-current) Green's functions \( G_e \) and \( F_e \), we now proceed to calculate the current response based on these results.

The general expression for the current is given by Eqs.~(\ref{general_J}) and (\ref{general_S}). Neglecting terms of order \( \mathcal{O}(q_0^4) \), the total current \( {\bf J}(t) \) can be decomposed into a linear term of order \( \mathcal{O}(q_0) \) and a nonlinear term of order \( \mathcal{O}(q_0^3) \):
\begin{eqnarray}
{\bf J}(t) &=& {\bf J}^{(1)}(t)+ {\bf J}^{(3)}(t) \nonumber \\
&=&\frac{\sigma_n}{8e} \int \Bigl\{ {\bf S}^{(1)}(\epsilon,t) + {\bf S}^{(3)}(\epsilon,t) \Bigr\} d\epsilon , 
\label{J}
\end{eqnarray}
where the integrands are defined as
\begin{eqnarray}
{\bf S}^{(1)}(\epsilon,t)&=&  {\rm Tr} \bigl[ \hat{\tau}_3 
\bigl\{ \hat{g}^{R}_e \circ (\hat{\partial} \circ \hat{g}^{K}_e) 
+ \hat{g}^K_e \circ  (\hat{\partial} \circ \hat{g}^{A}_e) \bigr\} \bigr]  , \label{S1} \\
{\bf S}^{(3)}(\epsilon,t)&=& {\rm Tr} \bigl[ \hat{\tau}_3 
\bigl\{ \hat{g}^{R}_e \circ (\hat{\partial} \circ \delta \hat{g}^{K}) 
+ \hat{g}^K_e \circ  (\hat{\partial} \circ \delta \hat{g}^{A}) \nonumber \\
&& + \delta \hat{g}^R \circ  (\hat{\partial} \circ \hat{g}_e^{K}) 
+ \delta \hat{g}^K \circ  (\hat{\partial} \circ \hat{g}^{A}_e) \bigr\} \bigr] . \label{S3}
\end{eqnarray}
Here, \( {\bf J}^{(1)} \) (or \( {\bf S}^{(1)} \) ) corresponds to the linear response term, 
while \( {\bf J}^{(3)} \) (or \( {\bf S}^{(3)} \) ) captures the leading-order nonlinear response induced by the ac field.

\subsubsection{Linear response}

Let us begin by deriving the well-known linear response formula for a dirty-limit superconductor subjected to an ac perturbation.  
We are considering the monochromatic ac field ${\bf q}(t) = {\bf q}_0 \cos (\omega_{\rm ac} t)$, which is associated with an electric field ${\bf E}(t) = {\bf E}_0 \sin (\omega_{\rm ac} t)$, where ${\bf E}_0 = (\hbar \omega_{\rm ac}/2|e|) {\bf q}_0$. 
This field can be expressed as ${\bf E}(t) = {\bf E}_{\rm 1H} e^{-i\omega_{\rm ac}t} + {\rm c.c.}$ with ${\bf E}_{\rm 1H} = (i/2) {\bf E}_0$, or equivalently in Fourier space as
\begin{eqnarray}
{\bf E}(\omega) = {\bf E}_{\rm 1H} 2\pi \delta(\omega-\omega_{\rm ac}) 
+  {\bf E}_{\rm 1H}^* 2\pi \delta(\omega+\omega_{\rm ac}). 
\end{eqnarray}
Since the linear response is proportional to ${\bf E}$, the Fourier transform of the linear current response, ${\bf J}^{(1)}(\omega)$, contains only the first harmonic at $\omega = \pm \omega_{\rm ac}$: 
\begin{eqnarray}
{\bf J}^{(1)}(\omega) &=& {\bf J}^{(1)}_{\rm 1H} 2\pi \delta (\omega - \omega_{\rm ac}) +
{\bf J}^{(1)*}_{\rm 1H} 2\pi \delta (\omega + \omega_{\rm ac}). 
\end{eqnarray}
The relation between ${\bf J}^{(1)}_{\rm 1H}$ and ${\bf E}_{\rm 1H}$ is determined by Eqs.~(\ref{J}) and (\ref{S1}) or its Fourier transform (see Appendix~\ref{appendix_linear}), and we obtain
\begin{eqnarray}
 {\bf J}^{(1)}_{\rm 1H}  = \sigma {\bf E}_{\rm 1H} , \label{MB1}
\end{eqnarray}
where
\begin{eqnarray}
\sigma &=& \sigma_n \int \frac{d\epsilon }{\hbar \omega_{\rm ac}}
\Bigl\{ ( {\rm Re} G  {\rm Re} G_2 +  {\rm Re} F  {\rm Re} F_2 ) (f -f_2) \nonumber \\
&&+ i ( {\rm Re} G  {\rm Im} G_2 +  {\rm Re} F  {\rm Im} F_2 ) (1 -2f) \Bigr\} .\label{MB2}
\end{eqnarray}
The resulting proportionality constant $\sigma(\omega_{\rm ac})$ is the complex conductivity, 
reproducing the well-known Mattis-Bardeen formula in the dirty limit~\cite{MB, Nam, Zimmermann, Gurevich_Kubo, 2022_Kubo}.

When the ac field is sufficiently weak, the third-order correction \( {\bf J}^{(3)} \sim \mathcal{O}(q_0^3) \) can be safely neglected.  
However, as the field strength increases, the linear-response approximation breaks down, and nonlinear contributions of order \( \mathcal{O}(q_0^3) \) begin to emerge.  
Given that \( {\bf E}^3(t) = {\bf E}_0^3 \sin^3(\omega_{\rm ac} t) \propto 3\sin(\omega_{\rm ac} t) - \sin(3\omega_{\rm ac} t) \),  
it follows that \( {\bf J}^{(3)}(t) \) contains both first- and third-harmonic components.  
In Fourier space, this is expressed as
\begin{eqnarray}
&&{\bf J}^{(3)}(\omega) = {\bf J}^{(3)}_{\rm 1H} 2\pi \delta (\omega - \omega_{\rm ac}) 
+  {\bf J}^{(3)*}_{\rm 1H} 2\pi \delta (\omega + \omega_{\rm ac}) \nonumber \\
&&+  {\bf J}^{(3)}_{\rm 3H} 2\pi \delta (\omega - 3\omega_{\rm ac}) 
+ {\bf J}^{(3)*}_{\rm 3H} 2\pi \delta (\omega + 3\omega_{\rm ac}) . \label{J3_Fourier}
\end{eqnarray}
Here, \( \mathbf{J}^{(3)}_{\rm 1H} \) represents the nonlinear correction to the linear first-harmonic component \( \mathbf{J}^{(1)}_{\rm 1H} \). 
The third-harmonic component \( \mathbf{J}^{(3)}_{\rm 3H} \) (in disordered superconductors) is known to exhibit a resonance feature associated with the Higgs mode, a hallmark frequently exploited in experimental studies.  
Analytical expressions for both components are derived below.

\subsubsection{Nonlinear correction to the first-harmonic current}

Starting from Eqs.~(\ref{J}) and (\ref{S3}), and after a lengthy analytic calculation (see Appendix~\ref{appendix_nonlinear}), we obtain the first harmonic-current generated by $\mathcal{O}(q_0^3)$ correction:
\begin{eqnarray}
J^{(3)}_{\rm 1H} &=& i \biggl( \frac{q_0}{q_{\xi}} \biggr)^{3} 
( I^{\rm qqq}_{\rm 1H} + I^{\rm Higgs}_{\rm 1H} + I^{\rm Eliash}_{\rm 1H} ) J_0 \nonumber \\
&=& \frac{2\sqrt{\pi} \sigma_n}{\hbar \omega_{\rm ac}/\Delta_0} \frac{s}{\Delta_0} ( I^{\rm qqq}_{\rm 1H} + I^{\rm Higgs}_{\rm 1H} + I^{\rm Eliash}_{\rm 1H} ) E_{\rm 1H}, \label{J1H}\\
I^{\rm qqq}_{\rm 1H} &=&  \frac{-1}{16\sqrt{\pi}} \int K_{\rm 1H} d\epsilon , \label{I1H_qqq}\\
I^{\rm Higgs}_{\rm 1H} &=& \frac{-1}{16\sqrt{\pi}} \Psi_{\rm 2H}  \int Z_{\rm 1H}^{\rm Higgs} d\epsilon , \label{I1H_Higgs}\\
I^{\rm Eliash}_{\rm 1H} &=& \frac{-1}{16\sqrt{\pi}} \Psi_{\rm 0H} \int Z_{\rm 1H}^{\rm Eliash} d\epsilon .\label{I1H_Eliash}
\end{eqnarray}
Here, $J_0=H_{c}/\lambda_0$, $H_{c0}=\sqrt{N_0/\mu_0}\Delta_0$ is the zero-temperature thermodynamic critical field, and $\lambda_0=\sqrt{\hbar/\pi\mu_0\Delta_0\sigma_n}$ is the zero-temperature London depth. 
The relation $J_0 \sqrt{s/\Delta_0}= \sqrt{\pi} \sigma_n (\Delta_0/\hbar \omega_{\rm ac}) E_0$ is used at the second line. 
$K_{\rm 1H}$, $Z_{\rm 1H}^{\rm Eliash}$, and $Z_{\rm 1H}^{\rm Higgs}$ are given by
\begin{eqnarray}
K_{\rm 1H}(\epsilon, \omega_{\rm ac}) = \sum_{i=1}^6 K_{{\rm 1H}, i} , \label{K1H}
\end{eqnarray}
\begin{eqnarray}
&&K_{{\rm 1H}, 1} = i \frac{  (F_{1}-F_{-3}) F_{-1} + (G_{1}-G_{-3}) G_{-1} }{4\hbar \omega_{\rm ac} (F_{1} +F_{-3})}   \nonumber \\
&& \times\Bigl[ \Bigl\{ (F_{1}+F_{-3}) G_{-1}^* + (G_{1}+G_{-3}) F_{-1}^* \Bigr\} (\mathcal{T}_{-1}-\mathcal{T}_{-3}) \nonumber \\
&& + \Bigl\{ (F_1+F_{-3})G_{-1} + (G_1+G_{-3})F_{-1} \Bigr\} \mathcal{T}_{-1} \Bigr] , \\
&&K_{{\rm 1H}, 2} =  i \frac{  (G_{3}-G_{-1}) G_{1} + (F_{3}-F_{-1}) F_{1}  }{4\hbar \omega_{\rm ac} (F_{3} +F_{-1})}  \nonumber \\
&& \times\bigl\{ (F_{3}+F_{-1}) G_{1} + (G_{3}+G_{-1}) F_{1} \bigr\} \mathcal{T}_{-1} , \\
&&K_{{\rm 1H}, 3} = -i \frac{  (F_{3}-F_{-1}^*) G_{1} + (G_{3}-G_{-1}^*) F_{1} }{4\hbar \omega_{\rm ac} (F_{3} -F_{-1}^*)} \nonumber \\
&&\times \biggl[ \Bigl\{  (G_{3}+G_{-1}^*) G_{1}^* +  (F_{3}+F_{-1}^*) F_{1}^* \Bigr\} (\mathcal{T}_{3}-\mathcal{T}_{1}) \nonumber \\
&& + \Bigl\{ (G_{3}+G_{-1}^*) G_{1} + (F_{3}+F_{-1}^*) F_{1} \Bigr\} (\mathcal{T}_{-1}-\mathcal{T}_{1}) \biggr] , 
\end{eqnarray}
\begin{eqnarray}
&&K_{{\rm 1H}, 4} = -i \frac{F_1 G_{-1} + G_1 F_{-1} }{2(\epsilon_{-1} G_{-1} - F_{-1} \Delta)} \nonumber \\
&&\times\bigl\{ (F_1+F_{-3})G_{-1} + (G_1 + G_{-3})F_{-1} \bigr\} \mathcal{T}_{-1} , \\
&&K_{{\rm 1H}, 5} =  -i \frac{(F_{3}+F_{-1})G_{1} + (G_3+G_{-1})F_{1} }{2(\epsilon_{1} G_{1} - F_{1} \Delta)} \nonumber \\
&&\times\Bigl\{  (G_1F_{-1}^* + F_1 G_{-1}^*) (\mathcal{T}_{-1} -\mathcal{T}_{1} ) \nonumber \\
&&+  (G_1F_{-1} + G_{-1}F_{1}) \mathcal{T}_{-1}  \Bigr\} , \\
&&K_{{\rm 1H}, 6} =  - \frac{G_{-1}^* {\rm Im}F_1 + F_{-1}^* {\rm Im} G_1}{\epsilon_{1} {\rm Im}G_{1} - {\rm Im} F_{1} \Delta} \nonumber \\
&& \times \Bigl\{ ( {\rm Re}G_3 {\rm Im}F_1 + {\rm Im} G_1 {\rm Re} F_3) (\mathcal{T}_3-\mathcal{T}_1)  \nonumber \\
&& +  ( {\rm Re}G_{-1} {\rm Im}F_1 + {\rm Im} G_1 {\rm Re} F_{-1}) (\mathcal{T}_{-1}-\mathcal{T}_1)  \Bigr\} ,
\end{eqnarray}
\begin{eqnarray}
Z^{\rm Eliash}_{\rm 1H}(\epsilon, \omega_{\rm ac}) = \sum_{i=1, 2}  Z^{\rm Eliash}_{{\rm 1H}, i}  , \label{Z1H_Eliash}
\end{eqnarray}
\begin{eqnarray}
&&Z^{\rm Eliash}_{{\rm 1H}, 1}  = 
\frac{4 G_{-1} ( G_1 F_{-1} + F_1 G_{-1} ) }{\epsilon_{-1} G_{-1} - F_{-1} \Delta}  \mathcal{T}_{-1} , \\
&&Z^{\rm Eliash}_{{\rm 1H}, 2}  = \frac{4G_1}{\epsilon_{1} G_{1} - F_{1} \Delta} 
\Bigl\{   (G_1F_{-1} + G_{-1}F_{1}) \mathcal{T}_{-1}  \nonumber \\
&& + (G_1F_{-1}^* + F_1 G_{-1}^*) (\mathcal{T}_{-1} -\mathcal{T}_{1} )   \Bigr\} ,
\end{eqnarray}
and
\begin{eqnarray}
Z^{\rm Higgs}_{\rm 1H}(\epsilon, \omega_{\rm ac}) = \sum_{i=1, 2, 3} Z^{\rm Higgs}_{{\rm 1H}, i} , \label{Z1H_Higgs}
\end{eqnarray}
\begin{eqnarray}
&&Z^{\rm Higgs}_{{\rm 1H}, 1}  = - \frac{ 2(F_1-F_{-3}) }{\hbar \omega_{\rm ac} (F_{1} +F_{-3})}   \nonumber \\
&& \times\Bigl[ \Bigl\{ (F_{1}+F_{-3}) G_{-1}^* + (G_{1}+G_{-3}) F_{-1}^* \Bigr\} (\mathcal{T}_{-1}-\mathcal{T}_{-3}) \nonumber \\
&& + \Bigl\{ (F_1+F_{-3})G_{-1} + (G_1+G_{-3})F_{-1} \Bigr\} \mathcal{T}_{-1} \Bigr] , \\
&&Z^{\rm Higgs}_{{\rm 1H}, 2}  = - \frac{  2(F_3-F_{-1})  }{\hbar \omega_{\rm ac} (F_{3} +F_{-1})}  \nonumber \\
&& \times\Bigl\{ (F_{3}+F_{-1}) G_{1} + (G_{3}+G_{-1}) F_{1} \Bigr\} \mathcal{T}_{-1} , \\
&&Z^{\rm Higgs}_{{\rm 1H}, 3}  = - \frac{ 2(F_3+F_{-1}^*) }{\hbar \omega_{\rm ac} (F_{3} -F_{-1}^*)} \nonumber \\
&&\times \Bigl\{  (F_{3}-F_{-1}^*) G_{1} +  (G_{3}-G_{-1}^*) F_{1} \Bigr\} (\mathcal{T}_{3}-\mathcal{T}_{-1}) 
\end{eqnarray}
Note $K_{\rm 1H}$ represents the $\mathcal{O}(q^3)$ direct action of the electromagnetic field to the first harmonic, while $Z^{\rm Eliash}_{\rm 1H}$ and $Z^{\rm Higgs}_{\rm 1H}$ represent the $\mathcal{O}(q^3)$  corrections to the first harmonic via the Eliashberg effect ($\delta\Delta_{\rm 0H}$) and the Higgs-mode ($\delta\Delta_{\rm 2H}$).

\begin{figure}[tb]
   \begin{center}
   \includegraphics[width=0.49\linewidth]{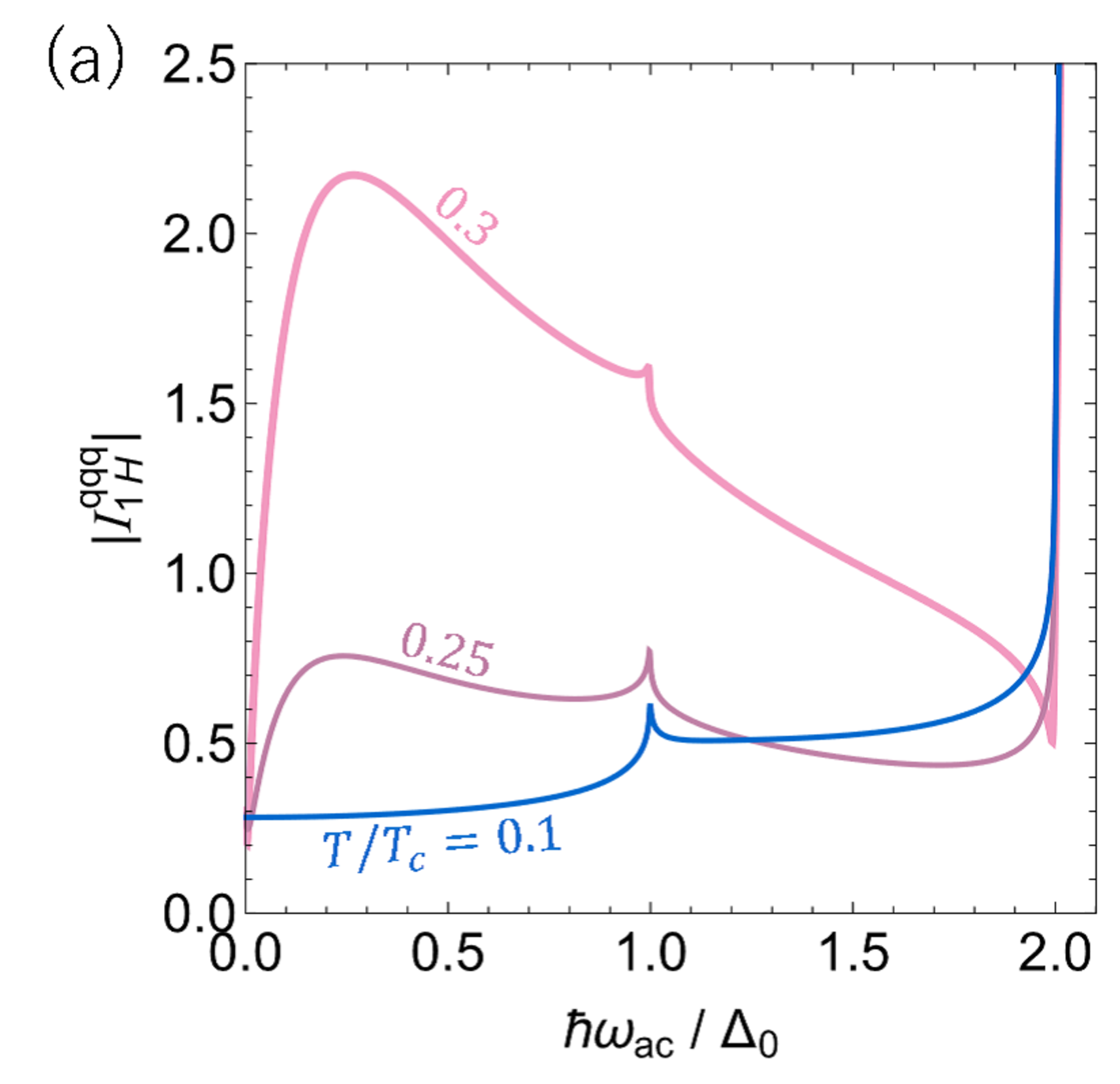}
   \includegraphics[width=0.49\linewidth]{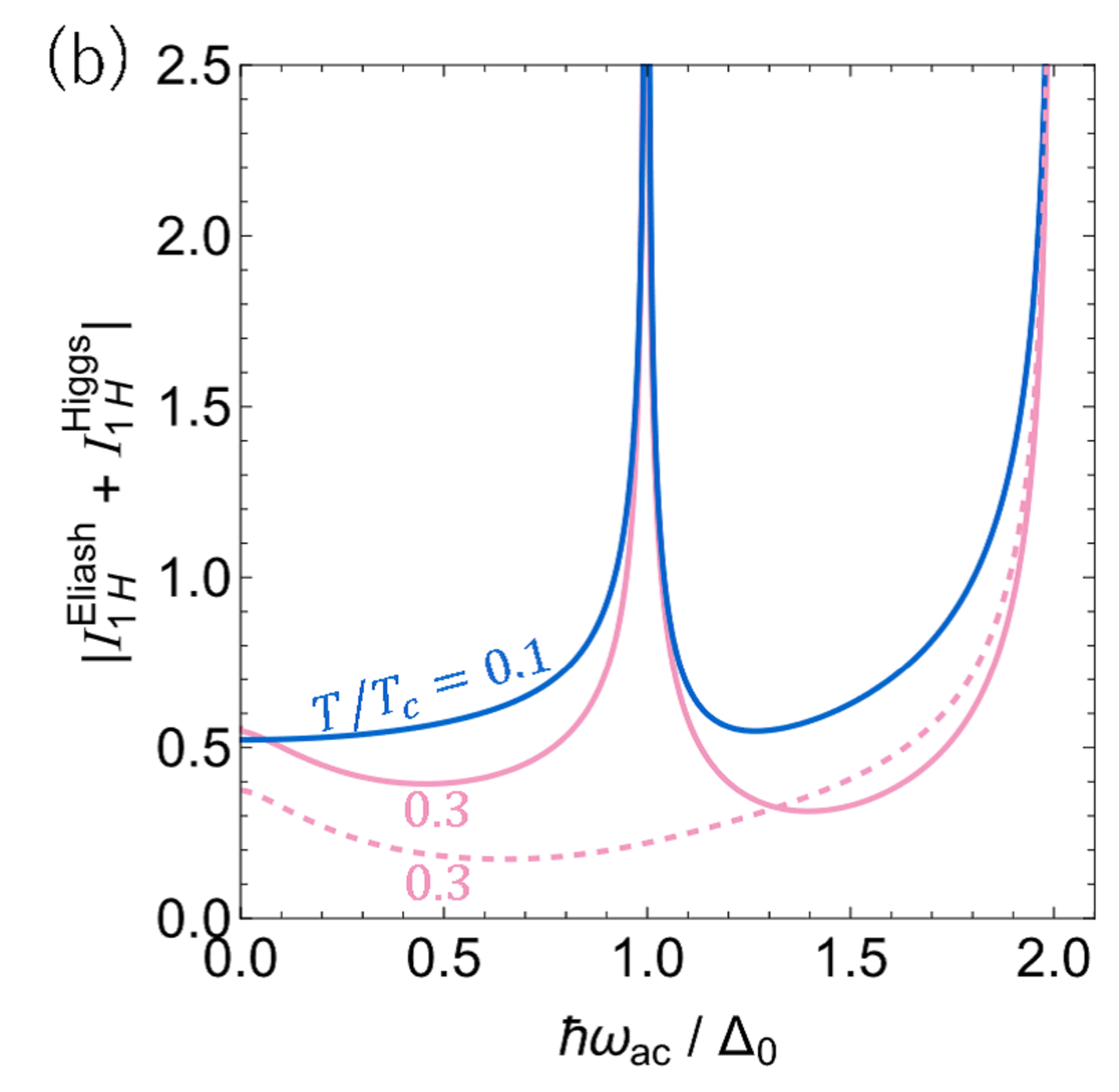}
   \includegraphics[width=0.49\linewidth]{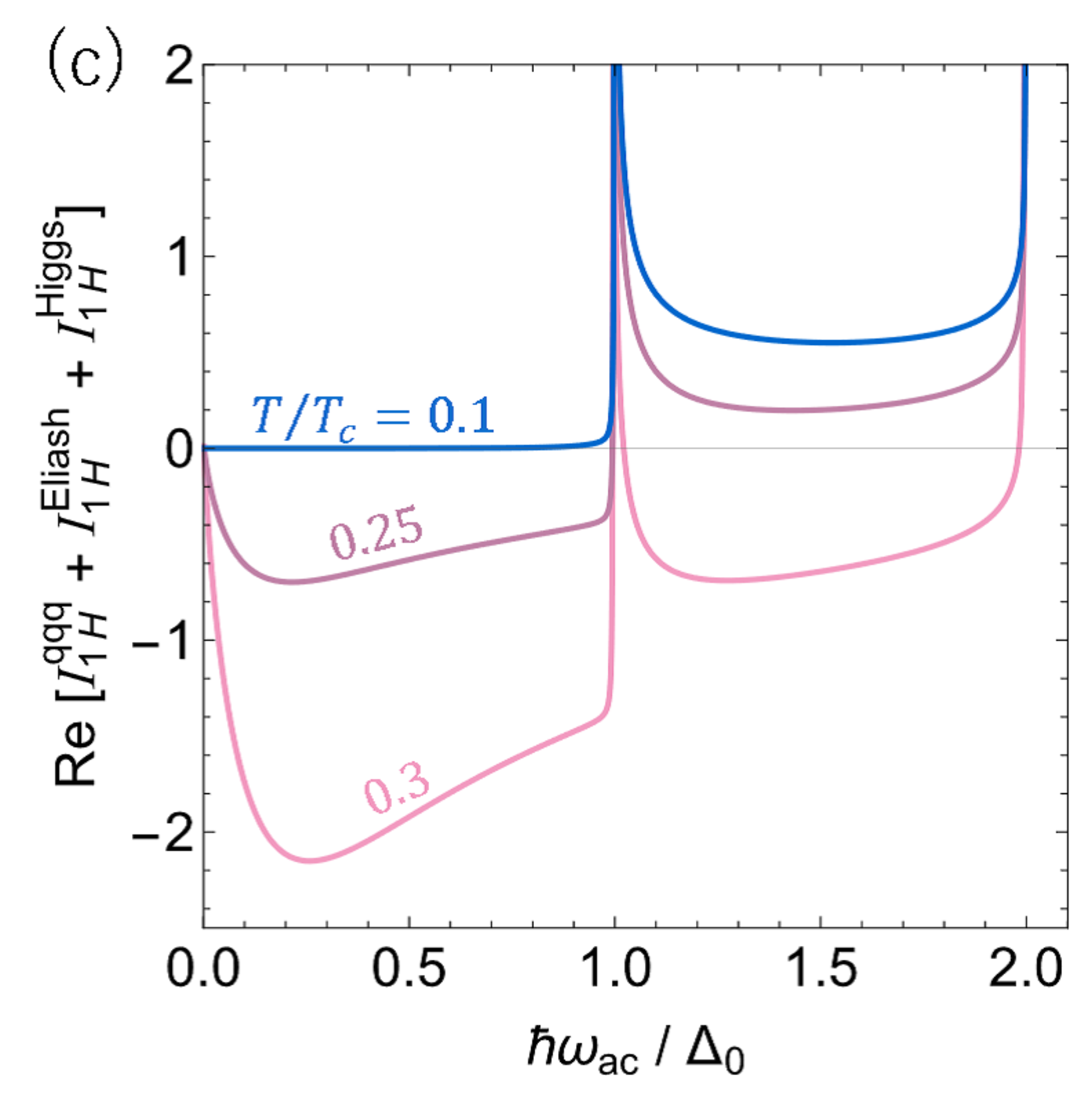}
   \includegraphics[width=0.49\linewidth]{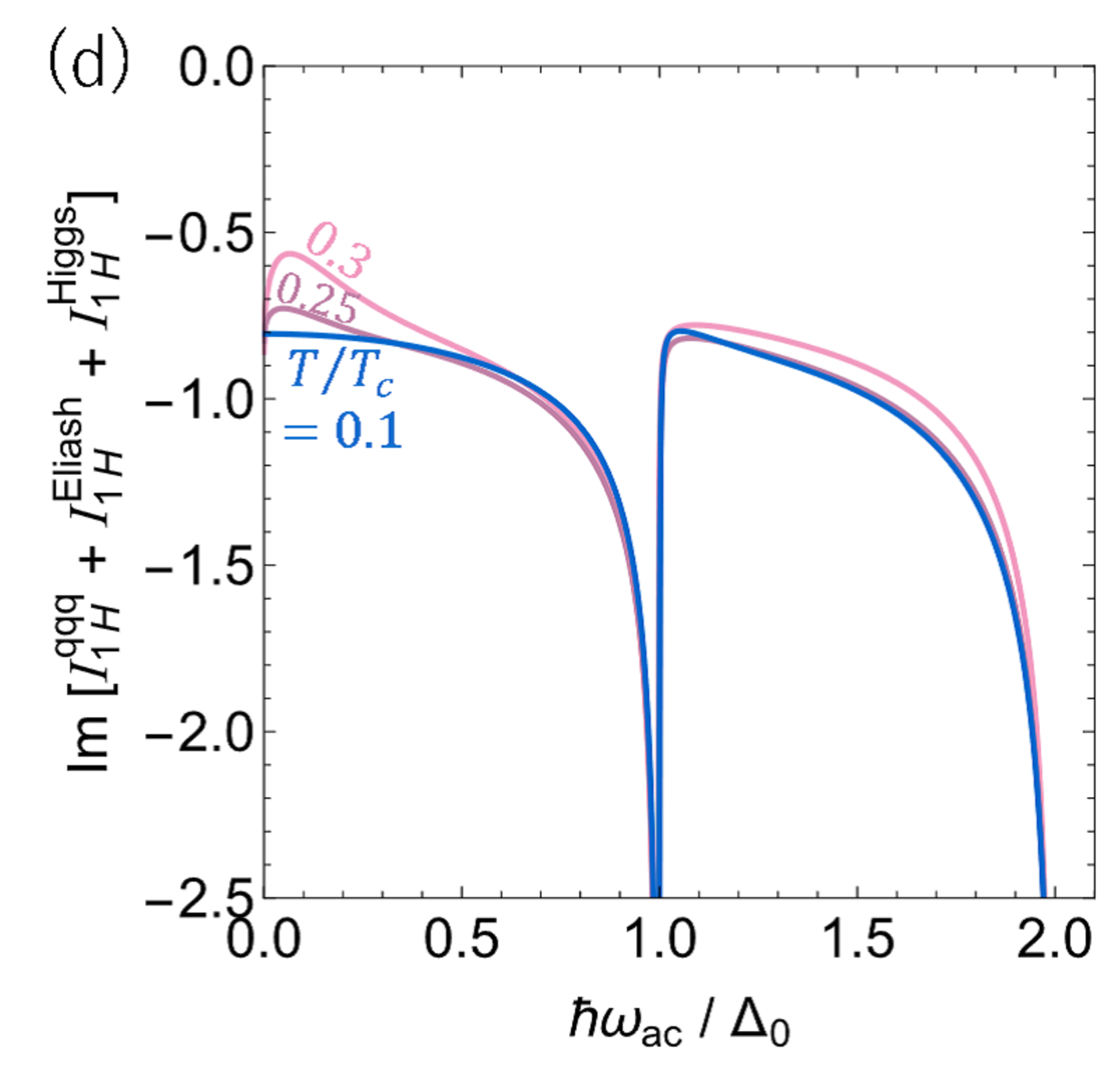}
\end{center}\vspace{0 cm}
   \caption{
Nonlinear corrections to the first-harmonic current.  
(a) Contribution from the direct nonlinear action of the electromagnetic field.  
(b) Sum of the contributions mediated by nonequilibrium pair-potential variations, \( \delta\Delta_{\mathrm{0H}} \) and \( \delta\Delta_{\mathrm{2H}} \). 
The dashed curve represents $|I_{\rm 1H}^{\rm Eliash}|$. 
(c) Real and (d) imaginary parts of the total nonlinear correction.
   }\label{fig2}
\end{figure}

Figure~\ref{fig2} summarizes the nonlinear corrections to the first-harmonic current response.  
Figure~\ref{fig2}(a) shows the direct nonlinear contribution from the electromagnetic field to \( J_{\mathrm{1H}}^{(3)} \),  
while Fig.~\ref{fig2}(b) presents the indirect contributions mediated by nonequilibrium pair-potential variations, \( \delta\Delta_{\mathrm{0H}} \) and \( \delta\Delta_{\mathrm{2H}} \).  
The dashed curve represents \( |I_{\mathrm{1H}}^{\mathrm{Eliash}}| \), indicating that the peak at \( \hbar \omega_{\mathrm{ac}} \simeq \Delta \) in Fig.~\ref{fig2}(b) originates from the resonance associated with the Higgs mode contribution \( I_{\mathrm{1H}}^{\mathrm{Higgs}} \).  
Figures~\ref{fig2}(c) and \ref{fig2}(d) show the real and imaginary parts of the total nonlinear correction,  
which exhibit a Higgs-induced peak and dip at \( \hbar \omega_{\mathrm{ac}} = \Delta \), respectively.

\subsubsection{Nonlinear generation of the third-harmonic current}

Starting from Eqs.~(\ref{J}) and (\ref{S3}), and following a procedure similar to that in the previous section involving lengthy analytical calculations (see Appendix~\ref{appendix_nonlinear}), we obtain the third-harmonic current arising from the \( \mathcal{O}(q_0^3) \) terms:
\begin{eqnarray}
J^{(3)}_{\rm 3H} &=& i \biggl( \frac{q_0}{q_{\xi}} \biggr)^{3} 
( I^{\rm qqq}_{\rm 3H} + I^{\rm Higgs}_{\rm 3H} ) J_0 \nonumber \\
&=& \frac{2\sqrt{\pi} \sigma_n}{\hbar \omega_{\rm ac}/\Delta_0} \frac{s}{\Delta_0} 
( I^{\rm qqq}_{\rm 3H} + I^{\rm Higgs}_{\rm 3H} ) E_{\rm 1H} , \label{J3H} \\
I^{\rm qqq}_{\rm 3H} &=& \frac{-1}{16\sqrt{\pi}} \int K_{\rm 3H} d\epsilon ,\\
I^{\rm Higgs}_{\rm 3H} &=& \frac{-1}{16\sqrt{\pi}} \Psi_{\rm 2H} \int Z_{\rm 3H}^{\rm Higgs} d\epsilon ,
\end{eqnarray}
where, $K_{\rm 3H}$ and $Z_{\rm 3H}^{\rm Higgs}$ are given by
\begin{eqnarray}
K_{\rm 3H}(\epsilon, \omega_{\rm ac}) = \sum_{i=1, 2, 3} K_{{\rm 3H}, i} , \label{K3H}
\end{eqnarray}
\begin{eqnarray}
&&K_{{\rm 3H}, 1} = i \frac{  (F_{1}-F_{-3}) F_{-1} + (G_{1}-G_{-3}) G_{-1} }{4\hbar \omega_{\rm ac} (F_{1} +F_{-3})} \nonumber \\
&& \times \Bigl\{ (F_{1}+F_{-3}) G_{3} + (G_{1}+G_{-3}) F_{3} \Bigr\} \mathcal{T}_{-3} , \\
&&K_{{\rm 3H}, 2} =  i \frac{  (F_{3}-F_{-1}) F_{1} + (G_{3}-G_{-1}) G_{1}  }{4\hbar \omega_{\omega_{\rm ac}} (F_{3} +F_{-1})} \nonumber \\
&& \times \biggl[ 
\Bigl\{ (F_{3}+F_{-1}) G_{-3}^* + (G_{3}+G_{-1}) F_{-3}^* \Bigr\} (\mathcal{T}_{-3}-\mathcal{T}_{-1}) \nonumber \\
&&+\Bigl\{ (F_{3}+F_{-1}) G_{-3} + (G_{3}+G_{-1}) F_{-3} \Bigr\} \mathcal{T}_{-3}   \biggr] , \\
&&K_{{\rm 3H}, 3} = i \frac{  (F_{3}-F_{-1}^*) G_{-3}^* + (G_{3}-G_{-1}^*) F_{-3}^*  }{4\hbar \omega_{\omega_{\rm ac}} (F_{3} -F_{-1}^*)} \nonumber \\
&& \times \biggl[ \Bigl\{  (G_{3}+G_{-1}^*) G_{1}^* +  (F_{3}+F_{-1}^*) F_{1}^* \Bigr\} (\mathcal{T}_{3}-\mathcal{T}_{1}) \nonumber \\
&& + \Bigl\{ (G_{3}+G_{-1}^*) G_{1} + (F_{3}+F_{-1}^*) F_{1} \Bigr\} (\mathcal{T}_{-1}-\mathcal{T}_{1}) \biggr] ,
\end{eqnarray}
and
\begin{eqnarray}
Z_{\rm 3H}^{\rm Higgs}(\epsilon, \omega_{\rm ac}) = \sum_{i=1,2,3} Z_{{\rm 3H}, i}^{\rm Higgs} , \label{Z3H_Higgs}
\end{eqnarray}
\begin{eqnarray}
&&Z_{{\rm 3H},1}^{\rm Higgs} = -2 \frac{ F_{1} -F_{-3} }{\hbar \omega_{\rm ac} (F_{1} +F_{-3})} \Bigl\{ (F_{1}+F_{-3}) G_{3} \nonumber \\
&& + (G_{1}+G_{-3}) F_{3} \Bigr\} \mathcal{T}_{-3} , \\
&&Z_{{\rm 3H},2}^{\rm Higgs} =-2 \frac{ F_{3} -F_{-1} }{\hbar \omega_{\rm ac} (F_{3} +F_{-1})} \nonumber \\
&& \times \biggl[ 
\Bigl\{ (F_{3}+F_{-1}) G_{-3}^* + (G_{3}+G_{-1}) F_{-3}^* \Bigr\} (\mathcal{T}_{-3}-\mathcal{T}_{-1}) \nonumber \\
&&+\Bigl\{ (F_{3}+F_{-1}) G_{-3} + (G_{3}+G_{-1}) F_{-3} \Bigr\} \mathcal{T}_{-3}  \biggr] , \\
&&Z_{{\rm 3H},3}^{\rm Higgs} = 2 \frac{ F_{3} + F_{-1}^*  }{\hbar \omega_r (F_{3} -F_{-1}^*)} 
\Bigl\{   (F_{3}-F_{-1}^*) G_{-3}^* \nonumber \\
&&  + (G_{3}-G_{-1}^*) F_{-3}^* \Bigr\} (\mathcal{T}_{3}-\mathcal{T}_{-1}) .
\end{eqnarray}
Here, $K_{\rm 3H}$ represents the $\mathcal{O}(q_0^3)$ direct action of the electromagnetic field to the third harmonic, while $Z^{\rm Higgs}_{\rm 3H}$ represents the $\mathcal{O}(q_0^3)$ contribution to the third harmonic via the Higgs-mode.

\begin{figure}[tb]
   \begin{center}
   \includegraphics[width=0.49\linewidth]{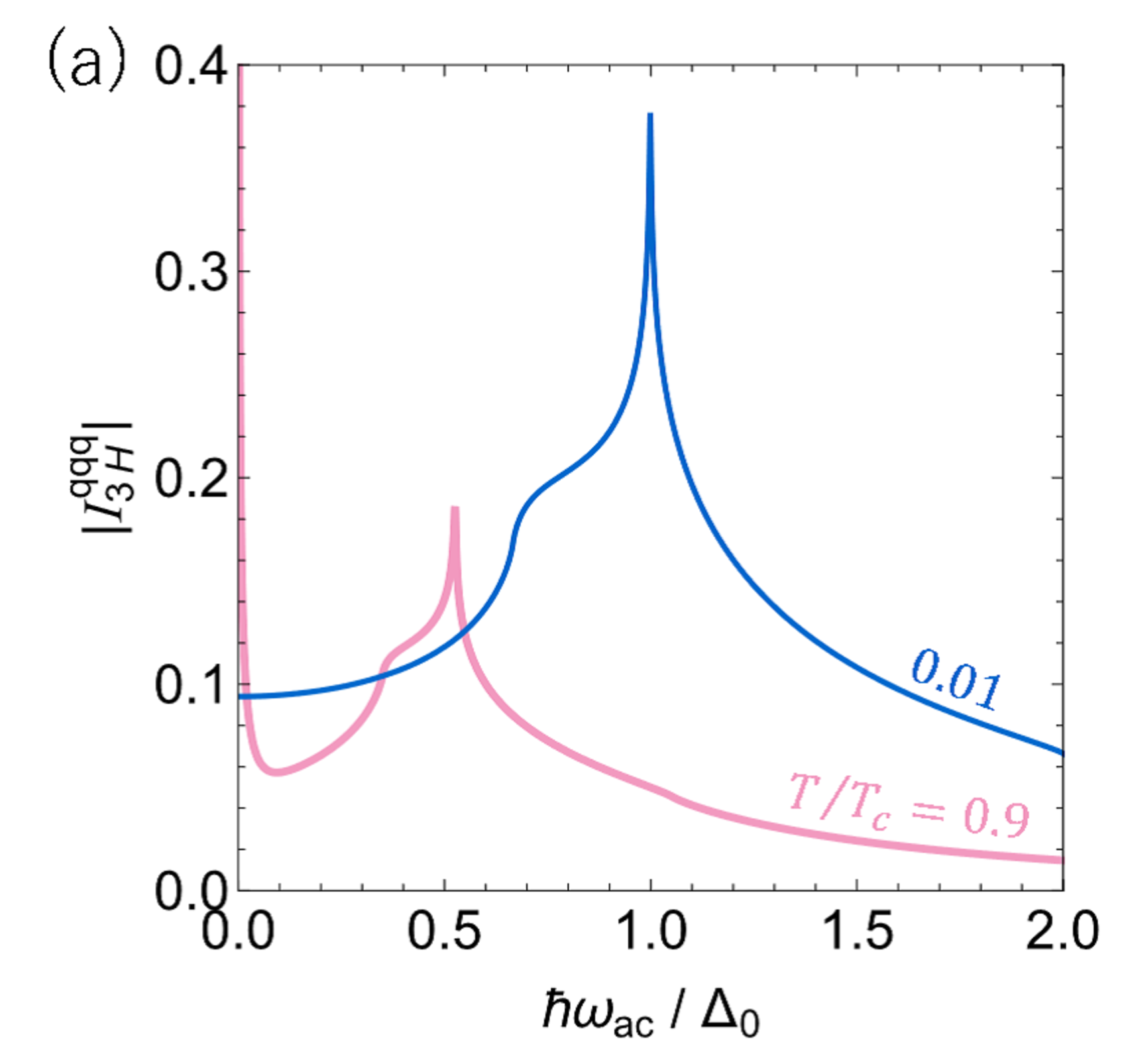}
   \includegraphics[width=0.49\linewidth]{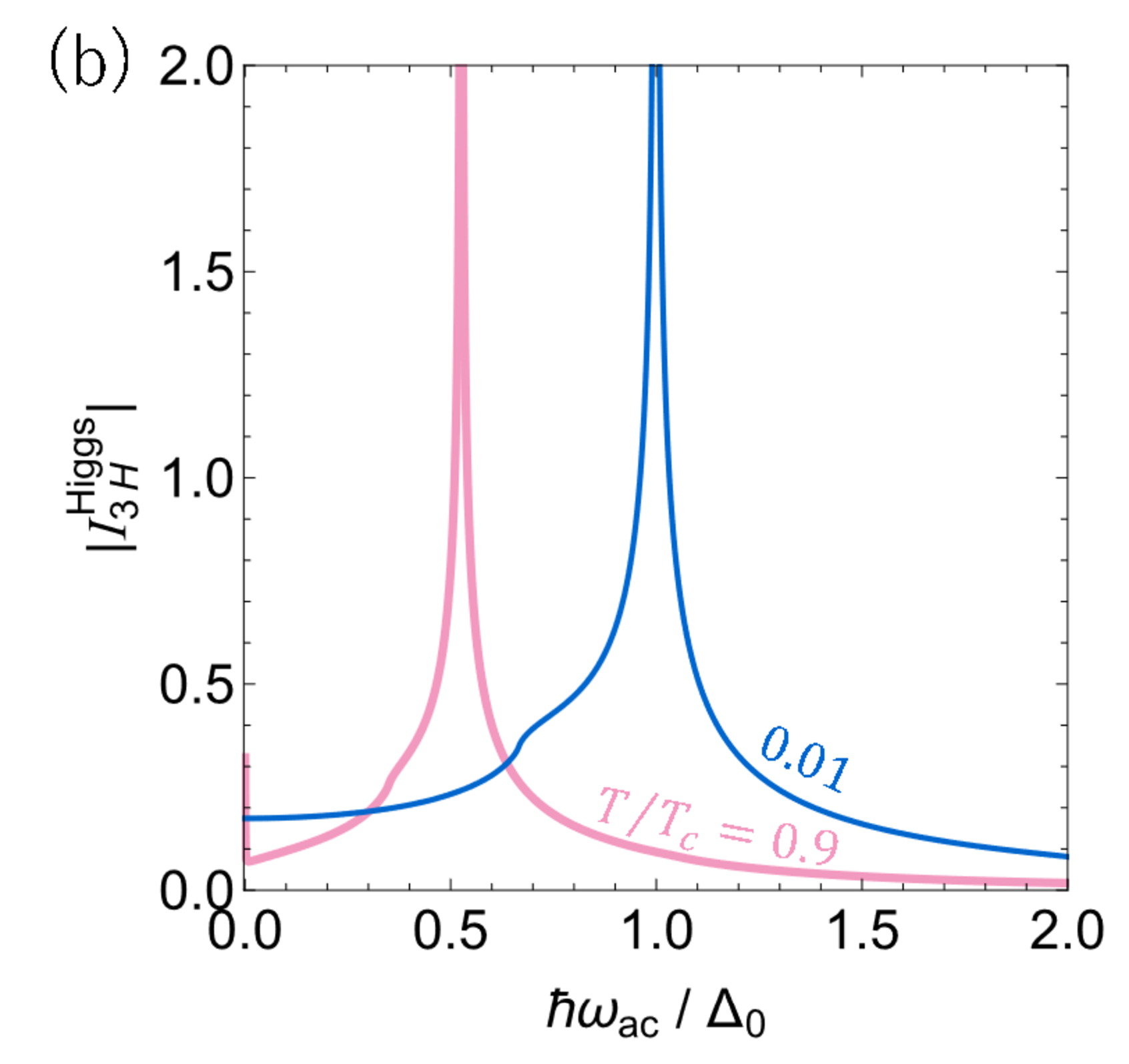}
   \includegraphics[width=0.49\linewidth]{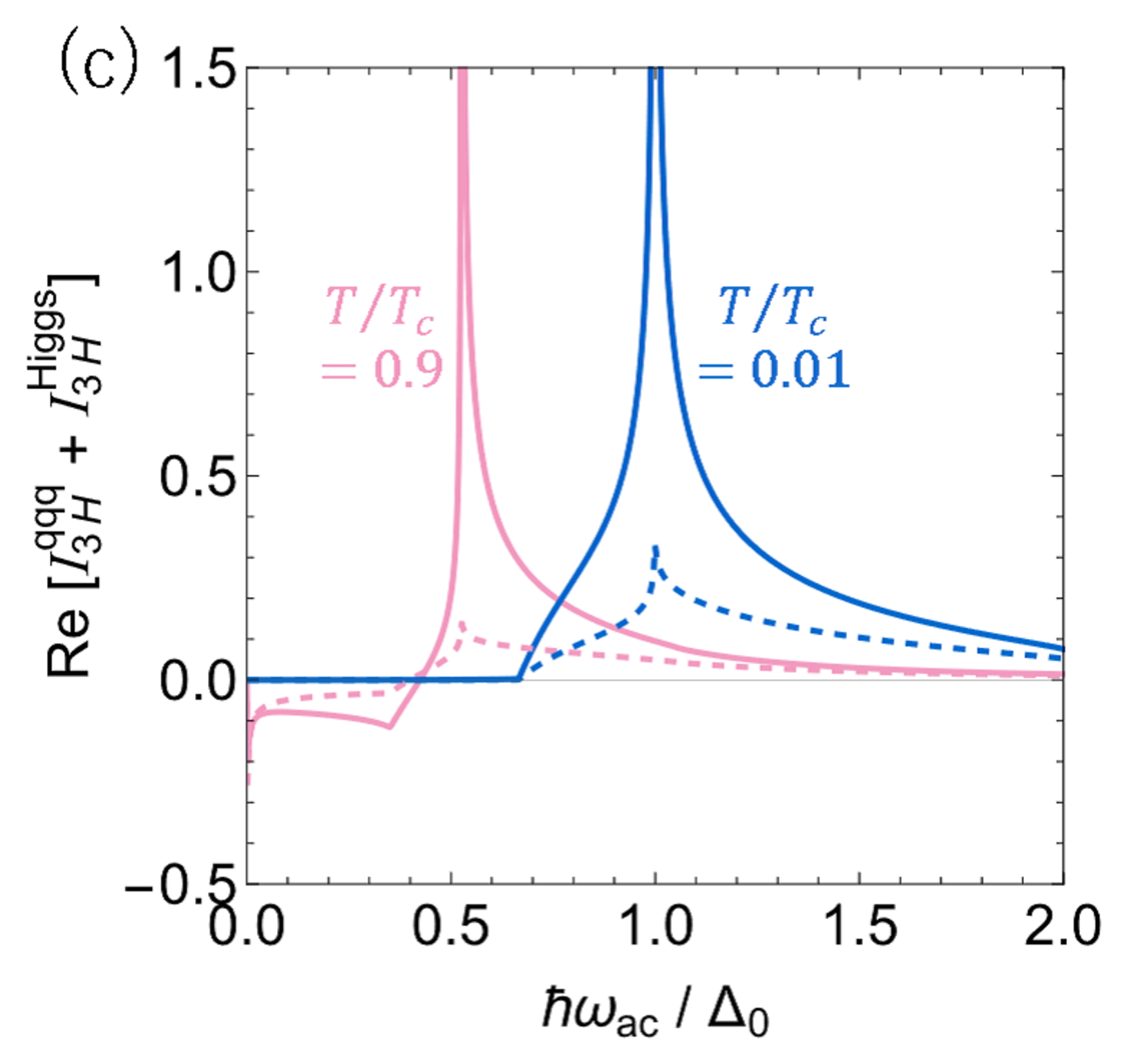}
   \includegraphics[width=0.49\linewidth]{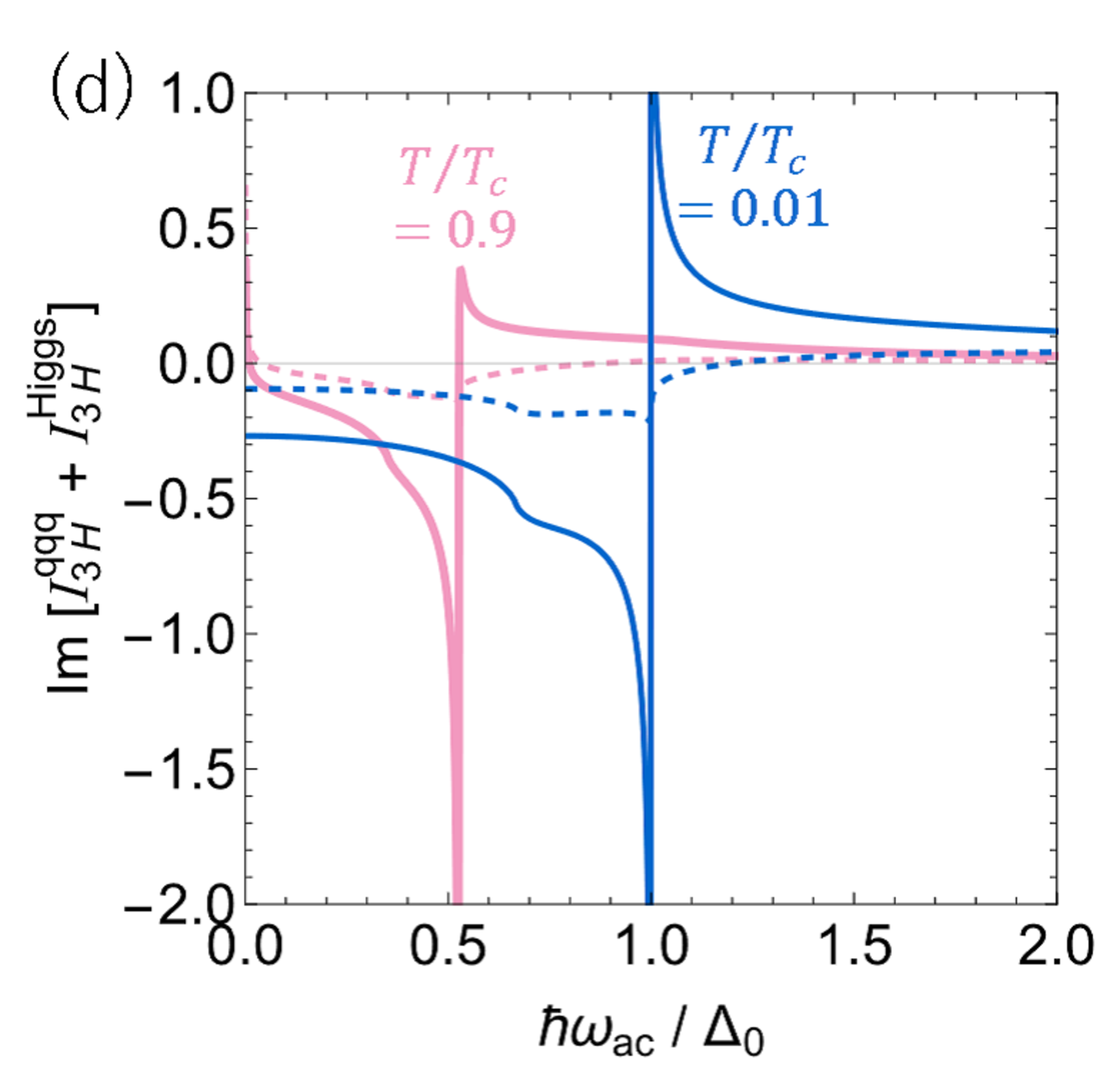}
   \end{center}\vspace{0 cm}
   \caption{
Third-harmonic current as a manifestation of the nonlinear current response.  
(a) Contribution from the direct nonlinear action of the electromagnetic field.  
(b) Contribution mediated by the Higgs mode, \( \delta\Delta_{\mathrm{2H}} \).  
(c) Real and (d) imaginary parts of the total third-harmonic current. The dashed curves represent the direct contribution \( I_{\mathrm{3H}}^{qqq} \).
   }\label{fig3}
\end{figure}

Figure~\ref{fig3} shows the third-harmonic current generated by  
(a) the direct nonlinear action of the electromagnetic field and  
(b) the contribution mediated by the Higgs mode.  
In both cases, a resonance peak appears near \( \hbar \omega_{\mathrm{ac}} \simeq \Delta \),  
where the superconducting gap takes the value \( \Delta(T) \simeq \Delta_0\) and \(0.526 \Delta_0 \) for \( T/T_c = 0.01\) and \( 0.9 \), respectively.  
Notably, the peak associated with the Higgs mode is several times stronger than that due to the direct nonlinear photon action,  
in agreement with previous studies~\cite{Silaev, Eremin}. 
See also Appendix~\ref{appendix_check_J3H} for comparison with previous studies~\cite{Silaev, Eremin}.

Figures~\ref{fig3}(c) and \ref{fig3}(d) show the real and imaginary parts of \( I_{\mathrm{3H}} \),  
which exhibit a Higgs-induced peak and dip at \( \hbar \omega_{\mathrm{ac}} = \Delta(T) \).  
The dashed curves represent the direct contribution \( I_{\mathrm{3H}}^{qqq} \), which shows a much smaller peak and dip compared to the Higgs-mediated response.  
It is worth noting that in the real part, a sharp onset appears at a frequency below the Higgs resonance peak,  
corresponding to the condition \( 3\hbar \omega_{\mathrm{ac}} = 2\Delta(T) \).

Before concluding this section, it is worth noting that in the limit \( \omega_{\rm ac} \to 0 \), all nonlinear contributions 
\( I_{\rm 1H}^{qqq} \), \( I_{\rm 1H}^{\rm Eliash} \), \( I_{\rm 1H}^{\rm Higgs} \), \( I_{\rm 3H}^{qqq} \), and \( I_{\rm 3H}^{\rm Higgs} \) 
merge into the nonlinear dc response (see Appendix~\ref{appendix_dc_current_response}), consistent with the well-established results obtained decades ago.  
This consistency between the nonlinear ac and dc responses, which to the best of my knowledge has not been explicitly confirmed in previous studies, 
supports not only our calculations of the nonlinear third-harmonic generation but also the present nonlinear correction to the first-harmonic current, which has received much less attention in the literature.

In this section, we have solved the Keldysh-Usadel equations and explicitly expressed the Eliashberg effect (\( \delta\Delta_{\rm 0H} \)), the Higgs mode (\( \delta\Delta_{\rm 2H} \)), and both the linear (\( J_{\rm 1H}^{(1)} \)) and nonlinear current responses (\( J_{\rm 1H}^{(3)} \), \( J_{\rm 3H}^{(3)} \)) in terms of the equilibrium Green's functions \( G_e \) and \( F_e \).  
In the following section, we employ these results to derive the amplitude-dependent nonlinear correction to the dissipative conductivity.

\section{Nonlinear dissipation}

In general, the electromagnetic power dissipation per unit volume can be written as 
\begin{eqnarray}
P = \overline{E(t) J(t)} , 
\end{eqnarray}
where the overline denotes time averaging, and the electric field is given by $E(t)=E_{\rm 1H} e^{-i\omega_{\rm ac}t} + {\rm c.c}$. 
Considering contributions up to third order in the ac amplitude, we have $J(t)=J^{(1)}(t) + J^{(3)}(t)$. 
The first term, arising from the linear response, yields only the first harmonic, whereas the third-order term contributes to both the first and third harmonics, as discussed in the previous section.  
Thus, we have $J(t)=J_{\rm 1H} e^{-i\omega_{\rm ac}t} + J_{\rm 3H} e^{-i 3\omega_{\rm ac}t} + {\rm c.c.}$,  with $J_{\rm 1H}=J_{\rm 1H}^{(1)}+J_{\rm 1H}^{(3)}$ and $J_{\rm 3H}=J_{\rm 3H}^{(3)}$, leading to the power dissipation $P=2{\rm Re}[ J_{\rm 1H} E_{\rm 1H}^*]$ or 
\begin{eqnarray}
&&P = P^{(1)} + P^{(3)}, \\
&&P^{(1)} = 2{\rm Re}[ J_{\rm 1H}^{(1)} E_{\rm 1H}^*] = \frac{1}{2} \sigma_1 E_0^2, \\
&&P^{(3)} = 2{\rm Re}[ J_{\rm 1H}^{(3)} E_{\rm 1H}^*] = \frac{1}{2} \delta \sigma_1 E_0^2 . 
\end{eqnarray}
Here, \( P^{(1)} \) represents the power dissipation in the linear-response regime,  
while \( P^{(3)} \) denotes the nonlinear contribution arising from the \( \mathcal{O}(q_0^3) \) correction to the first harmonic, \( J_{\rm 1H}^{(3)} \).  
The nonlinear conductivity correction \( \delta \sigma_1 \) is given by
\begin{eqnarray}
&&\delta \sigma_1 = \delta \sigma_1^{qqq} + \delta \sigma_1^{\rm Higgs}  + \delta \sigma_1^{\rm Eliash} ,\label{deltasigma} \\
&&\delta \sigma_1^{qqq} = 2\sqrt{\pi} \sigma_n {\rm Re}[ I_{\rm 1H}^{qqq} ] \frac{s}{\hbar\omega_{\rm ac}}, \\
&&\delta \sigma_1^{\rm Higgs} = 2\sqrt{\pi} \sigma_n {\rm Re} [ I_{\rm 1H}^{\rm Higgs} ] \frac{s}{\hbar\omega_{\rm ac}}, \\
&&\delta \sigma_1^{\rm Eliash} = 2\sqrt{\pi} \sigma_n {\rm Re} [ I_{\rm 1H}^{\rm Eliash} ] \frac{s}{\hbar\omega_{\rm ac}} . 
\end{eqnarray}
Here, $I_{\rm 1H}^{qqq}$, $I_{\rm 1H}^{\rm Higgs}$, and $I_{\rm 1H}^{\rm Eliash}$ are given by Eqs.~(\ref{I1H_qqq}), (\ref{I1H_Higgs}), and (\ref{I1H_Eliash}), respectively. 
It should be noted that the third-harmonic current, whose resonance peak is widely recognized as a hallmark of the Higgs mode in disordered superconductors and has been extensively investigated in the context of Higgs-mode physics, does not contribute to nonlinear dissipation due to its elimination through time averaging. 
Only the nonlinear correction to the {\it first-harmonic} current $J_{\rm 1H}^{(3)}$ remains relevant for the evaluation of nonlinear dissipation.

\begin{figure}[tb]
   \begin{center}
   \includegraphics[width=0.49\linewidth]{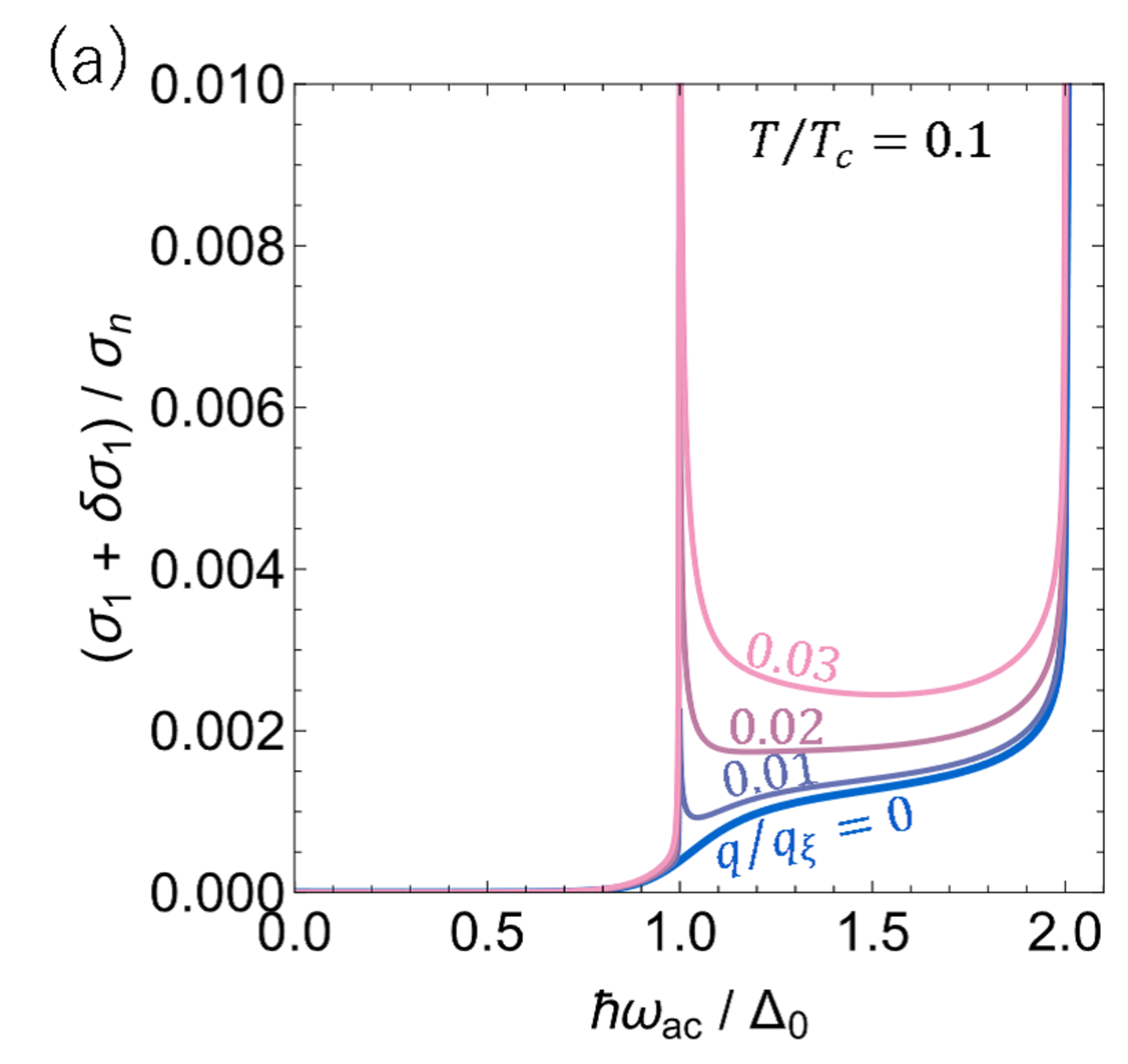}
   \includegraphics[width=0.49\linewidth]{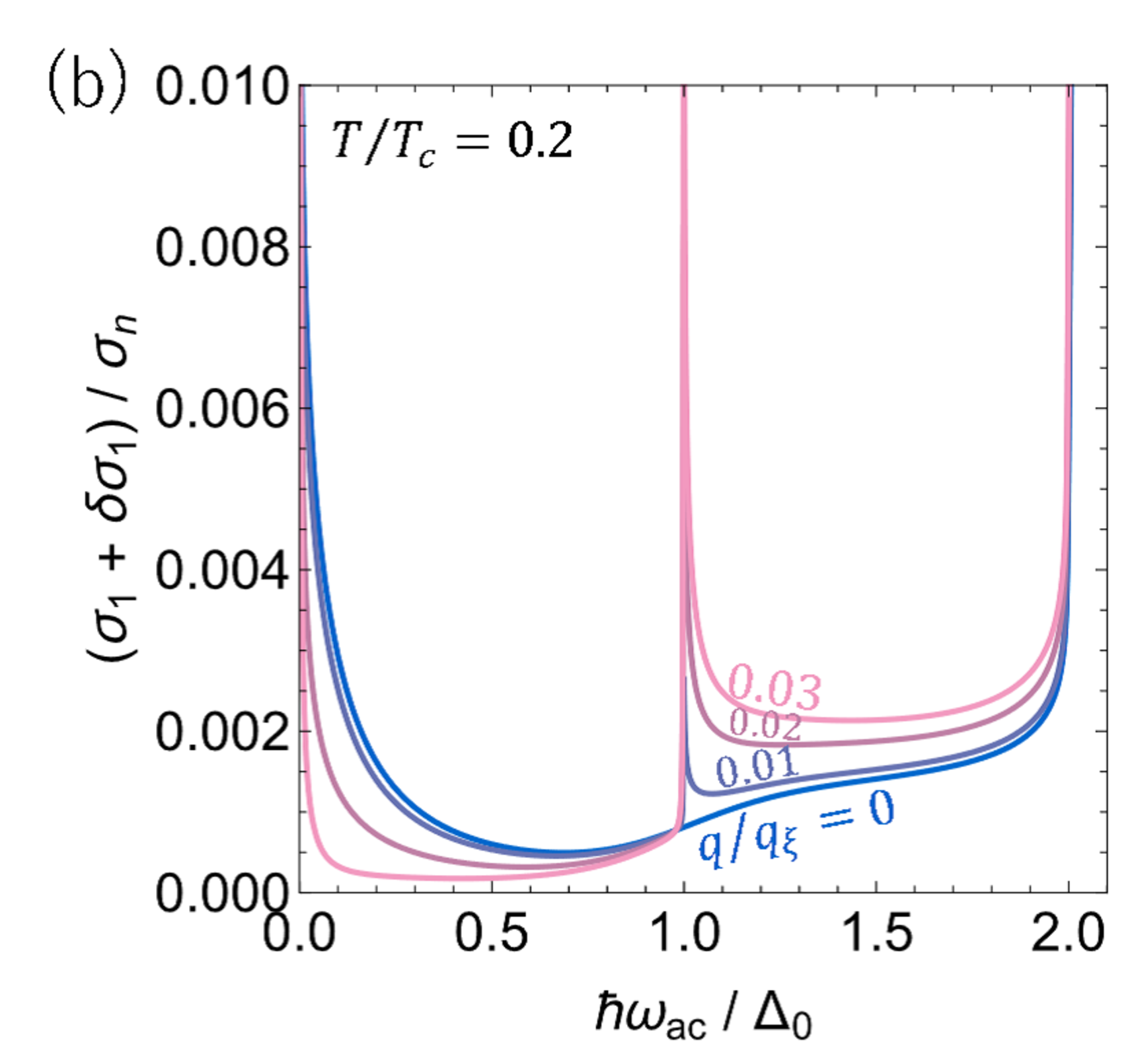}
   \includegraphics[width=0.49\linewidth]{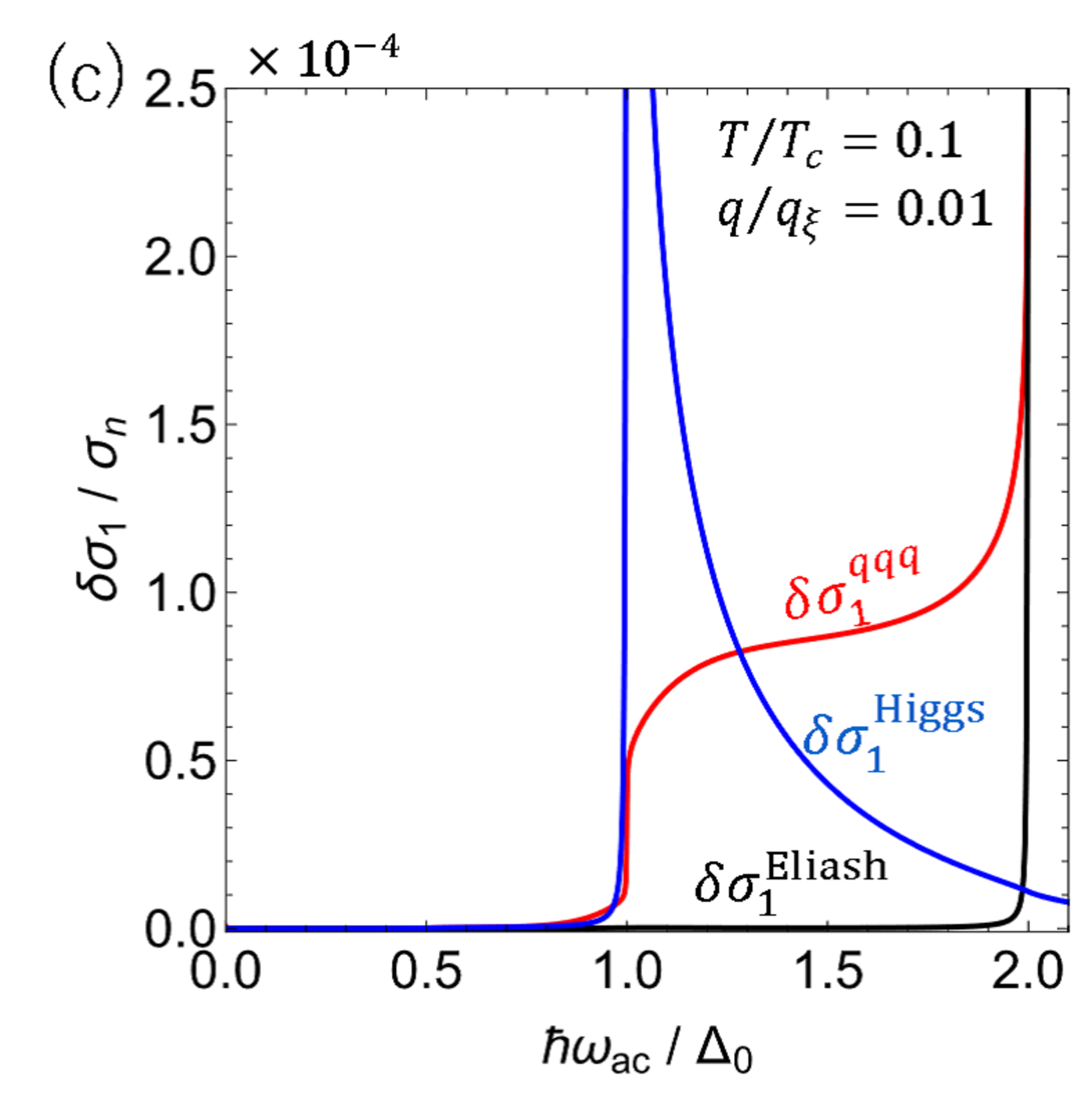}
   \includegraphics[width=0.49\linewidth]{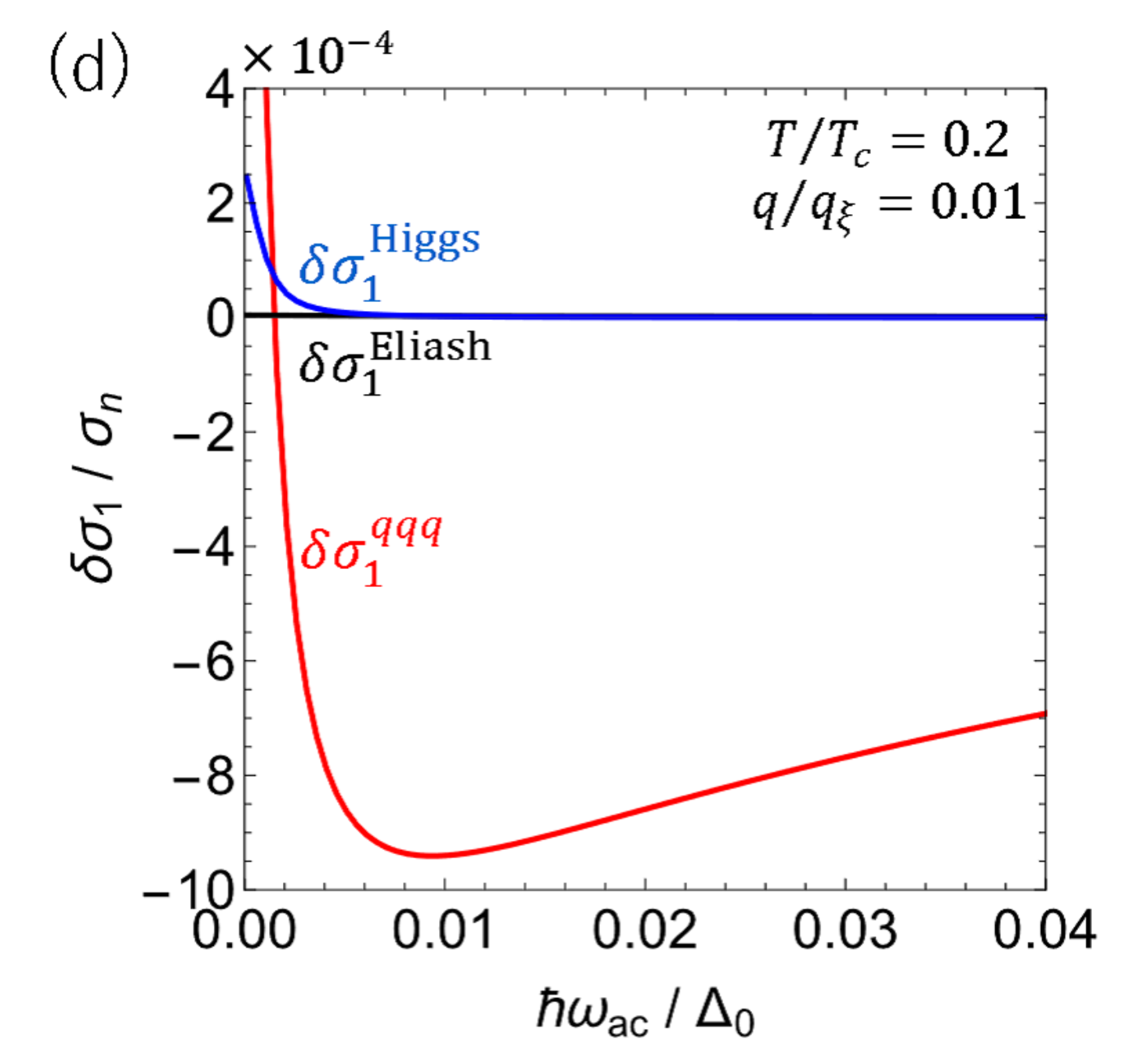}
   \end{center}\vspace{0 cm}
   \caption{
Nonlinear corrections to the real part of the complex conductivity.  
(a, b) Total nonlinear correction \( \delta \sigma_1 \) calculated at \( T/T_c = 0.1 \) and \( 0.2 \), respectively.  
(c, d) Decomposition of \( \delta \sigma_1 \) into individual contributions \( \delta \sigma_1^{qqq} \), \( \delta \sigma_1^{\rm Higgs} \) and \( \delta \sigma_1^{\rm Eliash} \). 
   }\label{fig4}
\end{figure}

Figures~\ref{fig4}(a, b) show the total dissipative conductivity \( \sigma_1 + \delta \sigma_1 \) as a function of the ac frequency \( \omega_{\rm ac} \) at \( T/T_c = 0.1 \) and \( 0.2 \).  
The blue curves represent the standard linear-response result \( \sigma_1(\omega_{\rm ac}) \),  
while the other curves include the nonlinear corrections.  
Figure~\ref{fig4}(c) presents the individual nonlinear components \( \delta \sigma_1^i \) (with \( i = qqq, \mathrm{Higgs}, \mathrm{Eliash} \)).  
This decomposition reveals that the peak near \( \hbar \omega_{\rm ac} \simeq \Delta_0 \) in Figs.~\ref{fig4}(a, b) originates from the Higgs-mode resonance,  
whereas the pronounced positive enhancement around \( \hbar \omega_{\rm ac} \simeq 2\Delta \) arises from the direct photon-driven term \( \delta \sigma_1^{qqq} \) and the Eliashberg contribution \( \delta \sigma_1^{\mathrm{Eliash}} \).  
These results demonstrate that the peak at \( \hbar \omega_{\rm ac} \simeq \Delta \) in the nonlinear dissipative conductivity under strong ac driving can serve as a signature of the Higgs mode, at least in dirty-limit superconductors.  
To fully understand the behavior in clean systems, a theoretical treatment within the Keldysh-Eilenberger framework is required.

It is also worth noting that the small shelf-like structure observed around \( \hbar \omega_{\rm ac} \gtrsim \Delta_0 \) in the blue curve of Fig.~\ref{fig4}(a), which corresponds to the linear response,  
is attributed to subgap states introduced by the small damping factor \( \Gamma / \Delta_0 = 10^{-3} \), and is known to vanish in the limit \( \Gamma \to 0 \).

To investigate finer features of the nonlinear corrections, particularly in the low-frequency regime \( \hbar \omega_{\rm ac} \ll \Delta_0 \), which is relevant to many superconducting device applications, Fig.~\ref{fig4}(d) presents a magnified view of the individual nonlinear contributions.
It shows that the direct nonlinear term \( \delta \sigma_1^{qqq} \) suppresses the dissipative conductivity in this regime. This suppression becomes increasingly pronounced with rising \( \omega_{\rm ac} \) within a certain frequency range.

\begin{figure}[tb]
   \begin{center}
   \includegraphics[width=0.49\linewidth]{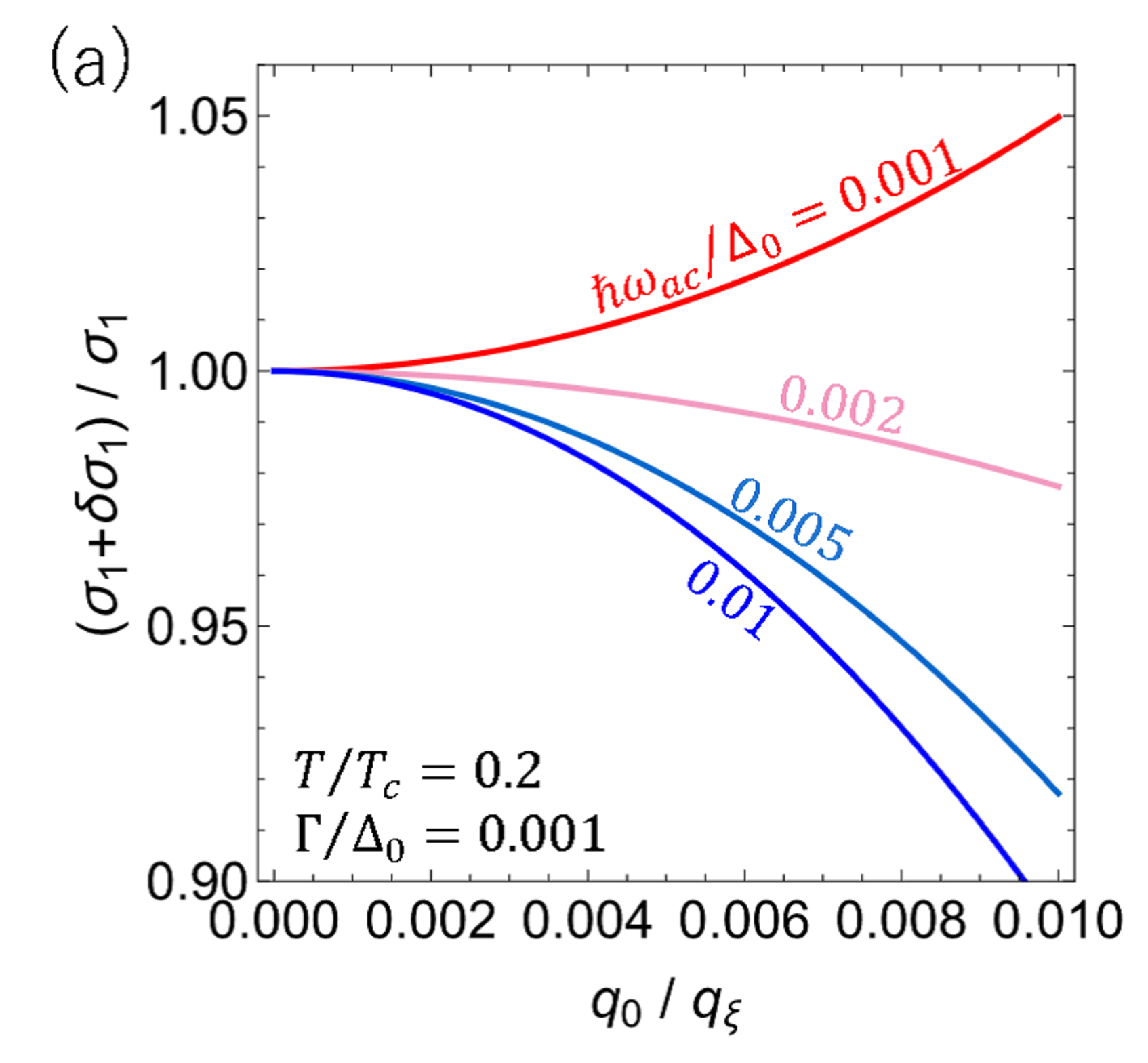}
   \includegraphics[width=0.49\linewidth]{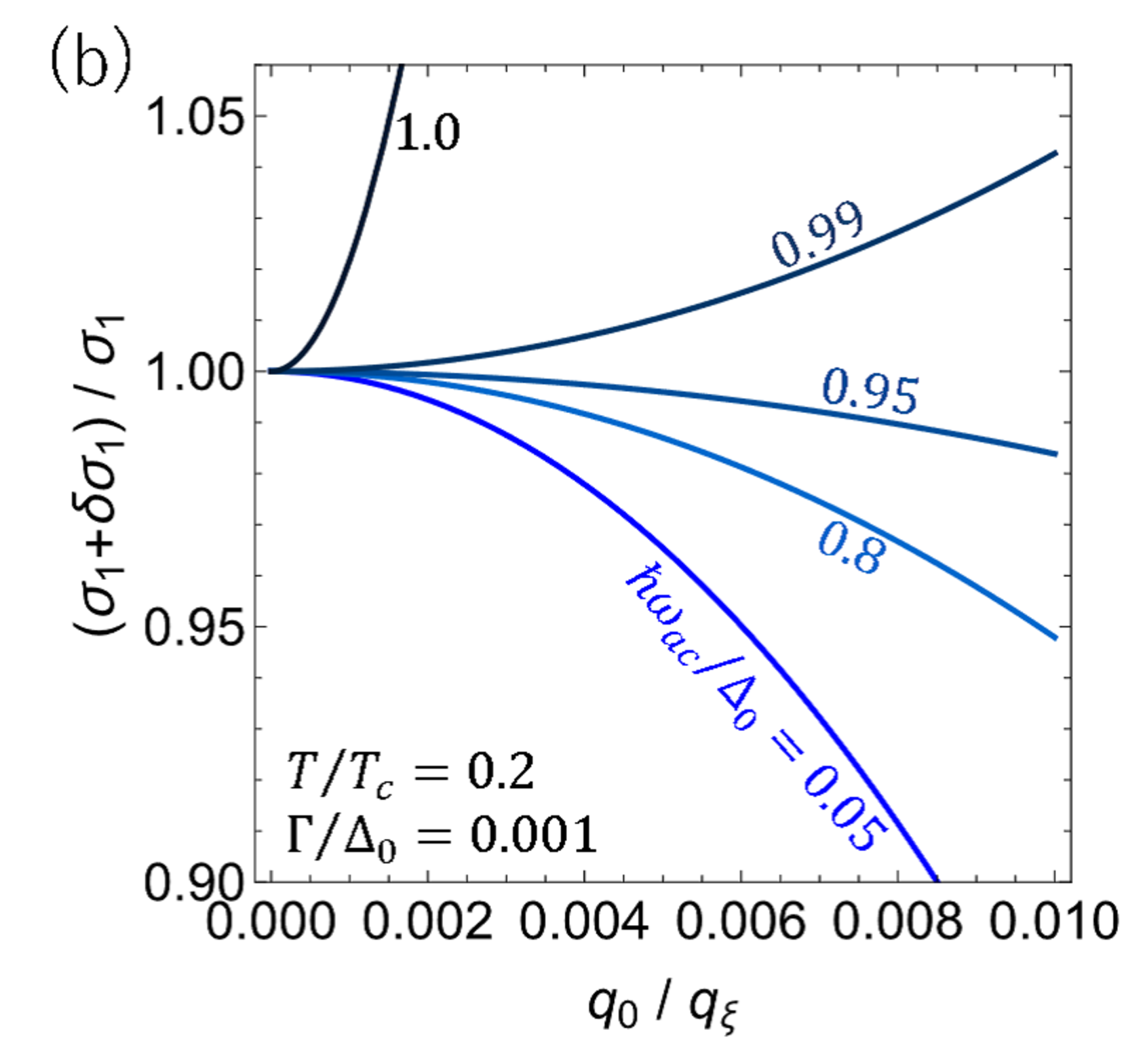}
   \includegraphics[width=0.48\linewidth]{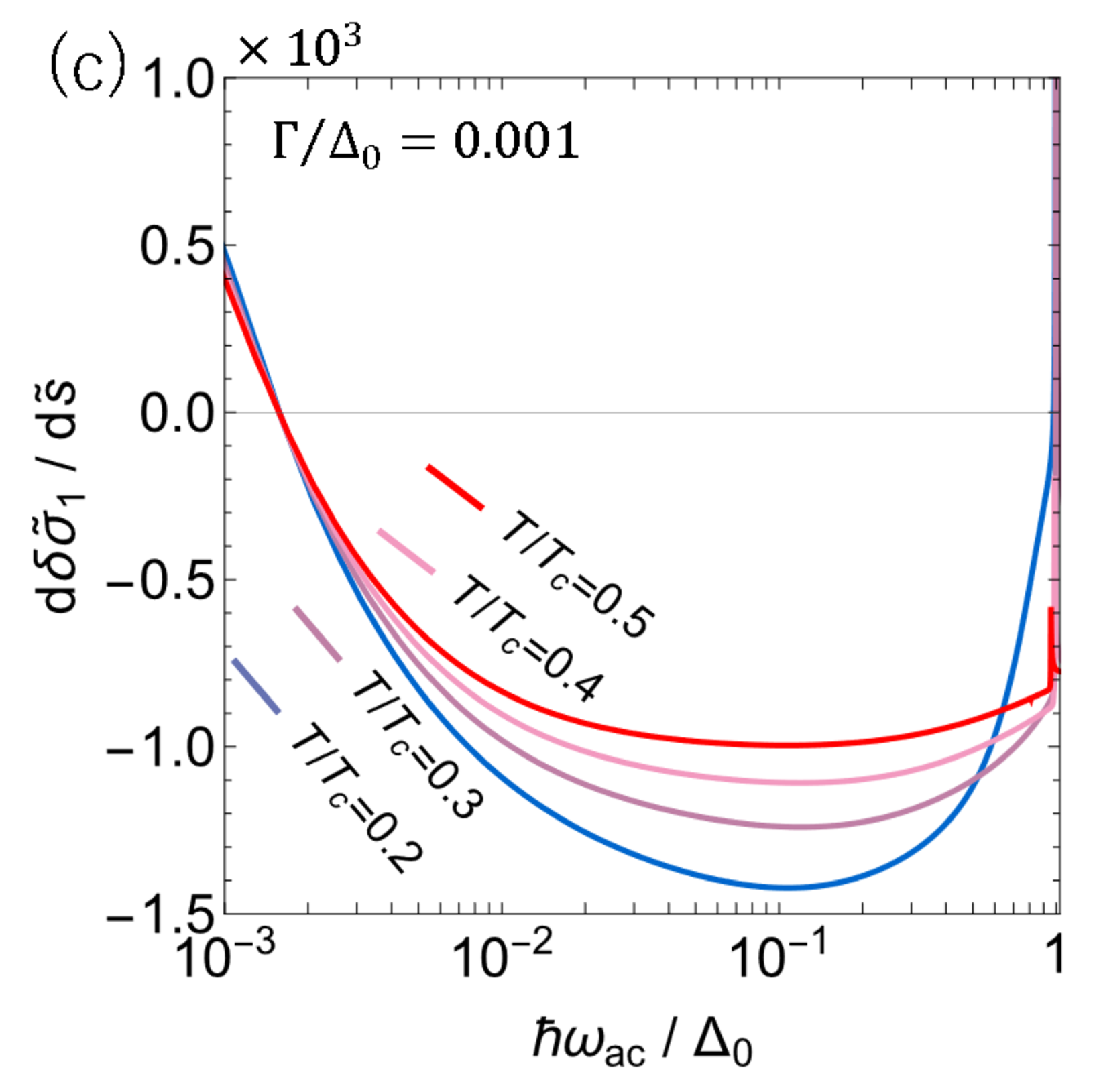}
   \includegraphics[width=0.5\linewidth]{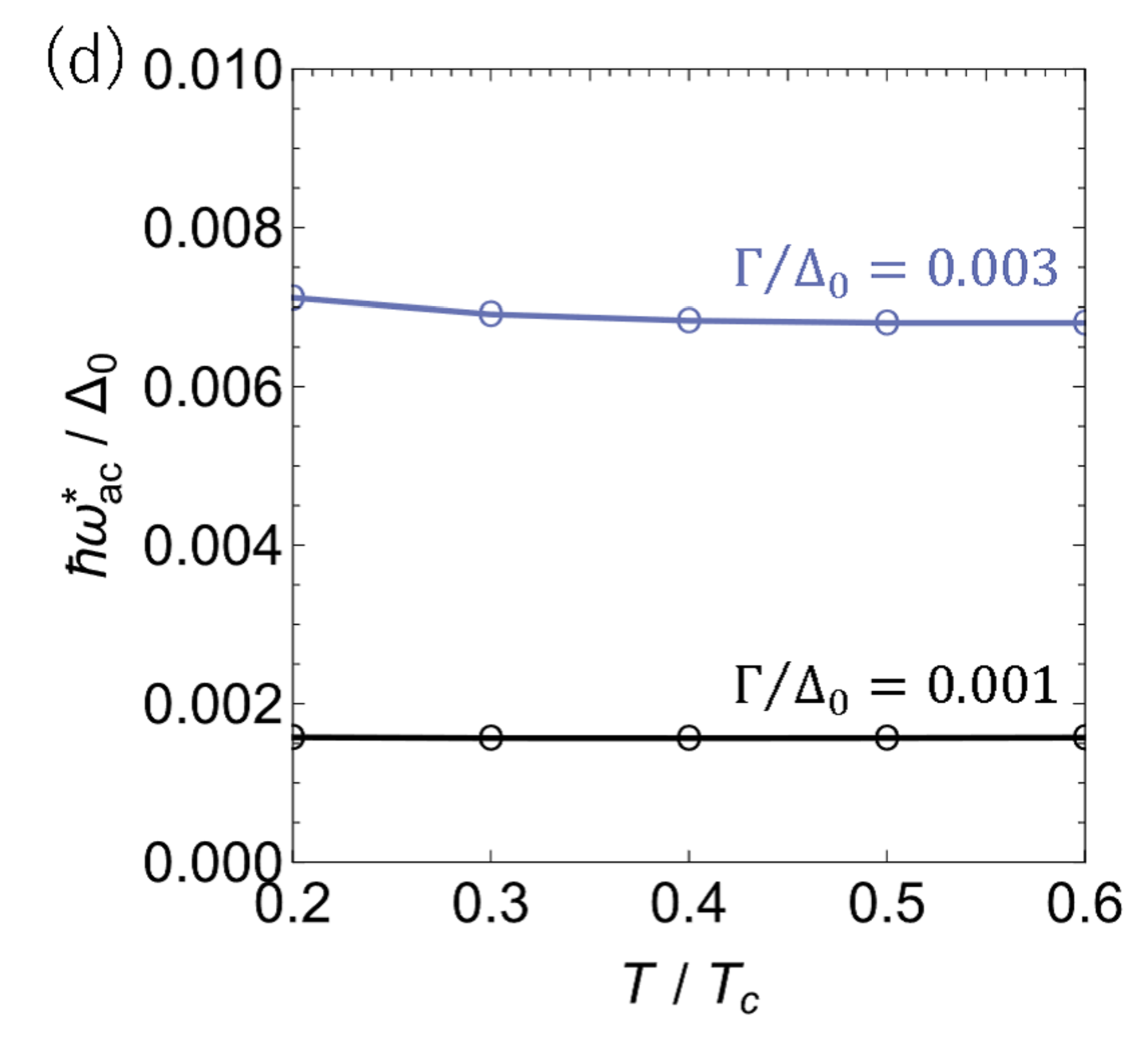}
   \end{center}\vspace{0 cm}
   \caption{
(a, b) Amplitude-dependent dissipative conductivity at various drive frequencies [cf. Fig.~4(b)].  
(c) Slope of the nonlinear correction, \( d\delta \sigma_1/ds \), as a function of frequency.  
(d) Switching frequency \( \omega_{\rm ac}^* \) as a function of temperature.  
   }\label{fig5}
\end{figure}

Figure~\ref{fig5}(a) displays the amplitude-dependent dissipative conductivity at \( T/T_c = 0.2 \) for various drive frequencies.  
At \( \hbar \omega_{\rm ac} / \Delta_0 = 0.001 \), the dissipative conductivity increases with ac amplitude, whereas it begins to decrease for \( \hbar \omega_{\rm ac} / \Delta_0 \gtrsim 0.002 \).  
This suppression becomes more pronounced as the frequency increases.  
As shown in Fig.~\ref{fig5}(b), this trend reverses at higher frequencies: for \( \hbar \omega_{\rm ac} / \Delta_0 \gtrsim 0.05 \), the suppression weakens, and by \( \hbar \omega_{\rm ac} / \Delta_0 = 0.99 \), the dissipative conductivity once again increases with amplitude. 
To interpret the behavior observed in Figs.~\ref{fig5}(a) and (b), it is helpful to revisit Fig.~\ref{fig4}(b),  
which shows that the dissipative conductivity decreases with increasing ac amplitude for most frequencies below \( \Delta \),  
while this trend is reversed near \( \Delta \), where the dissipative conductivity instead increases with ac amplitude.

A more direct way to capture this behavior is to examine the slope of the dissipative conductivity, defined as \( d \delta \tilde{\sigma}_1 / d\tilde{s} \),  
where \( \delta \tilde{\sigma}_1 := \delta \sigma_1 / \sigma_1 \) and \( \tilde{s} := s / \Delta_0 = (q_0 / q_{\xi})^2 \).  
Figure~\ref{fig5}(c) shows \( d \delta \tilde{\sigma}_1 / d\tilde{s} \) as a function of \( \omega_{\rm ac} \).  
At \( \hbar \omega_{\rm ac} / \Delta_0 \simeq 0.002 \), the derivative vanishes, indicating that below (above) this frequency,  
the dissipative conductivity increases (decreases) with increasing ac amplitude [see also Fig.~\ref{fig5}(a)].  
We define the switching frequency \( \omega_{\rm ac}^* \) such that $d\delta \sigma_1/ds=0$. 
Figure~\ref{fig5}(d) plots the switching frequency \( \omega_{\rm ac}^* \) as a function of temperature,  
indicating that \( \omega_{\rm ac}^* \) is largely insensitive to temperature variations.

\begin{figure}[tb]
   \begin{center}
   \includegraphics[width=0.49\linewidth]{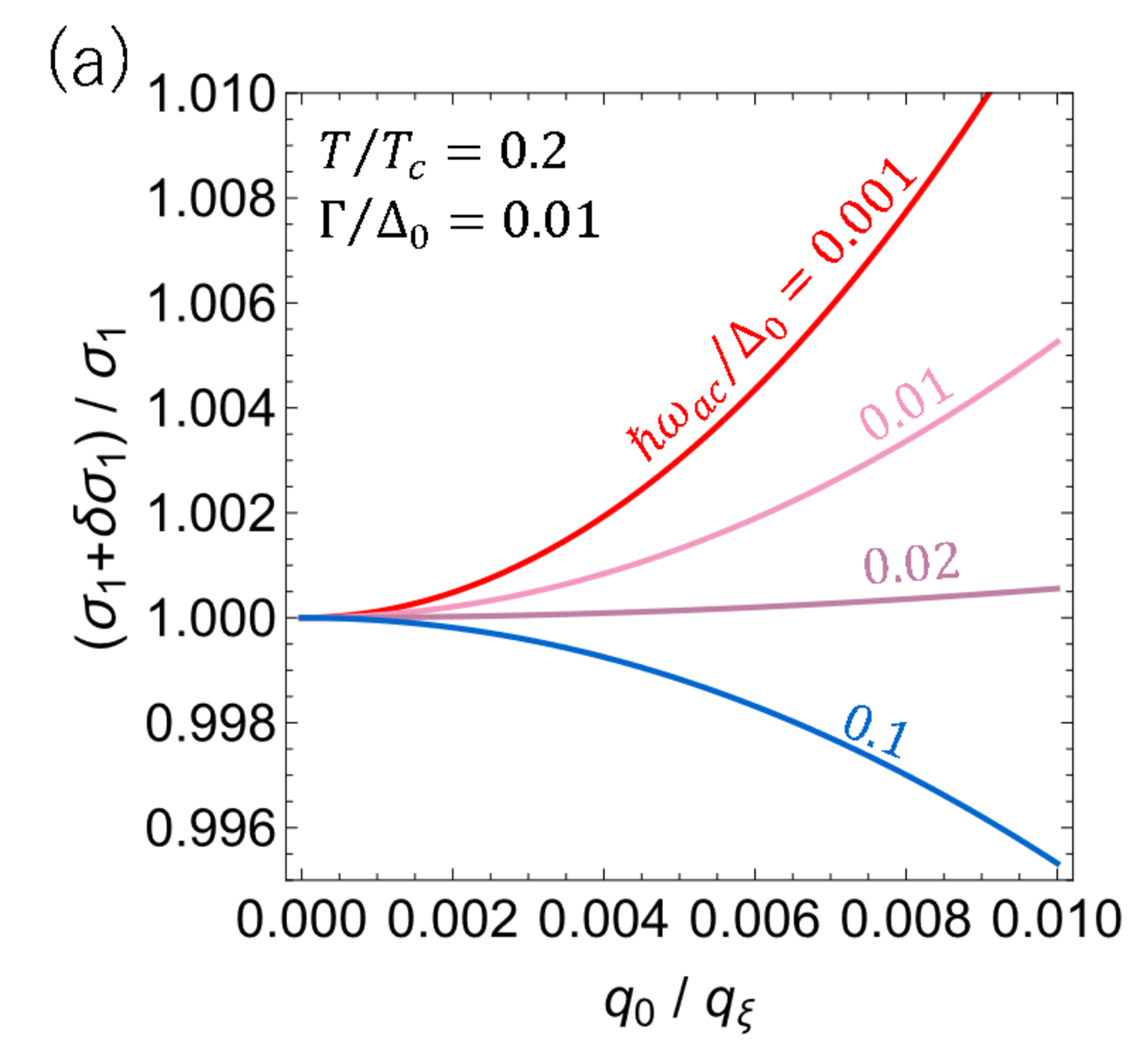}
   \includegraphics[width=0.49\linewidth]{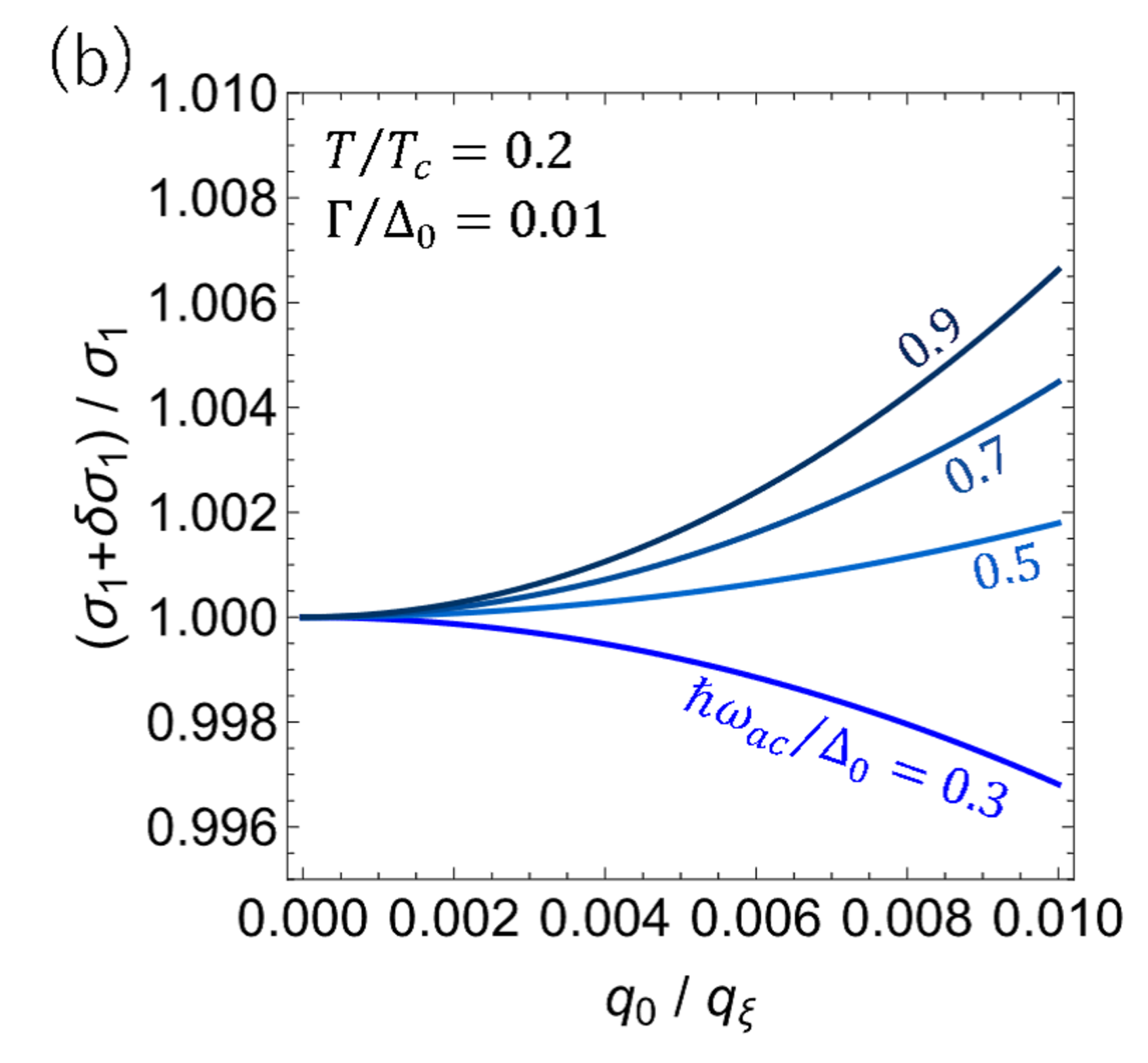}
   \includegraphics[width=0.49\linewidth]{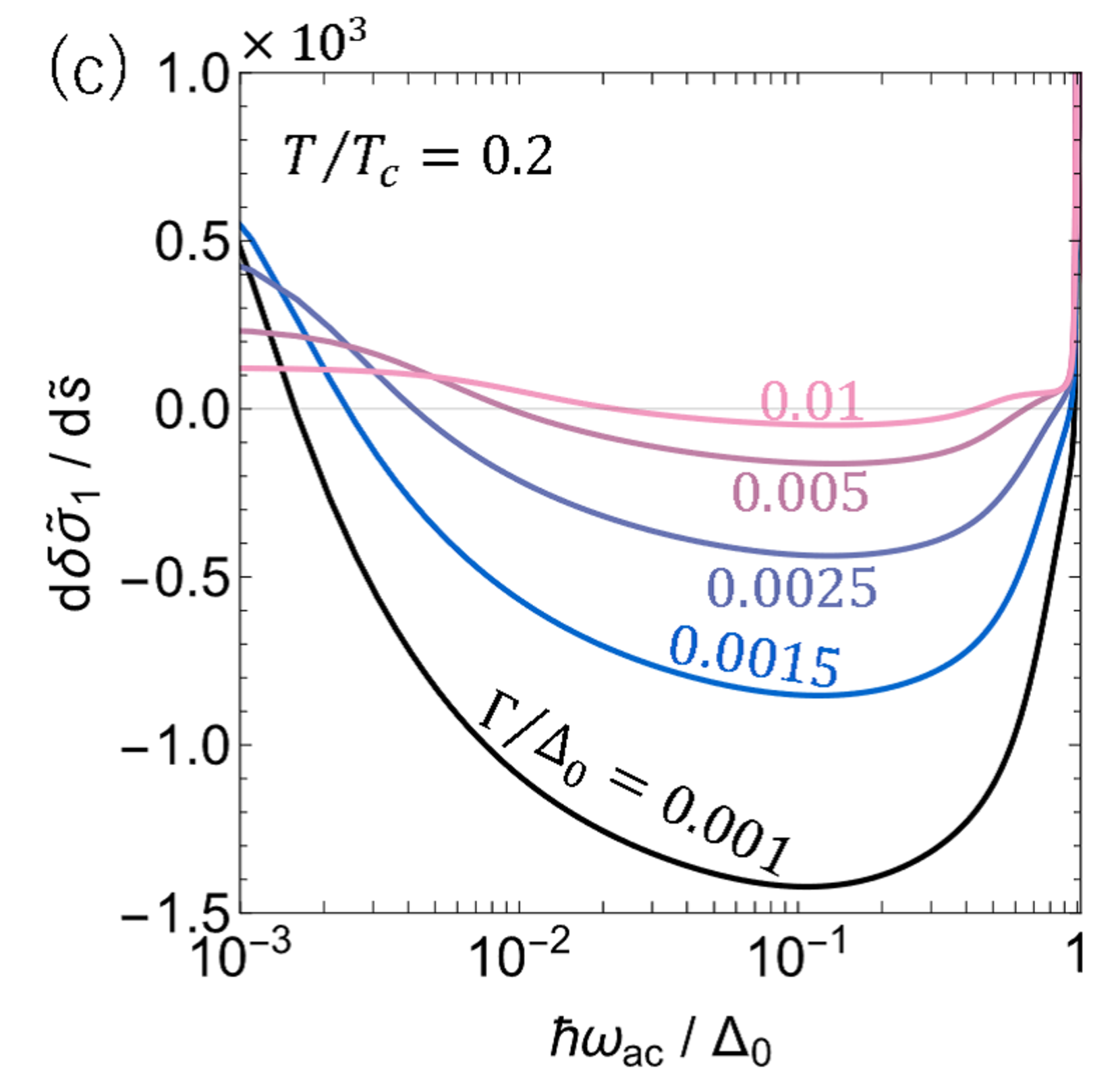}
   \includegraphics[width=0.49\linewidth]{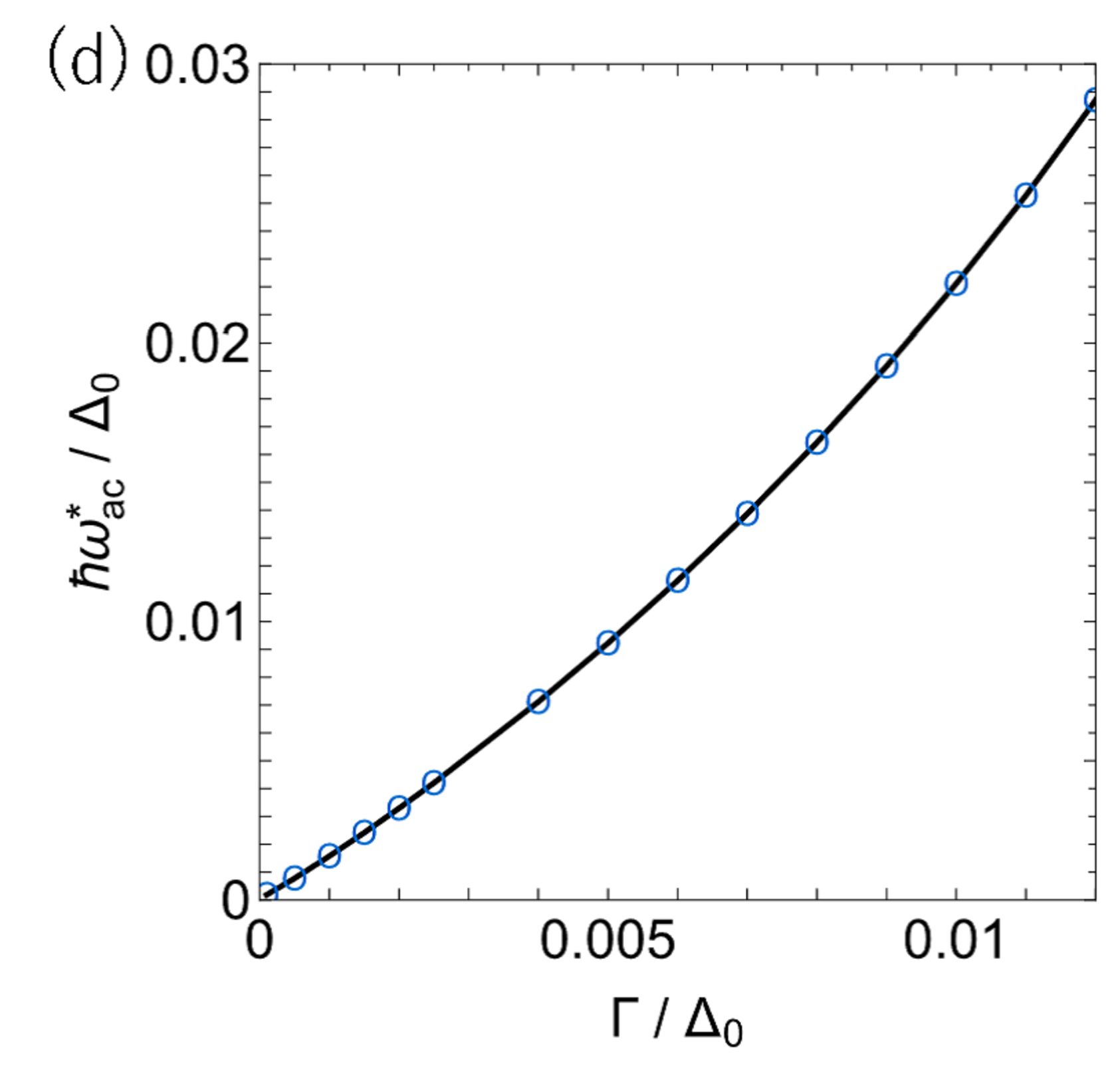}
   \end{center}\vspace{0 cm}
   \caption{
Effects of \( \Gamma \) on the amplitude-dependent dissipative conductivity.  
(a, b) Dissipative conductivity calculated for a moderately large damping factor, \( \Gamma/\Delta_0 = 0.01 \), at \( T/T_c = 0.2 \), shown for various drive frequencies.  
(c) Slope of the normalized nonlinear correction, \( d\delta\tilde{\sigma}_1/d\tilde{s} \), plotted as a function of frequency for different values of \( \Gamma \).  
(d) Switching frequency \( \omega_{\rm ac}^* \) as a function of \( \Gamma \) calculated for $T/T_c=0.2$. Note that the dependence \( \omega_{\rm ac}^*(\Gamma) \) exhibits little sensitivity to temperature [cf.~Fig.~\ref{fig5}(d)].
   }\label{fig6}
\end{figure}

It is well known that the damping factor \( \Gamma \), which corresponds to the inverse quasiparticle lifetime, influences the dissipative conductivity both in the linear-response regime~\cite{Gurevich_Kubo, 2022_Kubo} and beyond~\cite{Kubo_Gurevich}.  
The nonlinear correction \( \delta \sigma_1 \) is also significantly affected by \( \Gamma \).  
Figures~\ref{fig6}(a, b) present the results for \( \Gamma/\Delta_0 = 0.01 \) at \( T/T_c = 0.2 \), which is an order of magnitude larger than the value used in the previous figures.  
Overall, the amplitude dependence of the dissipative conductivity becomes weaker.  
As a result, the switching frequency shifts upward to \( \hbar \omega_{\rm ac}^* / \Delta_0 \simeq 0.02 \).  
This behavior is clearly illustrated in Fig.~\ref{fig6}(c), where \( d\delta\tilde{\sigma}_1/d\tilde{s} \) is plotted as a function of \( \omega_{\rm ac} \) for several values of \( \Gamma \).  
Figure~\ref{fig6}(d) shows the switching frequency \( \omega_{\rm ac}^* \) as a function of \( \Gamma \), indicating a strong dependence of \( \omega_{\rm ac}^* \) on the damping factor.
A note on the amplitude-dependent behavior at lower temperatures is given in Appendix~\ref{appendix_Gamma}.

To evaluate the surface resistance \( R_s \) or the quality factor \( Q \) of a superconducting cavity, it is necessary to account for the depth dependence of \( \delta \sigma_1 \) from the surface into the bulk over a length scale set by the London penetration depth. This spatial dependence arises due to the Meissner effect, which causes the ac magnetic field to decay exponentially into the superconductor; as a result, the field-dependent correction \( \delta \sigma_1 \) also exhibits exponential attenuation with depth.
After some calculations (see Appendix~\ref{appendix_Rs}), we obtain the following expression for the relative change in surface resistance in the limit \( \sigma_1 / \sigma_2 \ll 1 \):
\begin{equation}
\frac{\delta R_s}{R_s} = \frac{\delta \sigma_1/\sigma_1}{(q/q_\xi)^2} \left( \frac{B_{\mathrm{rf}}}{\sqrt{2\pi} B_c} \right)^2 ,
\end{equation}
where \( B_{\mathrm{rf}} \) is the amplitude of the rf magnetic field at the surface and \( B_c \) is the thermodynamic critical field.
Consequently, the plots in Figure~\ref{fig5} can be reinterpreted by replacing the horizontal axis as
\begin{equation}
\frac{q_0}{q_\xi} \longrightarrow \frac{B_{\mathrm{rf}}}{\sqrt{2\pi} B_c} ,
\end{equation}
and the vertical axis as either \( R_s(B_{\mathrm{rf}})/R_s(0) \) or \( Q^{-1}(B_{\mathrm{rf}})/Q^{-1}(0) \).

Note that the validity range of Eq.~(\ref{deltasigma}) is expected to be limited.  
As the ac amplitude increases, modifications of the quasiparticle density of states (DOS) induced by the ac current become significant, which cannot be captured by the present perturbative calculation.  
To estimate the threshold at which such nonperturbative effects become important, we consider the spectral gap under a dc current, given by \( \epsilon_g = (\Delta^{2/3} - s^{2/3})^{3/2} \), from which the gap reduction is approximately \( \delta \epsilon_g \simeq (3/2) \Delta_0^{1/3} s^{2/3} \).  
This modification becomes relevant when \( \delta \epsilon_g \gtrsim \hbar \omega_{\rm ac} \), or equivalently, when \( q_0 / q_\xi \gtrsim (2 \hbar \omega_{\rm ac} / 3 \Delta_0)^{3/4} \).  
For example, for \( \hbar \omega_{\rm ac}/\Delta_0 = 0.01 \), this condition gives \( q_0 / q_\xi \gtrsim 0.02 \), beyond which our perturbative approach is no longer valid and nonperturbative modifications of the quasiparticle spectrum must be taken into account.
Taking a Nb resonator as an example (\( \Delta_0^{\mathrm{Nb}} \simeq 360\,\mathrm{GHz}, \, B_c^{\rm Nb}= 200\,{\rm mT} \)),  
a dimensionless amplitude \( q_0 / q_\xi = 0.01 \) corresponds to a magnetic field amplitude \(  B_{\mathrm{rf}} \simeq 3.5\,\mathrm{mT} \).

\section{Discussion and conclusion}

We solved the Keldysh-Usadel equations under a monochromatic ac field [Eqs.~(\ref{RA2})-(\ref{gAno1})], obtaining nonequilibrium corrections to the pair potential and the nonlinear current response.  
The zero-harmonic component of the nonequilibrium pair-potential variation, \( \delta\Delta_{\rm 0H} \), which represents the (positive or negative) Eliashberg effect, is given by Eq.~(\ref{deltaDelta0H}) and plotted in Fig.~\ref{fig1}(a).  
The second-harmonic component, \( \delta\Delta_{\rm 2H} \), corresponding to the Higgs mode, is given by Eq.~(\ref{deltaDelta2H}) and shown in Fig.~\ref{fig1}(b).  
The resulting nonlinear current response includes both the first-harmonic correction \( J_{\rm 1H}^{(3)} \) [Eq.~(\ref{J1H}), Fig.~\ref{fig2}] and the third-harmonic component \( J_{\rm 3H}^{(3)} \) [Eq.~(\ref{J3H}), Fig.~\ref{fig3}].  
The third-harmonic current arises from two sources: the direct nonlinear action of the photon field and the contribution mediated by the Higgs mode.  
In contrast, the first-harmonic correction incorporates all three mechanisms: the direct photon action, the Higgs-mode contribution, and the Eliashberg effect.  
While the third-harmonic current is widely regarded as a hallmark of the Higgs resonance in disordered superconductors (see also Fig.~\ref{fig3}),  
Fig.~\ref{fig2} demonstrates that the nonlinear correction to the first-harmonic current also exhibits a peak and dip associated with the same resonance.

The expression for the amplitude-dependent correction to the dissipative conductivity is given by Eq.~(\ref{deltasigma}), which originates from the nonlinear correction to the \textit{first-harmonic} current.  
We found that the contribution mediated by the Higgs mode exhibits a resonance peak in the dissipative conductivity at \( \hbar \omega_{\rm ac} \simeq \Delta_0 \), and this peak becomes more pronounced with increasing ac amplitude (see Fig.~\ref{fig4}). 
This behavior may serve as an additional hallmark of the Higgs mode, at least in dirty-limit superconductors, complementing the well-known third-harmonic generation. Clean systems, on the other hand, require further investigation (Notably, Ref.~\cite{2025_Tsuji}, which appeared during the review process of this work, reports a similar resonance peak in both dirty and clean superconductors, attributed mainly to the Higgs mode and quasiparticles, respectively).
In contrast, in the low-frequency regime \( \hbar \omega_{\rm ac} \ll \Delta_0 \), which is most relevant for practical superconducting devices, the Eliashberg and Higgs-mode contributions are relatively weak.  
In this regime, the nonlinear conductivity is dominated by the direct term \( \delta \sigma_1^{qqq} \), as illustrated in Figs.~\ref{fig4}.  
Figure~\ref{fig5} presents the amplitude dependence of the dissipative conductivity for various drive frequencies and temperatures.  
A crossover from positive to negative amplitude dependence is observed around \( \hbar \omega_{\rm ac} / \Delta_0 \simeq 0.002 \) when $\Gamma/\Delta_0=10^{-3}$.  
Above this switching frequency $\omega_{\rm ac}^*$, the amplitude-induced reduction in the dissipative conductivity becomes more pronounced with increasing frequency or decreasing temperature.

Interestingly, the switching frequency \( \omega_{\mathrm{ac}}^* \) exhibits a remarkable insensitivity to temperature [see Figs.~\ref{fig5}(c, d)], while being predominantly governed by the damping factor \( \Gamma \) [see Figs.~\ref{fig6}(c, d)]. 
A deeper theoretical understanding of the origin of this behavior lies beyond the scope of the present work but represents an important avenue for future investigation.

It is instructive to interpret the results using concrete material parameters.  
Taking Nb as an example, \( T/T_c = 0.2 \), as used in Fig.~\ref{fig5}(a, b), corresponds to \( T \simeq 1.8\,\mathrm{K} \).  
Assuming \( \Delta_0^{\mathrm{Nb}} \simeq 360\,\mathrm{GHz} \), the frequencies shown in Fig.~\ref{fig5}(a), namely \( \hbar \omega_{\mathrm{ac}} / \Delta_0 = 0.001 \)-0.01, correspond to 360\,MHz-3.6\,GHz.  
In Fig.~\ref{fig5}(c), the amplitude-induced reduction in dissipative conductivity becomes slightly more pronounced as the temperature decreases from 4.6\,K to 1.8\,K, while the switching frequency \( \omega_{\rm ac}^* \) remains insensitive to temperature. This trend is consistent with experimental observations~\cite{Dhakal}.
In the superconducting radio-frequency (SRF) community, experiments have placed the switching frequency near 1\,GHz, corresponding to \( \hbar \omega_{\rm ac}^* / \Delta_0 \simeq 0.003 \)~\cite{Martinello, Dhakal}, and scanning tunneling spectroscopy measurements in nitrogen-doped Nb suggest \( \Gamma / \Delta_0 \lesssim 0.002 \)~\cite{Iavarone}. 
This combination of \( \omega_{\rm ac}^* \) and \( \Gamma \) agrees very well with the theoretical prediction shown in Fig.~\ref{fig6}(d).

Of course, our results apply to any conventional superconductor. Taking \( \mathrm{Nb_3Sn} \) as an example and assuming \( \Delta_0^{\mathrm{Nb_3Sn}} \simeq 770\,\mathrm{GHz} \), the switching frequency is given by \( \hbar \omega_{\mathrm{ac}}^* / \Delta_0 \simeq 0.0016 \), corresponding to \( f^* \simeq 1.2\,\mathrm{GHz} \) for \( \Gamma/\Delta_0 = 0.001 \), and \( f^* \simeq 12\,\mathrm{GHz} \) for \( \Gamma/\Delta_0 = 0.01 \).
This suggests that the widely used \( 1.3\,\mathrm{GHz} \) \( \mathrm{Nb_3Sn} \) accelerator cavities may exhibit the amplitude-dependent enhancement of the quality factor, known as the \emph{anti-}Q slope, provided that the cavity surface satisfies \( \Gamma/\Delta_0 = 0.001 \). In contrast, if the surface has \( \Gamma/\Delta_0 \simeq 0.01 \), the cavity is expected to show an amplitude-induced \emph{degradation} of the quality factor.
Earlier experimental studies reported values around \( \Gamma/\Delta_0 \gtrsim 0.01 \)~\cite{Becker}, which would require cavity frequencies exceeding \( f \simeq 12\,\mathrm{GHz} \) to observe the anti-Q slope for the same material quality. In other words, if the anti-Q slope is observed at a few GHz, it may be worth examining whether \( \Gamma \) has been significantly reduced in those samples.
It should be noted, however, that the present theory assumes the dirty limit, and experimental verification of these predictions may require tuning the electron mean free path to ensure consistency with this regime.

Taking aluminum as an example, the switching frequency corresponds to 70-\( 700\,\mathrm{MHz} \) for $\Gamma/\Delta_0 \simeq 0.001$-0.01, assuming \( \Delta_0^{\mathrm{Al}} \simeq 44\,\mathrm{GHz} \), which is significantly lower than the typical resonator frequencies used in superconducting devices.  
Therefore, conventional aluminum resonators operating in the several-gigahertz range are expected to exhibit clear amplitude-induced \( Q \)-enhancement.

Our analysis is based on the Keldysh-Usadel theory of nonequilibrium superconductivity with BCS coupling, which fully captures nonequilibrium quasiparticles and pair potential dynamics. The nonequilibrium distribution function is encoded in the Keldysh component of the Green's function, but phonon degrees of freedom are not explicitly included. Therefore, a promising direction for future work is to extend the present rigorous framework to incorporate phonon dynamics.

It should be noted that several previous studies have attempted to incorporate nonequilibrium dynamics of quasiparticles and phonons in order to explain the effects of strong microwave fields on superconducting devices.  
For instance, Refs.~\cite{2013_Goldie, 2014_Visser} obtained nonequilibrium distribution functions for both quasiparticles and phonons by solving kinetic equations.  
However, they applied these results by substituting the quasiparticle distribution into the Mattis--Bardeen formula in place of the equilibrium one.  
As demonstrated in the present work, such a naive extension is not theoretically justified in the nonlinear regime, as it neglects essential contributions, including the direct nonlinear response to the photon field.  
A more rigorous theoretical framework is therefore needed.

The present \textit{perturbative} theory does not account for \textit{nonperturbative} effects of the ac field, such as ac-induced modifications to the quasiparticle spectrum~\cite{2014_Gurevich, Kubo_Gurevich,  Semenov}.  
As a result, it becomes inapplicable at high field amplitudes, where pair breaking substantially alters the density of states and dominates the dissipative response.  
Extending the present framework to include such \textit{nonperturbative} effects remains an important direction for developing a comprehensive nonlinear \( Q \) theory that remains valid up to the depairing limit.

Nevertheless, for ac currents well below the depairing threshold, our theory remains fully applicable and offers a rigorous framework for understanding nonlinear dissipation and quality factor evolution in superconducting resonators.


\begin{acknowledgments}
I am profoundly grateful to my respectful sensei, Alex Gurevich, for his hospitality at Old Dominion University during my two-year sabbatical from 2017 to 2019, which was supported by the KEK long-term research abroad program and JSPS KAKENHI Grants No. JP17KK0100.  
I also extend my deepest gratitude to everyone who generously supported me during my three years of paternity leave from 2021 to 2024~\cite{ikuji}, which was enabled by the Act on Childcare Leave of Japan.  
These five years afforded me the invaluable opportunity to dedicate significant time to expanding my knowledge, during which I acquired all the theoretical techniques utilized in the present work.  
Without these periods, this work could never have been completed. 
This work is supported by JSPS KAKENHI Grants No. JP23H00125, JP25K01610, and JP25K23386. 
\end{acknowledgments}

\vspace{1cm}
\hspace{1.8cm} {\bf DATA AVAILABILITY}\\
The numerical data used to generate the figure plots are openly available~\cite{data}; 
embargo periods may apply.

\appendix

\section{Equations for zero and second harmonics} \label{appendix_1}

\subsection{Equations for the zero harmonic} 

Equation~(\ref{RA2}) with $(\eta, \eta') = (\omega_{\rm ac}, -\omega_{\rm ac})$ and $(-\omega_{\rm ac}, \omega_{\rm ac})$ yields the equations governing the $R$ and $A$ components of the zero harmonic ($\omega = 0$) in Fourier space:  
\begin{eqnarray}
&&-i\frac{s}{8}  \Bigl[ 
\hat{\tau_3} \hat{g}_e^r ( \epsilon - \hbar \omega_{\rm ac}) \hat{\tau}_3 \hat{g}_e^r ( \epsilon ) 
- \hat{g}_e^r ( \epsilon) \hat{\tau}_3 \hat{g}_e^r ( \epsilon + \hbar \omega_{\rm ac}) \hat{\tau}_3 
\nonumber \\
&&+ \hat{\tau_3} \hat{g}_e^r ( \epsilon + \hbar \omega_{\rm ac}) \hat{\tau}_3 \hat{g}_e^r ( \epsilon ) 
- \hat{g}_e^r ( \epsilon) \hat{\tau}_3 \hat{g}_e^r ( \epsilon - \hbar \omega_{\rm ac}) \hat{\tau}_3 
\Bigr] 
\nonumber \\
&=&  [\epsilon \hat{\tau}_3+ \hat{\Delta}_e , \delta \hat{g}^r_{\rm 0H}] 
+ [\delta \hat{\Delta}_{\rm 0H},  \hat{g}_e^r (\epsilon) ] 
 . \label{RA_0H}
\end{eqnarray}
The normalization condition, Eq.~(\ref{normRA2}), leads to  
\begin{eqnarray}
\hat{g}^{r}_{e} (\epsilon ) \delta \hat{g}^{r}_{\rm 0H} 
+ \delta \hat{g}^{r}_{\rm 0H} \hat{g}^{r}_{e}(\epsilon) =0 . \label{normRA_0H}
\end{eqnarray}
Similarly, Eq.~(\ref{K2}) provides the equations governing the $K$ components of the zero harmonic in Fourier space:  
\begin{eqnarray}
&&-i\frac{s}{8}\Bigl[ 
\hat{\tau_3} \hat{g}_e^R ( \epsilon - \hbar \omega_{\rm ac} ) \hat{\tau}_3 \hat{g}_e^K ( \epsilon ) 
- \hat{g}_e^R ( \epsilon ) \hat{\tau}_3 \hat{g}_e^K ( \epsilon + \hbar \omega_{\rm ac} ) \hat{\tau}_3 
\nonumber \\
&&+\hat{\tau_3} \hat{g}_e^K ( \epsilon - \hbar \omega_{\rm ac}) \hat{\tau}_3 \hat{g}_e^A ( \epsilon) 
- \hat{g}_e^K ( \epsilon) \hat{\tau}_3 \hat{g}_e^A ( \epsilon + \hbar \omega_{\rm ac}) \hat{\tau}_3 
\nonumber \\
&&\hat{\tau_3} \hat{g}_e^R ( \epsilon + \hbar \omega_{\rm ac} ) \hat{\tau}_3 \hat{g}_e^K ( \epsilon ) 
- \hat{g}_e^R ( \epsilon ) \hat{\tau}_3 \hat{g}_e^K ( \epsilon - \hbar \omega_{\rm ac} ) \hat{\tau}_3 
\nonumber \\
&&+\hat{\tau_3} \hat{g}_e^K ( \epsilon + \hbar \omega_{\rm ac}) \hat{\tau}_3 \hat{g}_e^A ( \epsilon) 
- \hat{g}_e^K ( \epsilon) \hat{\tau}_3 \hat{g}_e^A ( \epsilon - \hbar \omega_{\rm ac}) \hat{\tau}_3 
\Bigr] 
\nonumber \\
&=&  [\epsilon \hat{\tau}_3 + \hat{\Delta}_e , \delta \hat{g}^K_{\rm 0H}] 
+ [\delta \hat{\Delta}_{\rm 0H},  \hat{g}_e^K (\epsilon) ]   . \label{K_0H}
\end{eqnarray}
The corresponding normalization condition, Eq.~(\ref{normK2}), reduces to  
\begin{eqnarray}
&&\hat{g}_{e}^R(\epsilon ) \delta \hat{g}^K_{\rm 0H}  
+ \delta \hat{g}^K_{\rm 0H}  \hat{g}_{e}^A(\epsilon )  \nonumber \\
&&+ \hat{g}_{e}^K(\epsilon )  \delta \hat{g}^A_{\rm 0H}   
+\delta \hat{g}^R_{\rm 0H}  \hat{g}_{e}^K(\epsilon )   =0 .\nonumber \\
 \label{normK_0H}
\end{eqnarray}
Additionally, Eq.~(\ref{gAno1}) reduces to  
\begin{eqnarray}
\delta g^K_{\rm 0H} = (\delta g^R_{\rm 0H} - \delta g^A_{\rm 0H} ) \mathcal{T} + \delta g^a_{\rm 0H} . \label{gAno_0H}
\end{eqnarray}
Thus, Eqs.~(\ref{RA_0H})-(\ref{gAno_0H}) collectively define the governing equations for the zero harmonic corrections to the Green's functions.  
Solving these equations, we obtain Eqs.~(\ref{dG0H})-(\ref{kappaAno0H}).

\subsection{Equations for the second harmonic} 

Eq.~(\ref{RA2}) for $(\eta,\eta')=(-\omega_{\rm ac},-\omega_{\rm ac})$ 
yields the equations governing the $R$ and $A$ components of the second harmonic ($\omega = 2\omega_{\rm ac}$) in Fourier space:  
\begin{eqnarray}
&&-i\frac{s}{8}  \Bigl[ 
\hat{\tau_3} \hat{g}_e^r ( \epsilon) \hat{\tau}_3 \hat{g}_e^r ( \epsilon- \hbar \omega_{\rm ac} ) 
- \hat{g}_e^r ( \epsilon + \hbar \omega_{\rm ac}) \hat{\tau}_3 \hat{g}_e^r ( \epsilon) \hat{\tau}_3 
\Bigr] 
\nonumber \\
&=& (\epsilon + \hbar \omega_{\rm ac}) \hat{\tau}_3 \delta \hat{g}^r_{\rm 2H}
-  \delta \hat{g}^r_{\rm 2H} \hat{\tau}_3  (\epsilon - \hbar \omega_{\rm ac}) 
+ [ \hat{\Delta}_e , \delta \hat{g}^r_{\rm 2H}] 
\nonumber \\
&&+ \delta \hat{\Delta}_{\rm 2H} \hat{g}_e^r (\epsilon- \hbar \omega_{\rm ac}) 
- \hat{g}_e^r (\epsilon+\hbar \omega_{\rm ac})\delta \hat{\Delta}_{\rm 2H} 
 . \label{RA_2H}
\end{eqnarray}
The normalization condition, Eq.~(\ref{normRA2}), leads to 
\begin{eqnarray}
\hat{g}^{r}_{e} (\epsilon+ \hbar \omega_{\rm ac} ) \delta \hat{g}^{r}_{\rm 2H} 
+ \delta \hat{g}^{r}_{\rm 2H} \hat{g}^{r}_{e}(\epsilon- \hbar \omega_{\rm ac}) =0 , \label{normRA_2H}
\end{eqnarray}
Similarly, Eq.~(\ref{K2}) provides the equations governing the $K$ components of the second harmonic in Fourier space:  
\begin{eqnarray}
&&-i\frac{s}{8}\Bigl[ 
\hat{\tau_3} \hat{g}_e^R ( \epsilon ) \hat{\tau}_3 \hat{g}_e^K ( \epsilon- \hbar \omega_{\rm ac} ) 
- \hat{g}_e^R ( \epsilon + \hbar \omega_{\rm ac} ) \hat{\tau}_3 \hat{g}_e^K ( \epsilon  ) \hat{\tau}_3 
\nonumber \\
&&+\hat{\tau_3} \hat{g}_e^K ( \epsilon ) \hat{\tau}_3 \hat{g}_e^A ( \epsilon - \hbar \omega_{\rm ac}) 
- \hat{g}_e^K ( \epsilon + \hbar \omega_{\rm ac}) \hat{\tau}_3 \hat{g}_e^A ( \epsilon) \hat{\tau}_3 
\Bigr] 
\nonumber \\
&=&  [\hat{\Delta}_e , \delta \hat{g}^K_{\rm 2H}] 
+ \delta \hat{\Delta}_{\rm 2H} \hat{g}_e^K (\epsilon-\hbar \omega_{\rm ac}) 
- \hat{g}_e^K (\epsilon+\hbar \omega_{\rm ac}) \delta \hat{\Delta}_{\rm 2H} \nonumber \\
&&+ (\epsilon+\hbar\omega_{\rm ac}) \hat{\tau}_3 \delta \hat{g}^K_{\rm 2H} 
-  (\epsilon-\hbar\omega_{\rm ac})\delta \hat{g}^K_{\rm 2H}  \hat{\tau}_3   . \label{K_2H}
\end{eqnarray}
The corresponding normalization condition, Eq.~(\ref{normK2}), reduces to
\begin{eqnarray}
&&\hat{g}_{e}^R(\epsilon + \hbar \omega_{\rm ac}) \delta \hat{g}^K_{\rm 2H}  
+ \delta \hat{g}^K_{\rm 2H}  \hat{g}_{e}^A(\epsilon - \hbar \omega_{\rm ac})  \nonumber \\
&&+ \hat{g}_{e}^K(\epsilon + \hbar \omega_{\rm ac})  \delta \hat{g}^A_{\rm 2H}   
+\delta \hat{g}^R_{\rm 2H}  \hat{g}_{e}^K(\epsilon - \hbar \omega_{\rm ac})   =0 .
 \label{normK_2H}
\end{eqnarray}
Additionally, Eq.~(\ref{gAno1}) reduces to 
\begin{eqnarray}
\delta g^K_{\rm 2H} = \delta g^R_{\rm 2H} \mathcal{T}(\epsilon-\hbar\omega_{\rm ac}) - \delta g^A_{\rm 2H} \mathcal{T}(\epsilon+\hbar\omega_{\rm ac}) + \delta g^a_{\rm 2H} \nonumber \\
\label{gAno_2H}
\end{eqnarray}
Eqs.~(\ref{RA_2H})-(\ref{gAno_2H}) govern the second harmonic ($\omega = 2\omega_{\rm ac}$) corrections to the Green's functions. 
The corresponding equations for $\omega = -2\omega_{\rm ac}$ are derived in a similar manner. 
Solving these equations yields Eqs.~(\ref{dG2H})-(\ref{kappaAno2H}).

\section{Consistency check of the present calculation of $\delta\Delta_{\rm 0H, 2H}$} \label{appendix_check_deltaDelta}

\begin{figure}[tb]
   \begin{center}
   \includegraphics[height=0.46\linewidth]{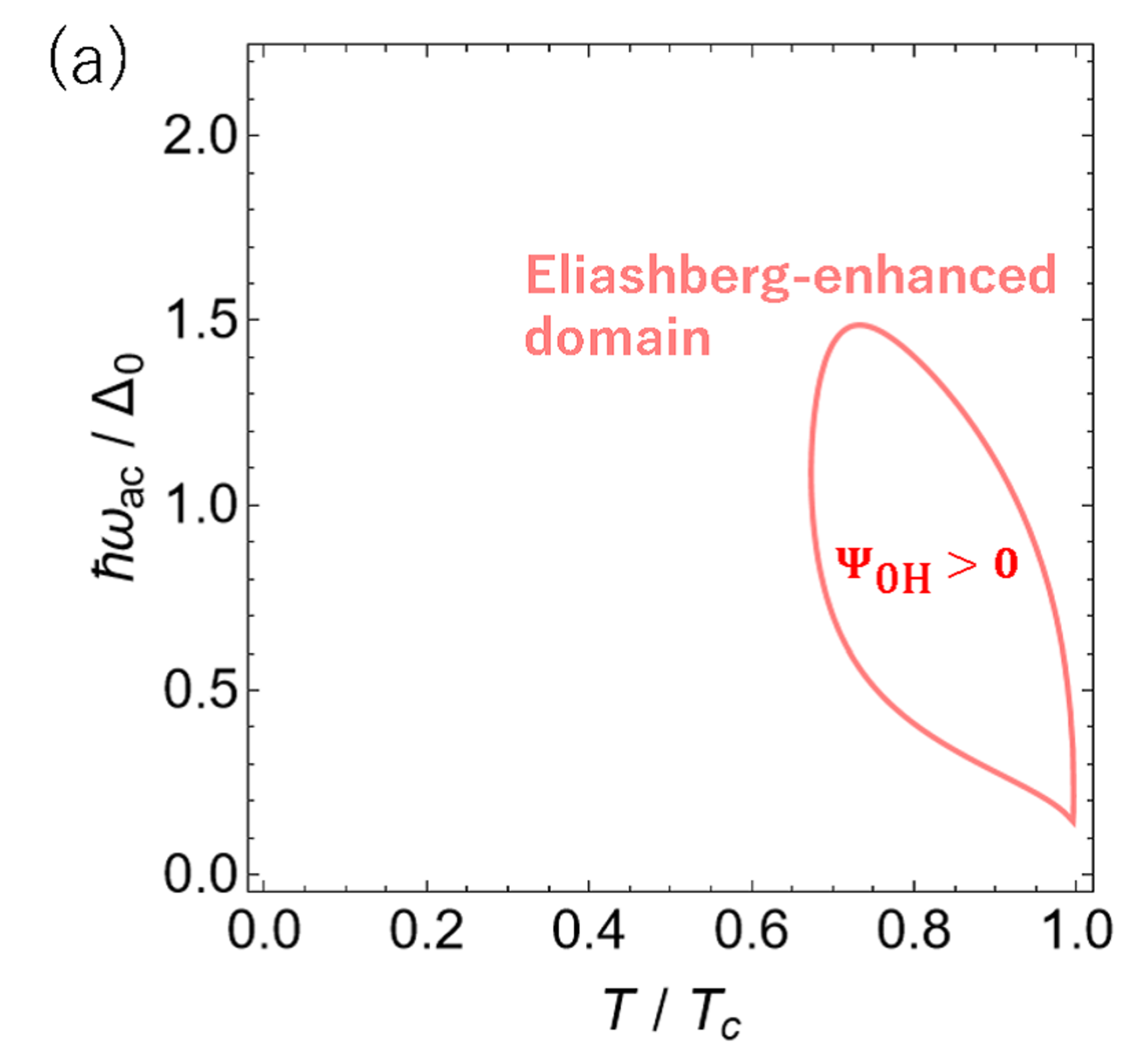}
   \includegraphics[height=0.49\linewidth]{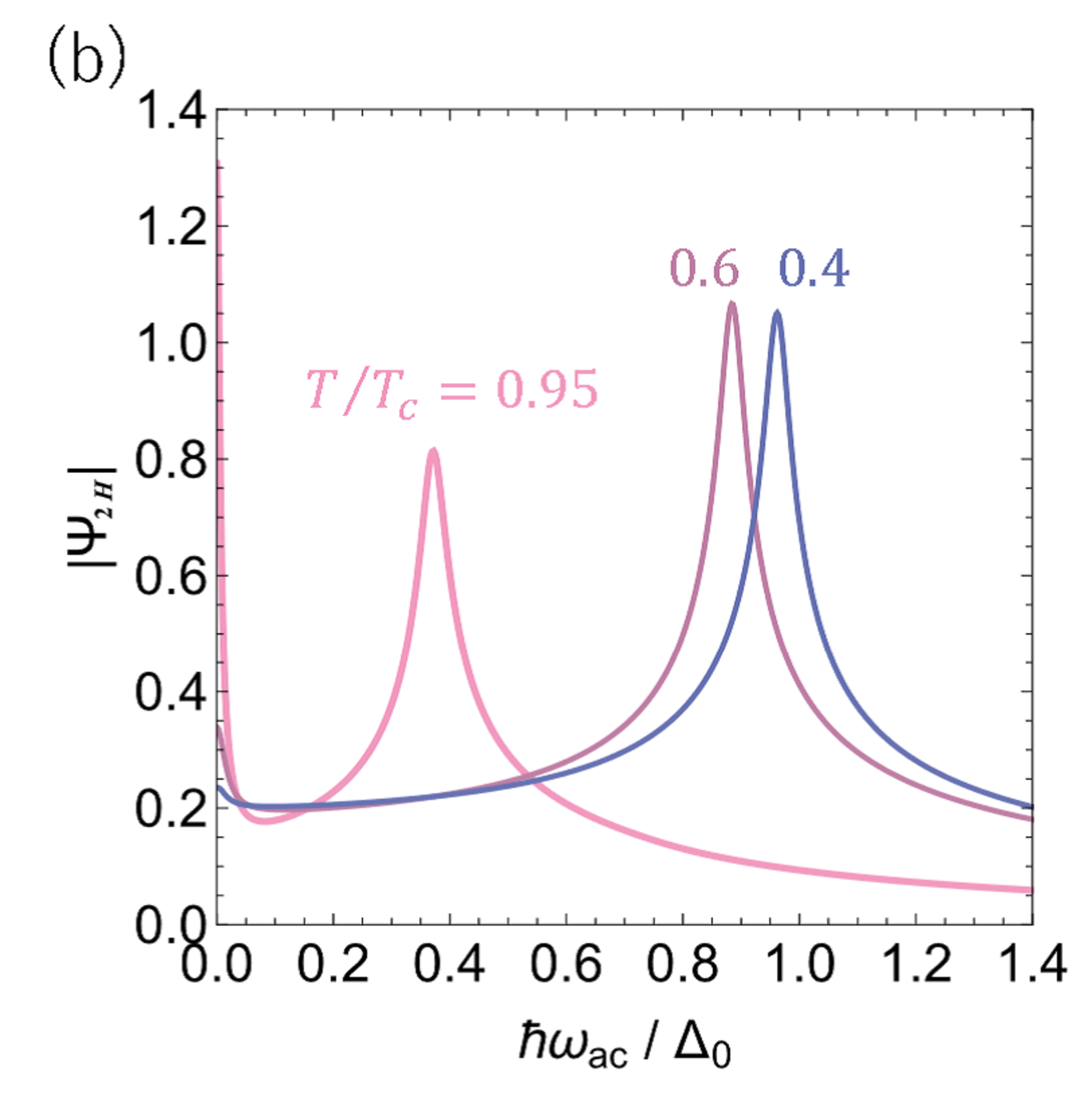}
   \end{center}\vspace{0 cm}
   \caption{
Consistency check: comparison of the present calculation with the previous study~\cite{Eremin}.  
(a) Zero-harmonic pair-potential variation \( \Psi_{\rm 0H} \), enhanced by the Eliashberg effect, shown in the \( (T, \omega_{\rm ac}) \) plane.  
(b) Second-harmonic pair-potential variation \( \Psi_{\rm 2H} \) as a function of \( \omega_{\rm ac} \). 
Both panels are calculated with \( \Gamma / \Delta_0 = 0.02 \), consistent with the parameter choice in Ref.~\cite{Eremin}.
}\label{fig7}
\end{figure}

\subsection{Zero Harmonic \( \delta \Delta_{\rm 0H} \) and the Eliashberg Effect} \label{appendix_0H}

We begin by verifying that our result for the nonequilibrium zero-harmonic correction to the pair potential is consistent with previous studies~\cite{Eremin}.

Figure~\ref{fig7}(a) shows the domain in the \( (T, \omega_{\rm ac}) \) plane where the Eliashberg effect is positive, i.e., \( \delta \Delta_{\rm 0H} > 0 \), calculated using the same damping parameter as in Ref.~\cite{Eremin}, namely \( \Gamma / \Delta_0 = 0.02 \).  
The result shows excellent agreement with Fig.~2 of Ref.~\cite{Eremin}.

A similar result was previously reported in Ref.~\cite{2018_Klapwijk}, where a damping parameter \( \Gamma / \Delta_0 = 0.02 / 1.76 \simeq 0.011 \) was used.  
This value is slightly smaller than that used in Ref.~\cite{Eremin} and in Fig.~\ref{fig7}(a), but still larger than the value adopted in Fig.~\ref{fig1}(a).  
As a consequence, the region of gap enhancement extends to \( T / T_c \gtrsim 0.47 \).  
If an even smaller \( \Gamma \) is used--corresponding to a longer quasiparticle lifetime--the enhancement region expands further.  
In Fig.~\ref{fig1}(a) of the main text, we take \( \Gamma / \Delta_0 = 10^{-3} \), which results in gap enhancement appearing already at \( T / T_c = 0.35 \).

\subsection{Second harmonic $\delta\Delta_{\rm 2H}$} \label{appendix_2H}

Next, we verify that our results for the nonequilibrium corrections to the second-harmonic components are consistent with the previous study~\cite{Eremin}.  

To this end, we first introduce the shorthand \( d(\epsilon) = i\sqrt{\Delta^2-(\epsilon+i\Gamma)^2} \) and \( d_{\pm n} = d(\epsilon \pm n\hbar \omega_{\rm ac} / 2) \) with \( n \in \mathbb{N} \).  
With this substitution, Eqs.~(\ref{zeta2H}), (\ref{kappa2H}), (\ref{zetaAno2H}), and (\ref{kappaAno2H}) reduce to the following expressions:
\begin{eqnarray}
\zeta_{2H} &=& -\frac{G_2 + G_{-2}}{F_2 + F_{-2}} \frac{F_2 - F_{-2}}{2\hbar \omega_{\rm ac}} \nonumber \\
&=& \frac{G_2 G_{-2} + F_2 F_{-2} +1 }{d_2 + d_{-2}}, \\
\kappa_{2H} &=& \frac{i}{8}  \frac{G_2 + G_{-2}}{F_2 + F_{-2}} \frac{ (G_2 - G_{-2})G + (F_2 - F_{-2})F }{2\hbar \omega_{\rm ac}} \nonumber \\
&=& -\frac{i}{8(d_2 + d_{-2})} \Bigl\{ (G_2 F_{-2} + F_2 G_{-2})G \nonumber \\
 && + (G_2 G_{-2} + F_2 F_{-2} +1)F \Bigr\} , \\
\zeta^a_{2H} &=& \frac{G_2 -G_{-2}^*}{F_2 -F_{-2}^*} \frac{F_2 + F_{-2}^*}{2\hbar \omega_{\rm ac}} (\mathcal{T}_{-2} - \mathcal{T}_2 ) \nonumber \\
&=& \frac{G_2 G_{-2}^* + F_2 F_{-2}^* -1}{d_2 - d_{-2}^*}  (\mathcal{T}_{-2} - \mathcal{T}_2 ) , \\
\kappa^a_{2H} &=& \frac{i}{16\hbar \omega_{\rm ac}} \frac{G_2 -G_{-2}^*}{F_2 -F_{-2}^*}  \nonumber \\
&&\times \Bigl[ \bigl\{ (G_2 + G_{-2}^*)G + (F_2 + F_{-2}^*)F \bigr\} (\mathcal{T} - \mathcal{T}_{-2} ) \nonumber \\
&& + \bigl\{ (G_2 + G_{-2}^*)G^* + (F_2 + F_{-2}^*)F^* \bigr\} (\mathcal{T} - \mathcal{T}_2) \Bigr] \nonumber \\
&=& \frac{i}{8 (d_2 - d_{-2}^*)}  \Bigl[ (G_2 F_{-2}^* + F_2 G_{-2}^*) \bigl\{  G (\mathcal{T} - \mathcal{T}_{-2} ) \nonumber \\
&&+ G^* (\mathcal{T} - \mathcal{T}_2 ) \bigr\} 
+ (G_2 G_{-2}^* + F_2 F_{-2}^* -1) \nonumber \\
&&\bigl\{  F (\mathcal{T} - \mathcal{T}_{-2} )  + F^* (\mathcal{T} - \mathcal{T}_2 ) \bigr\} \Bigr] . 
\end{eqnarray}
These results are in agreement with the corresponding formulas presented in Ref.~\cite{Eremin}.

Figure~\ref{fig7}(b) shows \( |\Psi_{\rm 2H}| \) as a function of \( \omega_{\rm ac} \), calculated with a damping parameter \( \Gamma / \Delta_0 = 0.02 \), which is the same value adopted in Ref.~\cite{Eremin} and significantly larger than the value used in the main text (\( \Gamma / \Delta_0 = 10^{-3} \)).  
Our results show good agreement with Fig.~3(a) of Ref.~\cite{Eremin}, further confirming the consistency of the present analysis.  
The peaks, which correspond to Higgs-mode excitation, appear at \( \hbar \omega_{\rm ac} = \Delta(T) \), where \( \Delta(T) / \Delta_0 = 0.371 \), \( 0.885 \), and \( 0.963 \) for \( T / T_c = 0.95 \), \( 0.6 \), and \( 0.4 \), respectively.

\section{Consistency check: $\delta\Delta$ in the dc limit}\label{appendix_dc_deltaDelta}

Let us confirm that the results presented in Section~\ref{analytical_solution}, including the anomalous low-frequency peak, are consistent with the well-established expression for the pair-potential variation induced by a dc supercurrent. To demonstrate this correspondence, we translate the preceding Fourier-space analysis into the time domain.
Taking the inverse Fourier transform of Eq.~(\ref{deltaDelta_spectrum}), the time-dependent pair-potential variation is given by
\begin{eqnarray}
\delta \Delta (t) = \delta \Delta_{\rm 0H} + \left( \delta\Delta_{\rm 2H} e^{-2i\omega_{\rm ac}t} + {\rm c.c.} \right).
\end{eqnarray}
In the limit \( \omega_{\rm ac} \to 0 \), and using the fact that \( \delta \Delta_{\rm 2H} \in \mathbb{R} \) in this limit, we obtain the static pair-potential variation:
\begin{eqnarray}
\delta \Delta_{\rm dc} = s( \Psi_{\rm 0H} + 2\Psi_{\rm 2H})|_{\omega_{\rm ac} \to 0}. \label{deltaDelta_dc1}
\end{eqnarray}
Here, \( \Psi_{\rm 0H} \) and \( \Psi_{\rm 2H} \) are given by Eqs.~(\ref{Psi0H}) and (\ref{Psi2H}), respectively. Taking the limit \( \omega_{\rm ac} \to 0 \), we find
\begin{eqnarray}
\Psi_{\rm 0H}\Bigr|_{\omega_{\rm ac} \to 0} 
&=& -\frac{1}{2} \frac{\sum_{M\ge 0} (\hbar\omega_M)^2 \Delta /d_M^4}{\sum_{M\ge 0} \Delta^2/d_M^3 }, \\
\Psi_{\rm 2H}\Bigr|_{\omega_{\rm ac} \to 0}
&=& \frac{1}{2} \Psi_{\rm 0H}\Bigr|_{\omega_{\rm ac} \to 0} ,
 \label{Psi0H_Psi2H}
\end{eqnarray}
where \( d_M = \sqrt{(\hbar \omega_M)^2 + \Delta^2} \) and \( \hbar \omega_M = 2\pi kT (M + 1/2) \) denotes the Matsubara frequency.
Substituting these expressions into Eq.~(\ref{deltaDelta_dc1}), we obtain
\begin{eqnarray}
\delta \Delta_{\rm dc} 
= -\frac{\sum_{M\ge 0} (\hbar\omega_M)^2 \Delta /d_M^4}{\sum_{M\ge 0} \Delta^2/d_M^3 } s ,
 \label{deltaDelta_dc}
\end{eqnarray}
which reproduces the well-known result for the pair-potential variation under a dc superflow at arbitrary temperature.

For \( T \to 0 \), the Matsubara summation reduces to an integral according to the replacement \( (2\pi kT/\Delta) \sum_M \to \int_0^{\infty} dw \), where \( w := \hbar \omega_M / \Delta \). This leads to the analytic results $\Psi_{\rm 0H}\to - \pi/8$ and $\Psi_{\rm 2H} \to -\pi/16$ at $(T, \omega_{\rm ac}) \to (0,0)$ and hence the static pair-potential variation at zero temperature becomes 
\begin{eqnarray}
\delta \Delta_{\rm dc} = -\frac{\pi}{4}s , \label{deltaDelta_dc_T=0}
\end{eqnarray}
as expected  (see, e.g., Refs.~\cite{2020_Kubo, 2020_Kubo_erratum, 2022_Kubo} and references therein).

\section{Consistency check of the present calculation of $J_{\rm 3H}$} \label{appendix_check_J3H}

\begin{figure}[tb]
   \begin{center}
   \includegraphics[width=0.5\linewidth]{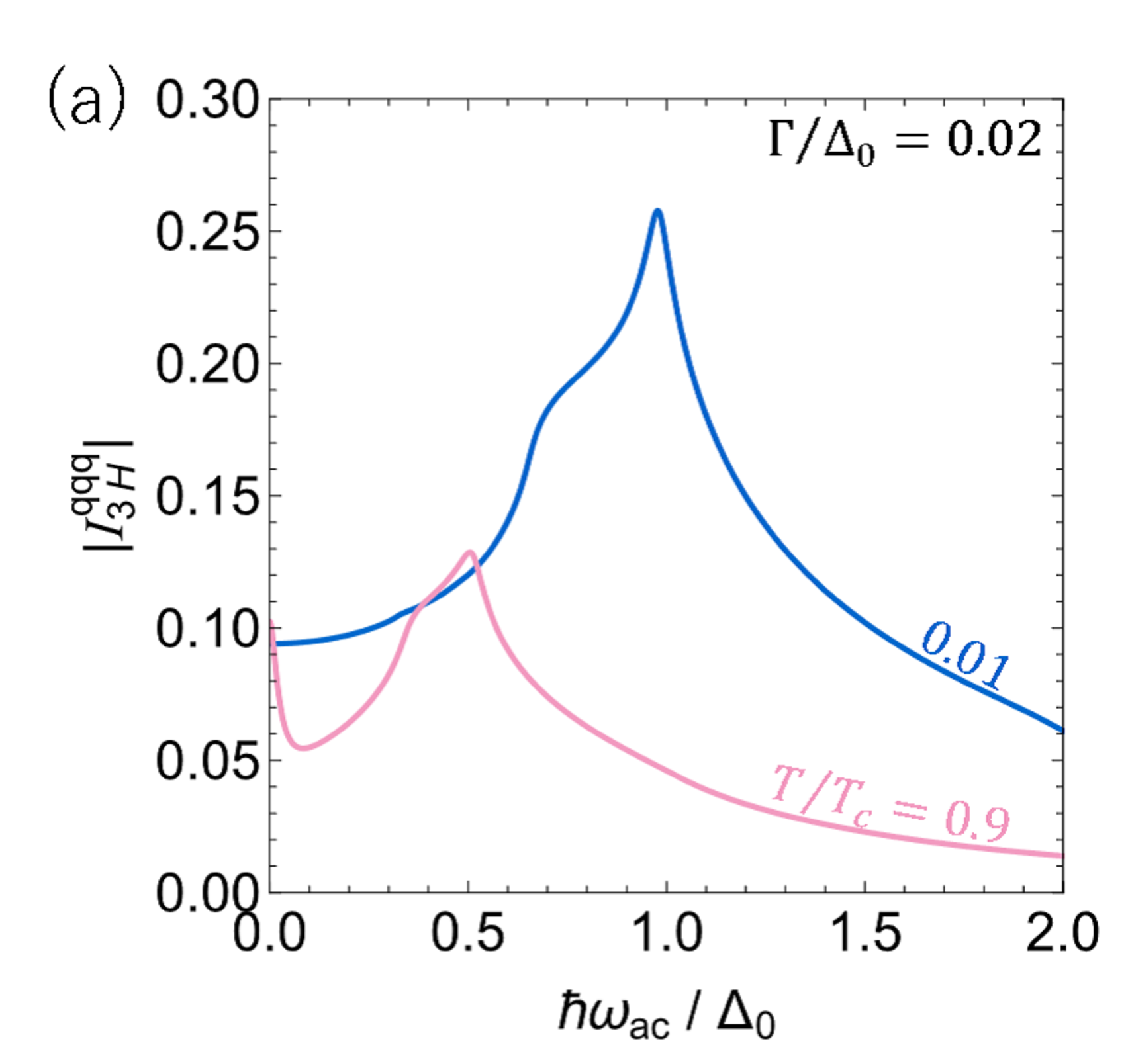}
   \includegraphics[width=0.48\linewidth]{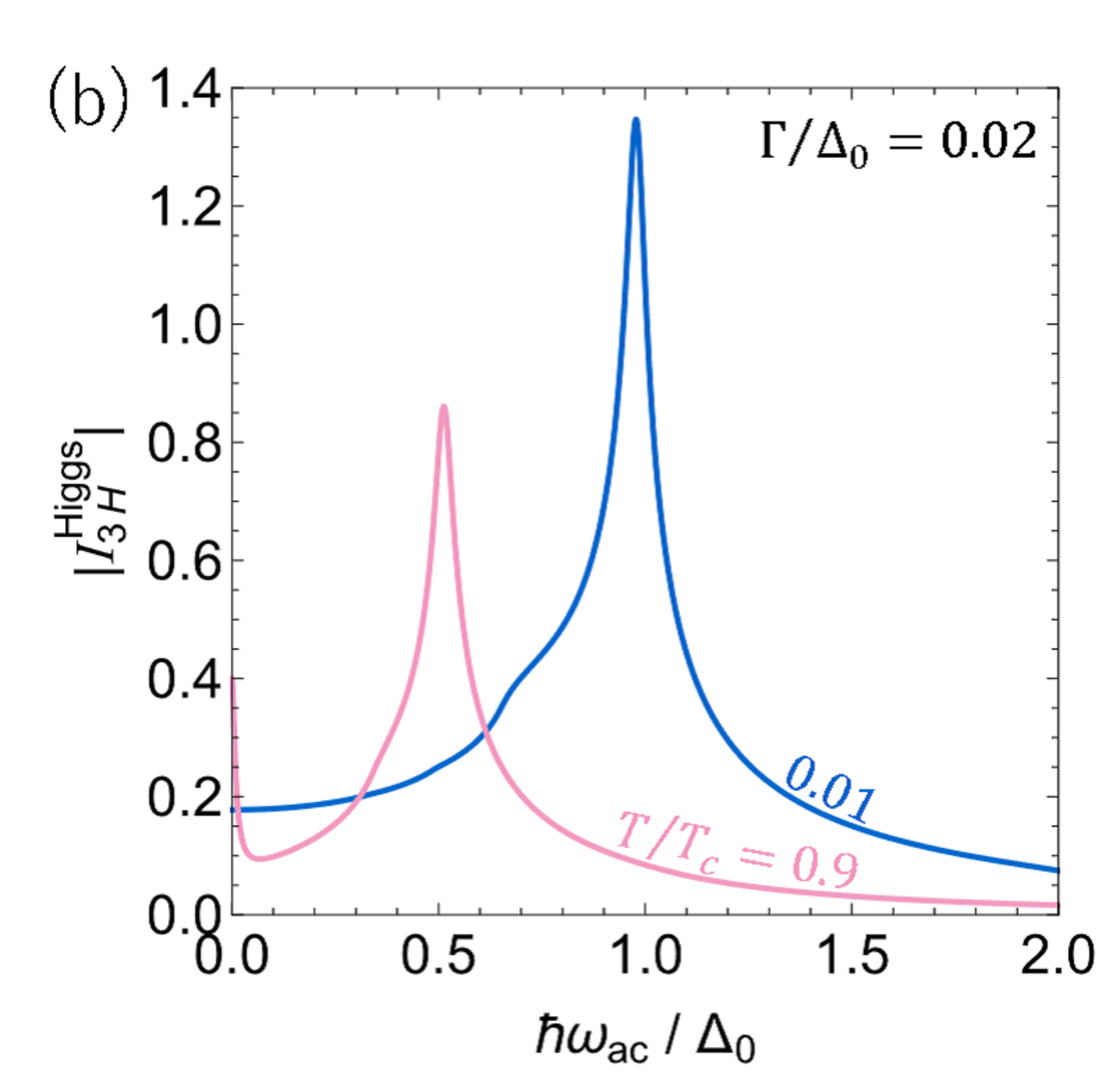}
   \includegraphics[width=0.5\linewidth]{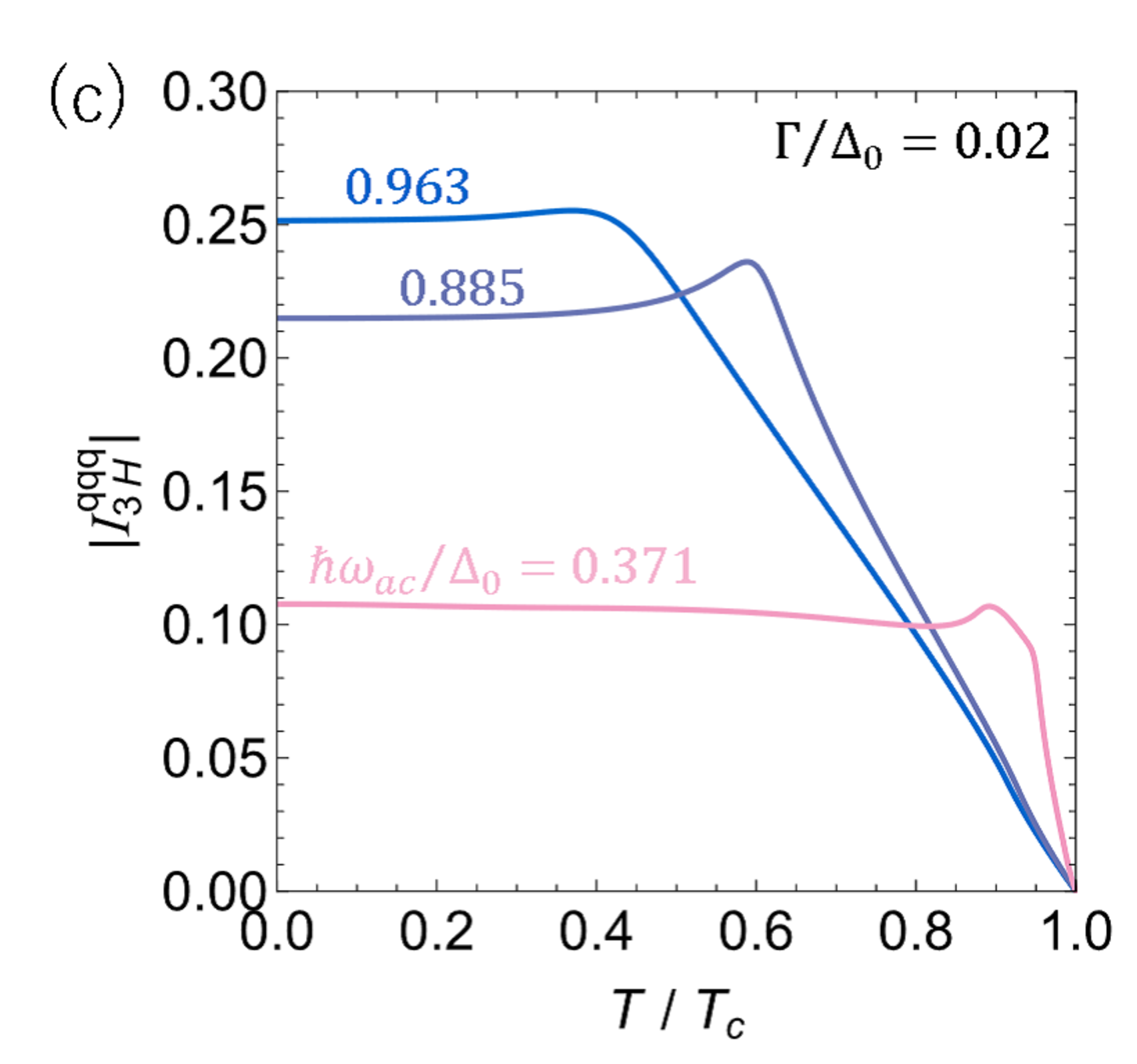}
   \includegraphics[width=0.48\linewidth]{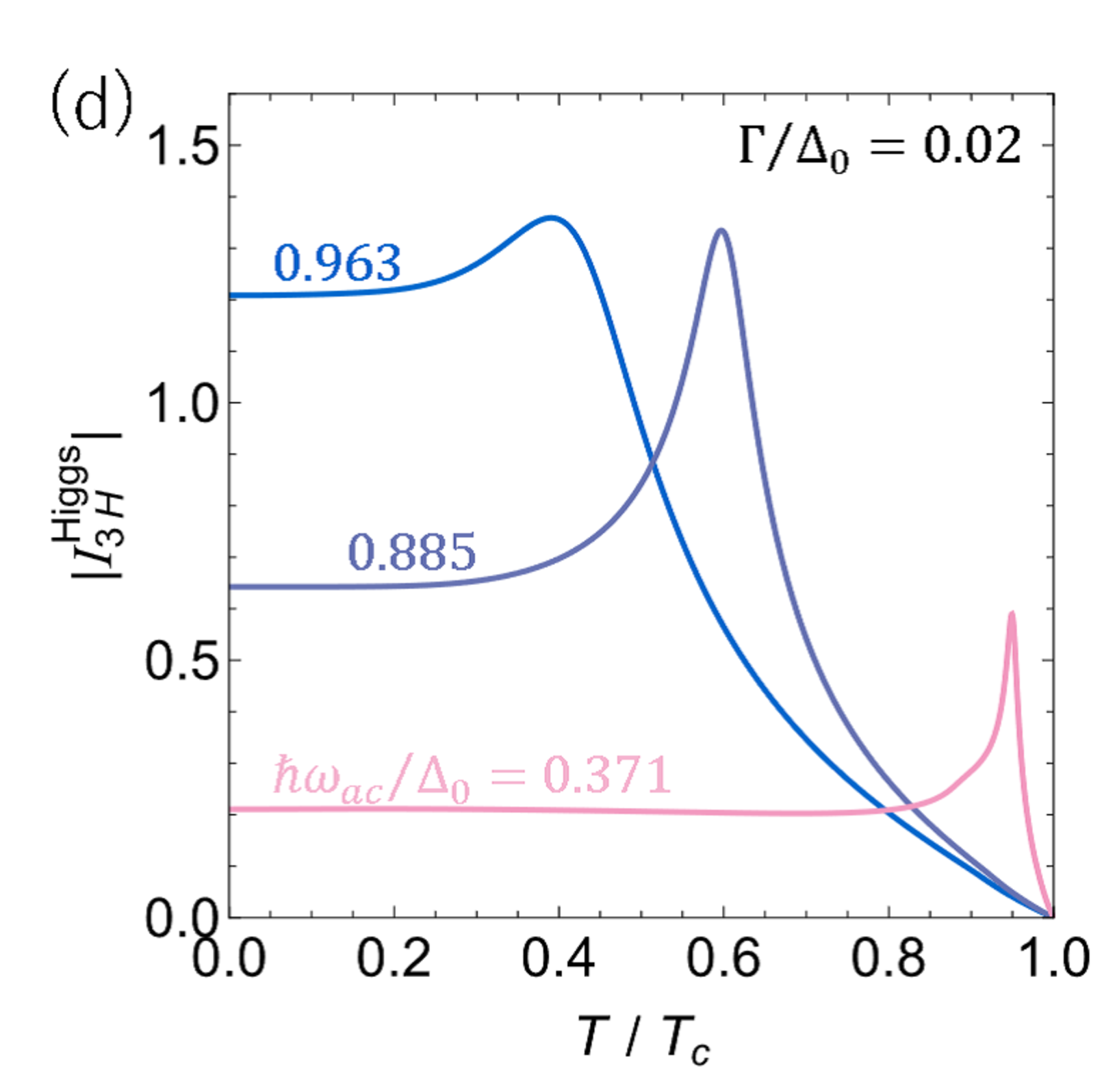}
   \end{center}\vspace{0 cm}
   \caption{
Consistency check: comparison between the present calculation of the third-harmonic current and previous results.  
(a, b) Contributions from the direct nonlinear photon action, \( |I_{\rm 3H}^{qqq}| \), and from the Higgs-mode-mediated process, \( |I_{\rm 3H}^{\rm Higgs}| \), plotted as functions of the ac frequency. 
(c, d) The same contributions plotted as functions of temperature. 
See also Fig.~8 of Ref.~\cite{Silaev} and Fig.~6 of Ref.~\cite{Eremin}. 
Similar plots incorporating the effect of paramagnetic impurities are shown in Ref.~\cite{2018_Jujo}.
   }\label{fig8}
\end{figure}

In this appendix, we present our calculation of the third-harmonic current for \( \Gamma/\Delta_0 = 0.02 \), the same value adopted in the previous study~\cite{Eremin}, in order to verify the consistency of our theoretical framework.  
Note that this value is significantly larger than that used in the main text, \( \Gamma/\Delta_0 = 10^{-3} \).

The third-harmonic current is given by Eq.~(\ref{J3H}) and consists of two contributions: the direct nonlinear photon action, \( I_{\rm 3H}^{qqq} \), and the Higgs-mode-mediated process, \( I_{\rm 3H}^{\rm Higgs} \).  
Figure~\ref{fig8} shows the frequency and temperature dependences of \( I_{\rm 3H}^{qqq} \) and \( I_{\rm 3H}^{\rm Higgs} \), confirming agreement with the results reported in Refs.~\cite{Silaev, Eremin}. 
Note that the peaks in Fig.~\ref{fig8}(d), which correspond to Higgs-mode excitation, appear at \( T / T_c = 0.4 \), \( 0.6 \), and \( 0.95 \), where the pair potential at each temperature matches the ac frequency: \( \hbar \omega_{\rm ac} =0.963\Delta_0= \Delta(0.4T_c) \), \( \hbar \omega_{\rm ac} =0.885\Delta_0=\Delta(0.6T_c) \), and \( \hbar \omega_{\rm ac} =  0.371\Delta_0=\Delta(0.95T_c)  \), respectively.

\section{Derivation of the linear-response contribution} \label{appendix_linear}

The linear-response contribution can be calculated in a straightforward manner.  
In Fourier space, Eq.~(\ref{S1}) reduces to
\begin{eqnarray}
&&{\bf S}^{(1)}(\epsilon,\omega) =  i\frac{{\bf q_0}}{4} \sum_{\eta=\pm \omega_{\rm ac}} 2\pi \delta (\omega + \eta) \times \nonumber \\
&&{\rm Tr} \biggl[
\hat{\tau}_3 \hat{g}_e^R \Bigl( \epsilon -\frac{\hbar \eta}{2} \Bigr) \hat{\tau}_3 \hat{g}_e^R \Bigl( \epsilon +\frac{\hbar \eta}{2} \Bigr) \mathcal{T}\Bigl( \epsilon +\frac{\hbar \eta}{2} \Bigr) 
\nonumber \\
&&-  \hat{\tau}_3 \hat{g}_e^A \Bigl( \epsilon -\frac{\hbar \eta}{2} \Bigr) \hat{\tau}_3 \hat{g}_e^A \Bigl( \epsilon +\frac{\hbar \eta}{2} \Bigr) \mathcal{T}\Bigl( \epsilon - \frac{\hbar \eta}{2} \Bigr) 
\nonumber \\
&& + \hat{\tau}_3 \hat{g}_e^R \Bigl( \epsilon -\frac{\hbar \eta}{2} \Bigr) \hat{\tau}_3 \hat{g}_e^A \Bigl( \epsilon +\frac{\hbar \eta}{2} \Bigr) \times \nonumber \\ 
&&\Bigl\{ \mathcal{T}\Bigl( \epsilon - \frac{\hbar \eta}{2} \Bigr) -\mathcal{T}\Bigl( \epsilon + \frac{\hbar \eta}{2} \Bigr)  \Bigr\} \biggr] .
\end{eqnarray}
By substituting the matrix elements into this expression, the Fourier transform of Eq.~(\ref{J}) yields Eqs.~(\ref{MB1}) and (\ref{MB2}).

\section{Derivation of nonlinear-response contributions} \label{appendix_nonlinear}

The calculation of the nonlinear-response contributions is significantly more involved than in the linear-response case, but can still be carried out in a straightforward manner.  
In Fourier space, Eq.~(\ref{S3}) reduces to
\begin{eqnarray}
&&{\bf S}^{(3)}(\epsilon, \omega) = i\frac{{\bf q}_0}{4} \sum_{\eta=\pm\omega_{\rm ac}} {\rm Tr} \biggl[ \hat{\tau}_3 g_e^R \Bigl(\epsilon+\frac{\hbar\omega}{2} \Bigr) \times \nonumber \\
&&\Bigl\{ \hat{\tau}_3 \delta \hat{g}^K \Bigl(\epsilon+\frac{\hbar\eta}{2} , \omega+\eta \Bigr)
- \delta \hat{g}^K \Bigl(\epsilon -\frac{\hbar\eta}{2} , \omega+\eta \Bigr) \hat{\tau}_3 \Bigr\} 
\nonumber \\
&&+ \hat{\tau}_3 g_e^K \Bigl(\epsilon+\frac{\hbar\omega}{2} \Bigr) \Bigl\{ \hat{\tau}_3 \delta \hat{g}^A \Bigl(\epsilon+\frac{\hbar\eta}{2} , \omega+\eta \Bigr) \nonumber \\
&&- \delta \hat{g}^A \Bigl(\epsilon -\frac{\hbar\eta}{2} , \omega+\eta \Bigr) \hat{\tau}_3 \Bigr\} 
+ \hat{\tau}_3 \delta \hat{g}^R \Bigl(\epsilon -\frac{\hbar\eta}{2} , \omega+\eta \Bigr) \times
\nonumber \\
&&\Bigl\{ \hat{\tau}_3 \hat{g}_e^K \Bigl(\epsilon -\frac{\hbar\omega}{2} \Bigr) 
- \hat{g}_e^K \Bigl(\epsilon -\frac{\hbar(\omega+2\eta)}{2} \Bigr) \hat{\tau}_3 \Bigr\}
\nonumber \\
&&+ \hat{\tau}_3 \delta \hat{g}^K \Bigl(\epsilon -\frac{\hbar\eta}{2} , \omega+\eta \Bigr) 
\Bigl\{ \hat{\tau}_3 \hat{g}_e^A \Bigl(\epsilon -\frac{\hbar\omega}{2} \Bigr) \nonumber \\
&&- \hat{g}_e^A \Bigl(\epsilon -\frac{\hbar(\omega+2\eta)}{2} \Bigr) \hat{\tau}_3 \Bigr\} . \label{S3_appendix}
\end{eqnarray}
Here, \( \delta \hat{g}^{R,A,K} \) can be decomposed into a sum of harmonics, as in Eq.~(\ref{dg0H2H}),  
which in this case given by
\begin{eqnarray}
&&\delta \hat{g}^{R,A,K}  \Bigl(\epsilon_{\pm 1} , \omega + \omega_{\rm ac} \Bigr) 
= \delta \hat{g}^{R,A,K}_{\rm 0H} (\epsilon_{\pm 1})  2\pi \delta (\omega + \omega_{\rm ac}) + \nonumber \\
&& \delta \hat{g}^{R,A,K}_{\rm 2H} (\epsilon_{\pm 1}) 2\pi \delta(\omega -\omega_{\rm ac})  
+ \delta \hat{g}^{R,A,K}_{\rm -2H}  (\epsilon_{\pm 1})  2\pi  \delta (\omega + 3 \omega_{\rm ac}) \nonumber \\
\end{eqnarray}
and
\begin{eqnarray}
&&\delta \hat{g}^{R,A,K}  \Bigl(\epsilon_{\pm 1} , \omega - \omega_{\rm ac} \Bigr) 
= \delta \hat{g}^{R,A,K}_{\rm 0H} (\epsilon_{\pm 1})  2\pi \delta (\omega - \omega_{\rm ac}) +\nonumber \\
&& \delta \hat{g}^{R,A,K}_{\rm 2H} (\epsilon_{\pm 1}) 2\pi \delta(\omega -3\omega_{\rm ac})  
+ \delta \hat{g}^{R,A,K}_{\rm -2H}  (\epsilon_{\pm 1})  2\pi  \delta (\omega + \omega_{\rm ac}) \nonumber \\
\end{eqnarray}
As in the main text, \( \epsilon_{\pm n} \) is defined as \( \epsilon \pm n(\hbar \omega_{\rm ac} / 2) \), with \( n \in \mathbb{N} \).
Substituting $\delta \hat{g}^{R,A,K} $ into Eq.~(\ref{S3_appendix}), we obtain
\begin{eqnarray}
&&{\bf S}^{(3)} (\epsilon, \omega) 
= {\bf S}^{(3)}_{\rm 1H} 2\pi \delta(\omega - \omega_{\rm ac}) + {\bf S}^{(3)}_{\rm -1H} 2\pi \delta(\omega + \omega_{\rm ac}) \nonumber \\
&&+ {\bf S}^{(3)}_{\rm 3H} 2\pi \delta(\omega - 3\omega_{\rm ac}) + {\bf S}^{(3)}_{\rm -3H} 2\pi \delta(\omega + 3\omega_{\rm ac}) 
\end{eqnarray}
Here,
\begin{eqnarray}
&&{\bf S}^{(3)}_{\rm 1H} = \nonumber \\
&& i \frac{{\bf q}_0}{4} {\rm Tr} \biggl[ 
\hat{\tau}_3 \hat{g}_e^R(\epsilon_{1}) \Bigl\{  
\hat{\tau}_3 \delta \hat{g}^K_{\rm 0H} (\epsilon_{-1}) - \delta \hat{g}^K_{\rm 0H} (\epsilon_{1}) \hat{\tau}_3 \nonumber \\
&&+ \hat{\tau}_3 \delta \hat{g}^K_{\rm 2H} (\epsilon_{1}) - \delta \hat{g}^K_{\rm 2H} (\epsilon_{-1}) \hat{\tau}_3 \Bigr\} \nonumber \\
&&+\hat{\tau}_3 \hat{g}_e^K(\epsilon_{1}) \Bigl\{  
\hat{\tau}_3 \delta \hat{g}^A_{\rm 0H} (\epsilon_{-1}) - \delta \hat{g}^A_{\rm 0H} (\epsilon_{1}) \hat{\tau}_3 \nonumber \\
&&+ \hat{\tau}_3 \delta \hat{g}^A_{\rm 2H} (\epsilon_{1}) - \delta \hat{g}^A_{\rm 2H} (\epsilon_{-1}) \hat{\tau}_3 \Bigr\} \nonumber \\
&&+ \hat{\tau}_3 \delta \hat{g}^R_{\rm 0H }(\epsilon_{1}) \Bigl\{  
\hat{\tau}_3 \hat{g}^K_e (\epsilon_{-1}) - \hat{g}^K_e (\epsilon_{1}) \hat{\tau}_3 \Bigr\} \nonumber \\
&&+ \hat{\tau}_3 \delta \hat{g}^R_{\rm 2H }(\epsilon_{-1}) \Bigl\{  
\hat{\tau}_3 \hat{g}^K_e (\epsilon_{-1}) - \hat{g}^K_e (\epsilon_{-3}) \hat{\tau}_3 \Bigr\} \nonumber \\
&&+ \hat{\tau}_3 \delta \hat{g}^K_{\rm 0H }(\epsilon_{1}) \Bigl\{  
\hat{\tau}_3 \hat{g}^A_e (\epsilon_{-1}) - \hat{g}^A_e (\epsilon_{1}) \hat{\tau}_3 \Bigr\} \nonumber \\
&&+ \hat{\tau}_3 \delta \hat{g}^K_{\rm 2H }(\epsilon_{-1}) \Bigl\{  
\hat{\tau}_3 \hat{g}^A_e (\epsilon_{-1}) - \hat{g}^A_e (\epsilon_{-3}) \hat{\tau}_3 \Bigr\}
\biggr] , \\
&& {\bf S}^{(3)}_{\rm -1H} =  {\bf S}^{(3)}_{\rm 1H}\Bigr|_{\omega_{\rm ac}\to -\omega_{\rm ac}} ,
\end{eqnarray}
and
\begin{eqnarray}
&&{\bf S}^{(3)}_{\rm 3H} = \nonumber \\
&& i \frac{{\bf q}_0}{4} {\rm Tr} \biggl[ 
\hat{\tau}_3 \hat{g}_e^R(\epsilon_{3}) \Bigl\{  
\hat{\tau}_3 \delta \hat{g}^K_{\rm 2H} (\epsilon_{-1}) - \delta \hat{g}^K_{\rm 2H} (\epsilon_{1}) \hat{\tau}_3  \Bigr\} \nonumber \\
&&+\hat{\tau}_3 \hat{g}_e^K(\epsilon_{3}) \Bigl\{  
\hat{\tau}_3 \delta \hat{g}^A_{\rm 2H} (\epsilon_{-1}) - \delta \hat{g}^A_{\rm 2H} (\epsilon_{1}) \hat{\tau}_3 \Bigr\} \nonumber \\
&&+ \hat{\tau}_3 \delta \hat{g}^R_{\rm 2H }(\epsilon_{1}) \Bigl\{  
\hat{\tau}_3 \hat{g}^K_e (\epsilon_{-3}) - \hat{g}^K_e (\epsilon_{-1}) \hat{\tau}_3 \Bigr\} \nonumber \\
&&+ \hat{\tau}_3 \delta \hat{g}^K_{\rm 2H }(\epsilon_{1}) \Bigl\{  
\hat{\tau}_3 \hat{g}^A_e (\epsilon_{-3}) - \hat{g}^A_e (\epsilon_{-1}) \hat{\tau}_3 \Bigr\}
\biggr] , \\
&& {\bf S}^{(3)}_{\rm -3H} =  {\bf S}^{(3)}_{\rm 3H}\Bigr|_{\omega_{\rm ac}\to -\omega_{\rm ac}} .
\end{eqnarray}
Substituting the matrix components into the above expressions, we get
\begin{eqnarray}
\int {\bf S}^{(3)}_{\rm 1H} d\epsilon 
&=& i\frac{{\bf q}_0}{4} \int \biggl\{ K_{\rm 1H} \frac{\hbar D q_0^2}{2} \nonumber \\
&&+ Z_{\rm 1H}^{\rm Eliash} \delta\Delta_{\rm 0H} 
+ Z_{\rm 1H}^{\rm Higgs} \delta\Delta_{\rm 2H} \biggr\} d\epsilon , 
\end{eqnarray}
and
\begin{eqnarray}
\int {\bf S}^{(3)}_{\rm 3H} d\epsilon = i\frac{{\bf q}_0}{4} \int \biggl\{ K_{\rm 3H} \frac{\hbar D q_0^2}{2} + Z_{\rm 3H}^{\rm Higgs} \delta\Delta_{\rm 2H} \biggr\} d\epsilon . \nonumber \\
\end{eqnarray}
Here, $K_{\rm 1H}$, $Z_{\rm 1H}^{\rm Eliash}$, $Z_{\rm 1H}^{\rm Higgs}$, $K_{\rm 3H}$, and $Z_{\rm 3H}^{\rm Higgs}$ are given by Eqs.~(\ref{K1H}), (\ref{Z1H_Eliash}), (\ref{Z1H_Higgs}), (\ref{K3H}), and (\ref{Z3H_Higgs}), respectively.

\section{Consistency check: current response in the dc limit}\label{appendix_dc_current_response}

\subsection{Derivation of $J^{(3)}_{\rm dc}$}

As a consistency check, we now confirm that taking the dc limit (\( \omega_{\rm ac} \to 0 \)) reproduces the well-known expression for the nonlinear dc response.

Taking the inverse Fourier transform of Eq.~(\ref{J3_Fourier}) yields
\begin{eqnarray}
J^{(3)}(t) = \Bigl( J^{(3)}_{\rm 1H} e^{-i\omega_{\rm ac}t} + J^{(3)}_{\rm 3H} e^{-i 3\omega_{\rm ac}t} \Bigr) + \text{c.c.}
\end{eqnarray}
In the limit \( \omega_{\rm ac} \to 0 \), the time-dependent components merge, resulting in a static nonlinear response
\begin{eqnarray}
J^{(3)}_{\rm dc}
&=& 2 (J^{(3)}_{\rm 1H}|_{\omega_{\rm ac}\to 0} +  J^{(3)}_{\rm 3H}|_{\omega_{\rm ac}\to0} ) \nonumber \\
&=& \biggl[ 2 \Bigl\{ \frac{\delta \Delta_{\rm 0H}}{\Delta_0} \frac{I_{\rm 1H}^{\rm Eliash}}{\Psi_{\rm 0H}} + \frac{\delta\Delta_{\rm 2H}}{\Delta_0} \Bigl( \frac{I_{\rm 1H}^{\rm Higgs}}{\Psi_{\rm 2H}} + \frac{ I_{\rm 3H}^{\rm Higgs}}{\Psi_{\rm 2H}} \Bigr) \Bigr\} \nonumber \\
&&+2 (I_{\rm 1H}^{qqq} + I_{\rm 3H}^{qqq}) \frac{s}{\Delta_0}  \biggr]_{\omega_{\rm ac}\to0} 
i \frac{q_0}{q_{\xi}}  J_0 . \label{J3dc_1}
\end{eqnarray}
As shown in Appendix~\ref{appendix_dc_I}, in the limit \( \omega_{\rm ac} \to 0 \) and \( T \to 0 \), we obtain 
$I_{\rm 1H}^{qqq}=3 I_{\rm 3H}^{qqq}=-i/2\sqrt{\pi}$ and 
$I_{\rm 1H}^{\rm Eliash}/\Psi_{\rm 0H}= I_{\rm 1H}^{\rm Higgs}/\Psi_{\rm 2H} = I_{\rm 3H}^{\rm Higgs}/\Psi_{\rm 2H}= i \sqrt{\pi}/2$. 
Here, $\Psi_{\rm 0H}=2\Psi_{\rm 2H}=-\pi/8$ for $(T, \omega_{\rm ac})=(0,0)$. 
Using the identity \( \delta\Delta_{\rm dc} = (\delta\Delta_{\rm 0H} + 2\delta\Delta_{\rm 2H})|_{\omega_{\rm ac}\to 0} \), 
Eq.~(\ref{J3dc_1}) reduces to
\begin{eqnarray}
J^{(3)}_{\rm dc} = \Bigl( \frac{\delta\Delta_{\rm dc}}{\Delta_0} - \frac{4}{3\pi} \frac{s}{\Delta_0} \Bigr)
\Bigl( -\frac{q_0}{q_{\xi}} \Bigr) \sqrt{\pi}  J_0
\end{eqnarray}
in the zero-temperature limit, where \( \delta\Delta_{\rm dc} \) is given by Eq.~(\ref{deltaDelta_dc_T=0}). 
This result agrees with the established expression for the \( \mathcal{O}(q_0^3) \) correction to the dc current response (see, e.g., Refs.~\cite{2020_Kubo, 2020_Kubo_erratum, 2022_Kubo} and references therein).

The analytical results at \( (T, \omega_{\mathrm{ac}}) = (0, 0) \),  
given by \( I_{\mathrm{1H}}^{qqq} = -i / (2\sqrt{\pi}) \simeq -0.282i \),  
\( I_{\mathrm{1H}}^{\mathrm{Eliash}} = -i\pi\sqrt{\pi} / 16 \simeq -0.348i \), and  
\( I_{\mathrm{1H}}^{\mathrm{Higgs}} = -i\pi\sqrt{\pi} / 32 \simeq -0.174i \),  
are in good agreement with the numerical results shown in Fig.~\ref{fig2}.
Similarly, for the third-harmonic current at \( (T, \omega_{\mathrm{ac}}) = (0, 0) \),  
the analytical values \( I_{\mathrm{3H}}^{qqq} = -i / (6\sqrt{\pi}) \simeq -0.094i \) and  
\( I_{\mathrm{3H}}^{\mathrm{Higgs}} = -i\pi\sqrt{\pi} / 32 \simeq -0.174i \)  
agree well with the numerical results shown in Fig.~\ref{fig3}.

\subsection{Analytical expressions of $I_{\rm 1H}$ and $I_{\rm 3H}$ for $\omega_{\rm ac}\to 0$}\label{appendix_dc_I}

We begin by eliminating the \( \omega_{\rm ac}^{-1} \) factor from \( K_{\rm 1H} \), \( Z_{\rm 1H}^{\rm Higgs} \), \( K_{\rm 3H} \), and \( Z_{\rm 3H}^{\rm Higgs} \), using the relations
\begin{eqnarray}
&&\frac{(F_{\pm 1} - F_{\mp 3})(F_{\pm 1} + F_{\mp 3})}{2\hbar\omega_{\rm ac}\Delta}
= \frac{(G_{\pm 1} - G_{\mp 3})(G_{\pm 1} + G_{\mp 3})}{2\hbar\omega_{\rm ac}\Delta}  \nonumber \\
&&= \mp \frac{G_{\pm 1} F_{\mp3} + F_{\pm 1} G_{\mp 3} }{d_{\pm 1} d_{\mp3}} , 
\end{eqnarray}
\begin{eqnarray}
&&\frac{(F_{\pm 1} - F_{\mp 3})(G_{\pm 1} + G_{\mp 3})}{2\hbar\omega_{\rm ac}\Delta}
= \mp \frac{G_{\pm 1} G_{\mp3} + F_{\pm 1} F_{\mp 3} +1 }{d_{\pm 1} d_{\mp3}} , \nonumber \\ \\
&&\frac{(G_{\pm 1} - G_{\mp 3})(F_{\pm 1} + F_{\mp 3})}{2\hbar\omega_{\rm ac}\Delta}  
= \mp \frac{G_{\pm 1} G_{\mp3} + F_{\pm 1} F_{\mp 3} -1 }{d_{\pm 1} d_{\mp3}} . \nonumber \\
\end{eqnarray}
Here, \( d(\epsilon) = i \sqrt{\Delta^2 - (\epsilon + i\Gamma)^2} \), and we define \( d_{\pm n} = d(\epsilon \pm n\hbar \omega_{\rm ac} / 2) \) with \( n \in \mathbb{N} \).  
Taking the limit \( \omega_{\rm ac} \to 0 \) then yields 
\begin{eqnarray}
K_{{\rm 1H}, i}\Bigr|_{\omega_{\rm ac} \to 0} = 
\begin{cases}
-2i \frac{G^2 F^2}{d} \mathcal{T} &  (i=1, 2)  \\
0                                             & (i=3, 6)  \\
-4i \frac{G^2 F^2}{d} \mathcal{T} & (i=4, 5) 
\end{cases} ,
\end{eqnarray}
\begin{eqnarray}
Z^{\rm Eliash}_{{\rm 1H}, i}\Bigr|_{\omega_{\rm ac} \to 0} = 
\begin{cases}
8 \frac{G^2 F}{d} \mathcal{T} & (i=1, 2)  \\
0 & (i=3) 
\end{cases} ,
\end{eqnarray}
\begin{eqnarray}
Z^{\rm Higgs}_{{\rm 1H}, i}\Bigr|_{\omega_{\rm ac} \to 0} = 
\begin{cases}
8 \frac{G^2 F}{d} \mathcal{T}  & (i=1, 2)  \\
0 & (i=3) 
\end{cases} ,
\end{eqnarray}
\begin{eqnarray}
K_{{\rm 3H}, i}\Bigr|_{\omega_{\rm ac} \to 0} = 
\begin{cases}
-2i \frac{G^2 F^2}{d} \mathcal{T} &  (i=1, 2)  \\
0                                             & (i=3) 
\end{cases} ,
\end{eqnarray}
\begin{eqnarray}
Z^{\rm Higgs}_{{\rm 3H}, i}\Bigr|_{\omega_{\rm ac} \to 0} = 
\begin{cases}
8 \frac{G^2 F}{d} \mathcal{T} & (i=1, 2)  \\
0 & (i=3) 
\end{cases} .
\end{eqnarray}
Then, we obtain
\begin{eqnarray}
I_{\rm 1H}^{qqq} \Bigr|_{\omega_{\rm ac}\to 0} 
&=& 3 I_{\rm 3H}^{qqq} \Bigr|_{\omega_{\rm ac}\to 0} 
= -\frac{3\pi kTi}{\sqrt{\pi}} \sum_{\omega_M >0} \frac{G_M^2 F_M^2}{ d_M }  , \nonumber \\ \\
\frac{I_{\rm 1H}^{\rm Eliash}}{\Psi_{\rm 0H}} \Bigr|_{\omega_{\rm ac}\to 0} 
&=& \frac{I_{\rm 1H}^{\rm Higgs}}{\Psi_{\rm 2H}} \Bigr|_{\omega_{\rm ac}\to 0} 
= \frac{I_{\rm 3H}^{\rm Higgs}}{\Psi_{\rm 2H}} \Bigr|_{\omega_{\rm ac}\to 0} \nonumber \\
&=& \frac{4\pi kTi}{\sqrt{\pi}} \sum_{\omega_M >0} \frac{G_M^2 F_M}{d_M}  ,
\end{eqnarray}
In particular, in the zero-temperature limit \( (T \to 0) \), the Matsubara sum \( (2\pi kT/\Delta) \sum_{\omega_M > 0} \) can be replaced by an integral \( \int_0^{\infty} dw \), where \( w := \hbar \omega_M / \Delta \), and we obtain
\begin{eqnarray}
&& I_{\rm 1H}^{qqq} \biggr|_{\substack{\omega_{\rm ac}\to 0 \\ T\to0}} 
= 3 I_{\rm 3H}^{qqq} \biggr|_{\substack{\omega_{\rm ac}\to 0 \\ T\to0}} = - \frac{i}{2\sqrt{\pi}} , \\
&& \frac{I_{\rm 1H}^{\rm Eliash}}{\Psi_{\rm 0H}} \biggr|_{\substack{\omega_{\rm ac}\to 0 \\ T\to0}} 
= \frac{I_{\rm 1H}^{\rm Higgs}}{\Psi_{\rm 2H}} \biggr|_{\substack{\omega_{\rm ac}\to 0 \\ T\to0}} 
= \frac{I_{\rm 3H}^{\rm Higgs}}{\Psi_{\rm 2H}} \biggr|_{\substack{\omega_{\rm ac}\to 0 \\ T\to0}} 
= \frac{i\sqrt{\pi}}{2} , \nonumber \\
\end{eqnarray}
where $\int_0^{\infty} w^2 (w^2+1)^{-2} dw = \pi/4$ and $\int_0^{\infty} w^2 (w^2+1)^{-5/2} dw= 1/3$ are used.

\section{Effects of $\Gamma$ at lower temperatures} \label{appendix_Gamma}

\begin{figure}[tb]
   \begin{center}
   \includegraphics[width=0.49\linewidth]{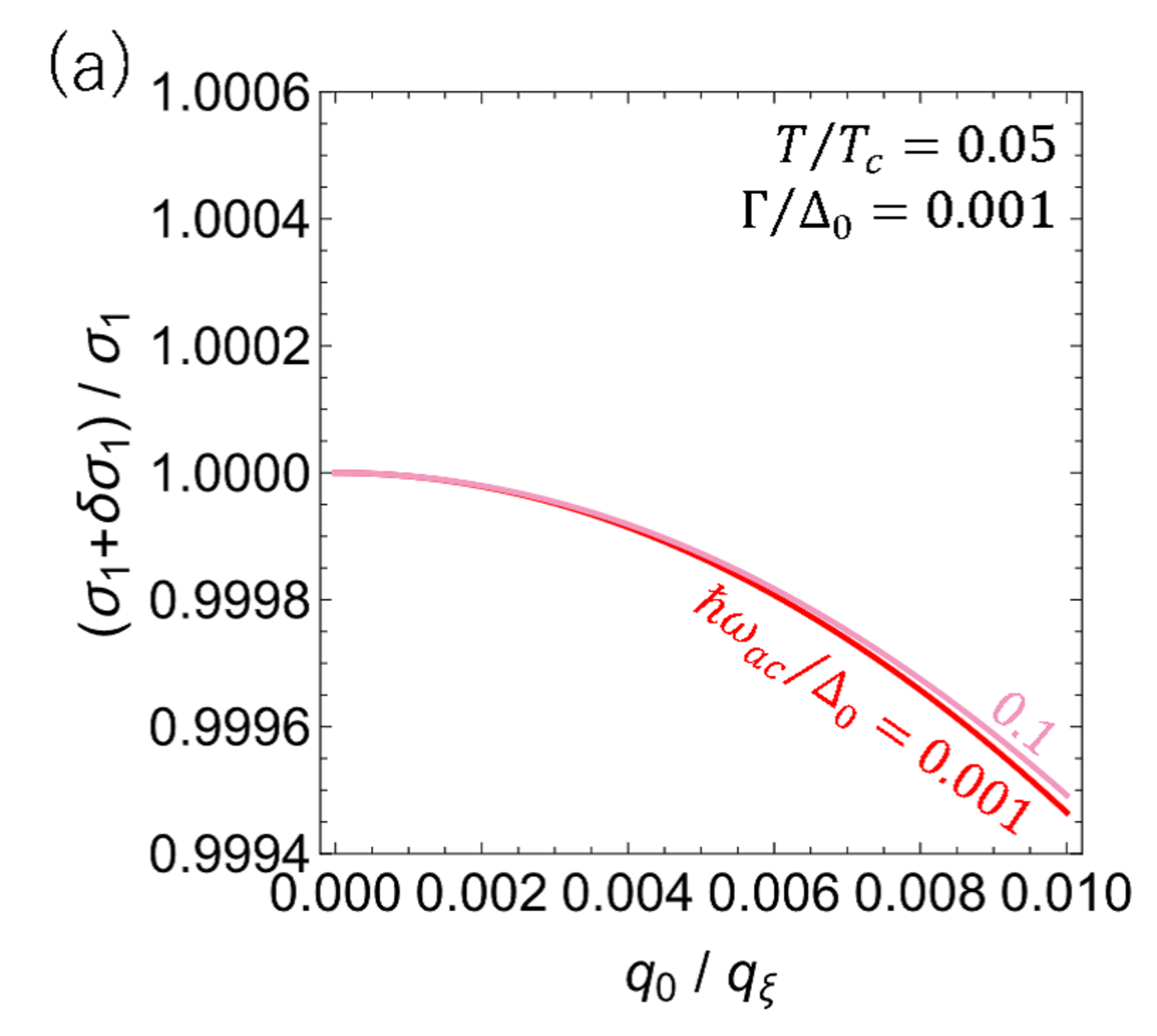}
   \includegraphics[width=0.49\linewidth]{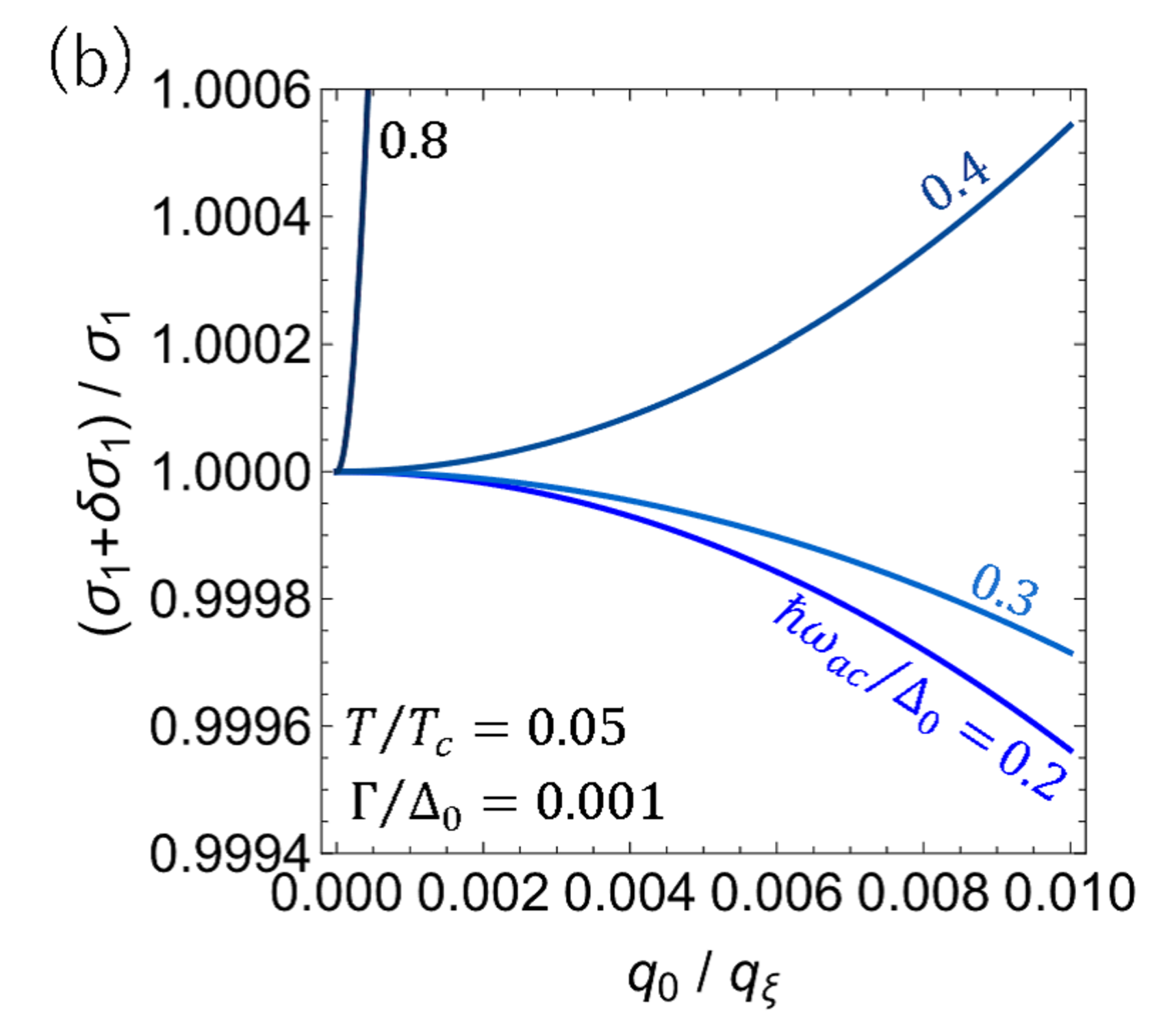}
   \end{center}\vspace{0 cm}
   \caption{
Results at a lower temperature \( T/T_c = 0.05 \).
   }\label{fig9}
\end{figure}

At lower temperatures, quasiparticles occupying \( \Gamma \)-induced subgap states begin to dominate the dissipative conductivity, even when \( \Gamma/\Delta_0 = 10^{-3} \).  
As shown in Figures~\ref{fig9}, this leads to a significant suppression of amplitude dependence in this regime.
However, the nature of these phenomenological subgap states depends on the specific value of \( \Gamma \), which differs between materials.
Moreover, subgap states in real materials are not always accurately described by a constant \( \Gamma \).  
Therefore, understanding the amplitude-dependent behavior of dissipative conductivity at low temperatures requires a more careful investigation.

\section{Nonlinear correction to the surface resistance}\label{appendix_Rs}

We express the corrections to $R_s$ in terms of the nonlinear contribution $\delta\sigma_1$.  
Because the field amplitude varies with depth due to Meissner screening, 
the nonlinear complex conductivity is likewise depth dependent.  
Accordingly, the first step is to formulate the surface impedance for a depth-dependent complex conductivity.

From Maxwell's equations one obtains $\partial_{\eta} E = i\mu_0 \omega_{\rm ac} H$ and $\partial_{\eta} H = - \sigma E$,  
where $\eta$ denotes the depth from the surface, $E$ and $H$ are the field components parallel to the surface, 
and $\sigma(\eta) = \sigma_{\infty} + \delta \sigma(\eta)$.  
The nonlinear correction $\delta \sigma$ is confined near the surface, since both the fields and the induced currents are localized there due to the Meissner screening effect.  
Defining the local surface impedance as $\zeta(\eta) = E(\eta)/H(\eta)$, one finds
$d\zeta/d\eta = i\mu_0 \omega_{\rm ac} + \sigma \zeta^2$.  
Introducing $\zeta(\eta) = Z_{\infty} + \delta \zeta(\eta)$, we arrive at
\begin{eqnarray}
\left( \frac{d}{d\eta} - \frac{2}{\Lambda} \right) \delta\zeta 
= - \mu_0^2 \omega_{\rm ac}^2 \Lambda^2 \delta \sigma .
\end{eqnarray}
Here, $\Lambda^{-2}:=-i\mu_0 \omega_{\rm ac}\sigma$. 
The solution is
\begin{eqnarray}
\delta Z_s = \delta\zeta(0)
=  \mu_0^2 \omega^2 \Lambda^2 \int_0^{\infty} d\eta \,
e^{-2\eta/\Lambda} \, \delta \sigma(\eta).
\end{eqnarray}
In the special case $\sigma_1/\sigma_2 \ll 1$, 
$\Lambda$ reduces to the penetration depth $\lambda$, and one obtains
\begin{eqnarray}
\delta R_s =  \mu_0^2 \omega^2 \lambda^2 
\int_0^{\infty} d\eta \, e^{-2\eta/\lambda} \, \delta \sigma_1(\eta). \label{DeltaRs}
\end{eqnarray}
for $\sigma_1/\sigma_2 \ll 1$.
This expression is consistent with the formulas for the surface resistance of inhomogeneous superconductors considered in Refs.~\cite{Gurevich_Kubo, Kubo_Gurevich}.

The next step is to substitute the explicit form of $\delta\sigma_1(\eta)$ into the above expression.  
Since the amplitude decays as $q_0 \propto E \propto e^{-\eta/\lambda}$, we have $\delta \sigma_1 \propto e^{-2\eta/\lambda}$, and Eq.~(\ref{DeltaRs}) reduces to
\begin{eqnarray}
\delta R_s
=  \frac{1}{2}\mu_0^2 \omega^2 \lambda^3 \sigma_n \frac{\sum_{i} {\rm Re} I_{\rm 1H}^{i}}{\sqrt{\pi} \hbar\omega/\Delta_0} \biggl( \frac{B_{\rm rf}}{B_c} \biggr)^2, 
\end{eqnarray}
where $i= qqq, {\rm Eliash}, {\rm Higgs}$. 
This is the leading order nonlinear correction to the surface resistance.

\end{document}